%% file: main.tex
\documentclass[%
 aip,
 amsmath,amssymb,
 reprint,%
]{revtex4-2}

\usepackage{graphicx}
\usepackage{dcolumn}
\usepackage{bm}

\usepackage[colorlinks, citecolor=blue, linkcolor=blue]{hyperref}
\usepackage[utf8]{inputenc}
\usepackage[T1]{fontenc}
\usepackage{mathptmx}
\usepackage{etoolbox}
\usepackage{xspace}
\usepackage{physics}
\usepackage{siunitx}
\usepackage{nicefrac}

\newcommand{\BFO}{BiFeO\textsubscript{3}\xspace}
\newcommand{\CO}{Cr\textsubscript{2}O\textsubscript{3}\xspace}
\newcommand{\hem}{$\alpha$-Fe\textsubscript{2}O\textsubscript{3}\xspace}

\makeatletter
\def\@email#1#2{%
 \endgroup
 \patchcmd{\titleblock@produce}
  {\frontmatter@RRAPformat}
  {\frontmatter@RRAPformat{\produce@RRAP{*#1\href{mailto:#2}{#2}}}\frontmatter@RRAPformat}
  {}{}
}%
\makeatother

\begin{document}

\preprint{AIP/123-QED}

\title[Quantum sensing applied to antiferromagnetism]{Single spin magnetometry and relaxometry applied to antiferromagnetic materials}
\author{Aurore Finco}
\affiliation{Laboratoire Charles Coulomb, Université de Montpellier, CNRS, 34095 Montpellier, France}
\email{aurore.finco@umontpellier.fr}
\author{Vincent Jacques}%
\affiliation{Laboratoire Charles Coulomb, Université de Montpellier, CNRS, 34095 Montpellier, France}
\email{vincent.jacques@umontpellier.fr}

\date{\today}

\begin{abstract}
  Despite the considerable interest for antiferromagnets which appeared with the perspective of using them for spintronics, their experimental study, including the imaging of antiferromagnetic textures, remains a challenge. To address this issue, quantum sensors, and in particular the nitrogen-vacancy (NV) defects in diamond have become a widespread technical solution. We review here the recent applications of single NV centers to study a large variety of antiferromagnetic materials, from quantitative imaging of antiferromagnetic domains and non-collinear states, to the detection of spin waves confined in antiferromagnetic textures and the non-perturbative measurement of spin transport properties. We conclude with recent developments improving further the magnetic sensitivity of scanning NV microscopy, opening the way to detailed investigations of the internal texture of antiferromagnetic objects.
\end{abstract}

\maketitle

\section{Introduction}

Imaging the antiferromagnetic order with nanoscale spatial resolution is a notoriously difficult task~\cite{cheongSeeingBelievingVisualization2020} but it is also a general requirement to understand the physics of antiferromagnetic materials and to use them for spintronics~\cite{jungwirthAntiferromagneticSpintronics2016,
  baltzAntiferromagneticSpintronics2018}. Two approaches can be followed for magnetic imaging: probing directly the magnetization direction, or probing the magnetic stray field generated by the material. Since this stray field is very weak in antiferromagnets, most of the available imaging techniques measure the magnetization direction. Optical techniques rely on birefringence~\cite{roth1960neutron}, nonreciprocal reflection~\cite{krichevtsov1993spontaneous}, nonlinear processes like second harmonic generation (SHG)~\cite{fiebigDomainTopographyAntiferromagnetic1995} or nonreciprocal directional dichroism~\cite{figotin2001nonreciprocal}. The spatial resolution of these optical methods is however limited by diffraction at the micron scale. Synchrotron X-ray-based techniques like PhotoElectron Emission Microsocopy combined with X-ray Magnetic Linear Dichroism (XMLD-PEEM) are very powerful to reach nanoscale resolution~\cite{chmiel2018observation}, but remain scarcely employed owing to the required large scale experimental facilities. Scanning probe microscopies like spin-polarized scanning tunneling microscopy~\cite{bodeAtomicSpinStructure2006} or magnetic exchange force microscopy~\cite{kaiserMagneticExchangeForce2007} offer an ultimate atomic resolution for conducting and insulating materials, respectively. Nevertheless, these techniques are only suited for the study of nearly perfect surfaces in ultrahigh vacuum, thus precluding the study of technologically-relevant antiferromagnetic materials.

In this review, we will present another technique which uses the second path, probing the weak stray field coming from antiferromagnetic textures, with the help of a quantum sensor: the nitrogen-vacancy (NV) center in diamond. Single NV centers can be used to detect static magnetic fields with a sensitivity in the \si{\micro\tesla} range at a distance of a fews tens of nanometers from the sample when integrated in a scanning probe microscope. In the last years, this technique was employed to study many phenomena in condensed matter physics~\cite{rondinMagnetometryNitrogenvacancyDefects2014,
  casolaProbingCondensedMatter2018, xuRecentAdvancesApplications2023}, including non-collinear magnetic states in ferromagnets~\cite{rondinStrayfieldImagingMagnetic2013, tetienneNanoscaleImagingControl2014, tetienneNatureDomainWalls2015, dovzhenkoMagnetostaticTwistsRoomtemperature2018, grossSkyrmionMorphologyUltrathin2018}, superconducting vortices~\cite{pelliccioneScannedProbeImaging2016, thielQuantitativeNanoscaleVortex2016} or spin waves~\cite{wolfeOffresonantManipulationSpins2014, duControlLocalMeasurement2017, vandersarNanometrescaleProbingSpin2015}.
In this review, we will focus only the applications of sensing with single NV centers to antiferromagnetic materials, from the imaging of complex magnetic textures~\cite{grossRealspaceImagingNoncollinear2017, kosubPurelyAntiferromagneticMagnetoelectric2017} to the investigation of intrinsic spin transport phenomena through magnetic fluctuations~\cite{flebusQuantumImpurityRelaxometryMagnetization2018, wangNoninvasiveMeasurementsSpin2022}.
We will start with an explanation of the principle of magnetic field measurements with NV centers and some technical details about diamond probes in section~\ref{sec:NV}, before rewieving applications of scanning NV magnetometry to the quantitative imaging of antiferromagnetic textures in section~\ref{sec:quantitative}. We will then focus on relaxometry experiments, which rely on the detection of magnetic fluctuations, in section~\ref{sec:relaxo}. To conclude, we will highlight recent developments which provide significant improvement of the magnetic sensitivity through the conversion of field gradients into oscillating signals in section~\ref{sec:gradio}.    

\section{\label{sec:NV} Probing magnetism with NV centers}

\input{nv.tex}

\section{\label{sec:quantitative} Quantitative  imaging of antiferromagnets with scanning NV microscopy}

\input{quantitative.tex}

\section{\label{sec:relaxo} Spin waves and spin transport studied in antiferromagnets with NV center relaxometry}

\input{relaxometry.tex}

\section{\label{sec:gradio} Improving the measurement sensitivity with pulsed techniques}

\input{gradiometry.tex}

\section{Conclusion}

Within less than a decade, single NV centers have become an essential tool for probing antiferromagnetism at the nanoscale. Owing to its high magnetic sensitivity and its operation under ambiant conditions, scanning NV center magnetometry is now routinely being used to image the magnetic state of many antiferromagnetic materials, from oxides like \CO~\cite{kosubPurelyAntiferromagneticMagnetoelectric2017, appelNanomagnetismMagnetoelectricGranular2019, hedrichNanoscaleMechanicsAntiferromagnetic2021, wornleCoexistenceBlochEel2021, ericksonNanoscaleImagingAntiferromagnetic2022, makushkoFlexomagnetismVerticallyGraded2022}, \hem~\cite{guoCurrentinducedSwitchingThin2023, tanRevealingEmergentMagnetic2023, welterFastScanningNitrogenvacancy2022}, \BFO~\cite{grossRealspaceImagingNoncollinear2017, chauleauElectricAntiferromagneticChiral2020, haykalAntiferromagneticTexturesBiFeO2020, zhongQuantitativeImagingExotic2022, fincoImagingTopologicalDefects2022} to synthetic~\cite{fincoImagingNoncollinearAntiferromagnetic2021}, non-collinear~\cite{yanQuantumSensingImaging2022, liNanoscaleMagneticDomains2023} or van der Waals~\cite{thielProbingMagnetism2D2019} antiferromagnets. We have reviewed here the main requirements to perform these measurements and analyze the obtained data, highlighting the need to carefully calibrate the NV sensor in order to extract quantitative information about the magnetization configuration.

Recent developments relying on the detection of spin fluctuations open further perspectives to measure non-invasively spin waves~\cite{fincoImagingNoncollinearAntiferromagnetic2021} and spin transport properties~\cite{wangNoninvasiveMeasurementsSpin2022}, suggesting that NV centers will also be of great help in the future to address antiferromagnetic magnonics.
Finally, integrating AC sensing inside scanning NV center microscopy~\cite{huxterScanningGradiometrySingle2022} leads to a significant improvement of the magnetic sensitivity, which will enable the imaging of perfectly compensated antiferromagnetic textures, so far inaccessible. This approach also opens the way to scanning NV center electrometry~\cite{huxterImagingFerroelectricDomains2023a} for the imaging of ferroelectric states and further beyond to combined measurements of antiferromagnetic and ferroelectric textures at the nanoscale in multiferroics.

\section*{Acknowledgments}

We are thankful for support from the French Agence Nationale de la Recherche (ANR) through the project TATOO (ANR-21-CE09-0033-01) and the European Union’s Horizon 2020 research and innovation programme under Grant Agreements No. 964931 (TSAR) and No. 866267 (EXAFONIS).

\section*{References}

%

\end{document}

%% file: nv.tex
\subsection{\label{sec:meas_principle}Principle of the magnetic field measurement}

\begin{figure}[h]
  \centering
  \includegraphics{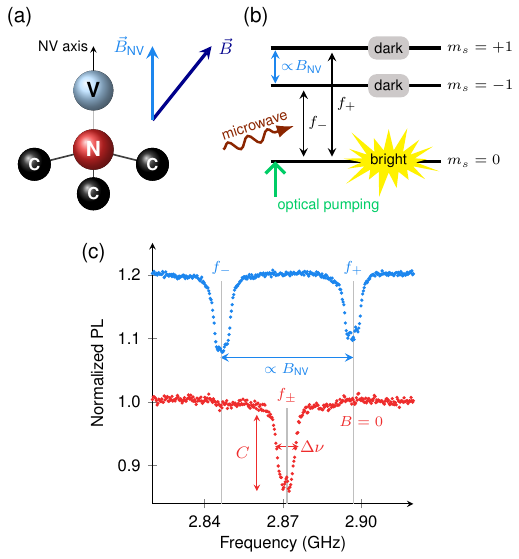}
  \caption{(a) Sketch defining the structure of the NV center and its quantification axis. (b) Schematic representation of the spin sublevels in the ground state of the NV center, with spin $S=1$. (c) Optically detected magnetic resonance spectra measured on a single NV center in an all-diamond probe, with and without the presence of an external magnetic field. The top curve has been vertically shifted for readability.
  }
  \label{fig:NV}
\end{figure}

NV centers are point defects in the carbon lattice of diamond made of a substitutional nitrogen atom next to a vacancy (Fig.~\ref{fig:NV}(a)). In their negatively-charged state NV$^-$ (which will be the only charge state considered here), they exhibit a spin triplet ground state which can be initialized by optical pumping, manipulated with long coherence times, and read out optically even at room temperature. These properties make them powerful quantum sensors~\cite{degenQuantumSensing2017}. The direction defined by the positions of the nitrogen atom and the vacancy defines a quantification axis for the electronic spin, which is designated as the NV axis (see Fig.~\ref{fig:NV}(a)). This axis can be oriented along 4 different directions, corresponding to the $[111]$ crystallographic direction in the diamond lattice. NV-based sensing can be performed using ensembles of NV centers~\cite{acostaDiamondsHighDensity2009}, which are thus randomly oriented along the 4 equivalent directions, or single NV defects. In this review, we will focus only on the use of single NV spins for sensing the properties of antiferromagnets, and we will write as $m_s$ the projection of the spin along the NV axis.

Under optical illumination, NV centers produce a stable spin-dependent photoluminescence which is usually detected using a confocal microscope~\cite{gruberScanningConfocalOptical1997}. A stronger photoluminescence is coming from the system in the state $m_s=0$ (bright state) than from the states $m_s = \pm 1$ (dark states)~\cite{rondinMagnetometryNitrogenvacancyDefects2014} (see Fig.~\ref{fig:NV}(b)). In addition, under continuous green laser illumination, optical pumping polarizes NV centers in the bright state $m_s=0$.  From these two properties, it is possible to measure optically the position of the electron spin resonance (ESR) frequencies of NV centers: continuous optical excitation polarizes the spin in the bright $m_s=0$ state, and the application of a resonant microwave field drives the transition to the darker $m_s=\pm1$ states. As a result, one gets the Optically Detected Magnetic Resonance (ODMR) spectrum presented in Fig.~\ref{fig:NV}(c). In the absence of external magnetic field, the states $m_s=1$ and $m_s=-1$ are degenerate, and a single dip is found in the spectrum. However, the internal electric field in the diamond usually introduces a small splitting~\cite{mittigaImagingLocalCharge2018}, as visible in the red curve in Fig.~\ref{fig:NV}(c). 

When a magnetic field is applied to the NV center, Zeeman effect splits the states $m_s=-1$ and $m_s=+1$, and two dips at $f_-$ and $f_+$ are present in the spectrum. In the low field regime ($B<\SI{5}{\milli\tesla}$), which is going to be the relevant one to study antiferromagnets, this splitting depends linearly on the $B_\text{NV}$ component of the field along the NV axis~\cite{rondinMagnetometryNitrogenvacancyDefects2014}.

Measurements are usually performed under the application of a small bias magnetic field aligned along the NV axis for two reasons. The first one is that otherwise, the sign of the measured magnetic field cannot be determined. The second reason is related to the electric-field-induced splitting usually measured at zero field~\cite{mittigaImagingLocalCharge2018}. This effect prevents the detection of very small field variations around $B=0$ and therefore deteriorates the sensitivity of the measurement~\cite{rondinMagnetometryNitrogenvacancyDefects2014}. Applying a bias field allows us to rather work in a regime for which the Zeeman shift evolves linearly with the magnetic field. In this regime, the magnetic field sensitivity $\eta_B$ of the measurement can be expressed as~\cite{rondinMagnetometryNitrogenvacancyDefects2014,barrySensitivityOptimizationNVdiamond2020}:
\begin{equation}
  \label{eq:sensi}
  \eta_B \simeq \frac{h}{g \mu_B}\frac{\Delta \nu}{C\sqrt{R}}
\end{equation}
where $h$ is the Planck constant, $g \simeq 2$ is the Landé $g$-factor, $\mu_B$ is the Bohr magneton, $C$ is the ODMR contrast, $\Delta \nu$ the linewidth of the resonance (see Fig.~\ref{fig:NV}(c)) and $R$ is the photoluminescence rate of the NV center polarized in the bright state. Under continuous laser and microwave illumination, the achieved sensitivity is usually on the order of a few $\si{\micro\tesla\per\sqrt\Hz}$, which is sufficient to detect the stray field produced by many antiferromagnets, as discussed in section~\ref{sec:quantitative}.

In order to combine this high magnetic field sensitivity with nanoscale spatial resolution, a single NV center can be integrated at the apex of the tip of a scanning probe microscope~\cite{chernobrodSpinMicroscopeBased2004, balasubramanianNanoscaleImagingMagnetometry2008}. An ODMR spectrum is then recorded at each pixel of the scan in order to extract the value of the local magnetic field. In the linear low field regime, measuring the Zeeman shift of a single magnetic resonance is sufficient. Historically, scanning NV microscopy was achieved by attaching a nanodiamond hosting a single NV center on a standard atomic force microscopy (AFM) tip~\cite{balasubramanianNanoscaleImagingMagnetometry2008, rondinNanoscaleMagneticField2012}. However, this method is rather difficult to implement and only a low photoluminescence signal from the NV center can be collected. To circumvent these drawbacks, all-diamond probes have been developed~\cite{maletinskyRobustScanningDiamond2012, appelFabricationAllDiamond2016, hedrichParabolicDiamondScanning2020}, providing reliable and now commercially available probes for scanning NV center magnetometry.

\subsection{Probes for scanning NV microscopy}

\subsubsection{All-diamond probes}

\begin{figure}[h]
  \centering
  \includegraphics{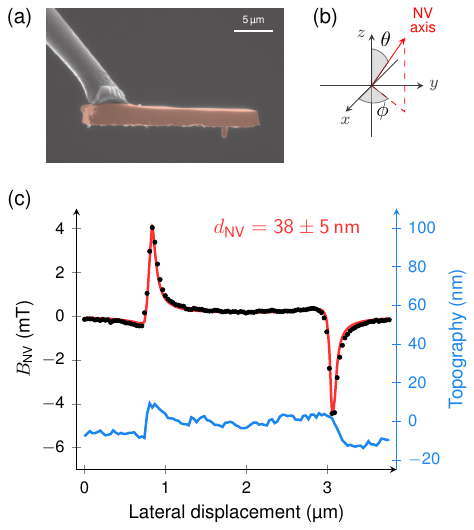}
  \caption{(a) Scanning electron microscopy image of an all-diamond probe attached at the end of a quartz capillary tip. Adapted with permission from ref.~\onlinecite{maletinskyRobustScanningDiamond2012}. (b) Sketch defining the polar and azimuthal angles $\theta$ and $\phi$ of the NV center quantification axis. (c) Stray field line profile measured across a \SI{2}{\micro\meter}-wide ferromagnetic stripe made of Ta/CoFeB(\SI{1}{\nano\meter})/MgO using a commercial all-diamond probe. This data is used to extract $d_\text{NV}$ from a fit to the analytical expression of the stray field generated at the edges of an out-of-plane magnetized stripe.}
  \label{fig:tips}
\end{figure}

All-diamond probes for scanning NV center microscopy have been introduced about ten years ago~\cite{maletinskyRobustScanningDiamond2012} to overcome the limitations of the use of nanodiamonds attached to AFM tips. In particular, NV centers in nanodiamonds have short spin coherent time because of the low quality of the diamond matrix and the vicinity to the diamond surface. In addition, the collection of the photoluminescence emitted by NV centers in nanodiamonds is inefficient, yielding a small signal and thus reducing the magnetic field sensitivity (see Eq.~(\ref{eq:sensi})). Finally, nanodiamonds can easily detach from the AFM tips, making the probes mechanically fragile. The solution which was proposed is to attach a diamond cantilever to a quartz capillary, which is itself glued to a tuning fork used as the AFM probe. The diamond cantilever possesses a small pillar, with a diameter on the range of \SI{100}{\nano\meter} at the end of which a single NV center is implanted (see Fig.~\ref{fig:tips}(a)). The bottom of the diamond pillar is then placed very close to the sample, to perform an AFM scan. These probes are fabricated from high purity diamond in order to ensure good spin coherence properties, and the pillar guides the light emitted by the NV center in order to improve the collection efficiency~\cite{appelFabricationAllDiamond2016}. Recently, further improvement have been made with the design of parabolic pillars, to increase even more the light collection efficiency~\cite{hedrichParabolicDiamondScanning2020}.

\subsubsection{Orientation of the NV center}

As mentioned in the previous section, the orientation of the NV center is a very important parameter to determine before doing any measurement, as only the component of the magnetic field along the NV axis is measured. The direction of the NV axis can be described with its polar axis $\theta$ and its azimuthal angle $\phi$, as defined in Fig.~\ref{fig:tips}(b). These angles can be measured independently by applying a calibrated magnetic field of fixed intensity but variable orientation to the NV center and monitoring the ODMR spectrum~\cite{rondinMagnetometryNitrogenvacancyDefects2014}.

Diamond probes are usually made from (001) diamond, and as a result, NV centers are pointing along a tilted direction, with $\theta \sim \ang{54}$ from the sample normal. A difficulty arises from the fact that the bias field applied during the measurement should be as much as possible aligned with the NV center axis, to avoid electron spin mixing. Such a mixing of the spin states would indeed reduce both the photoluminescence signal and the ESR contrast~\cite{tetienneMagneticfielddependentPhotodynamicsSingle2012}. Aligning the field along this tilted direction can be experimentally challenging and even lead to unexpected effects, as such a tilted magnetic field can perturb the sample behavior~\cite{akhtarCurrentInducedNucleationDynamics2019}. To solve this issue, recent developments have been made to fabricate diamond probes from (111)~\cite{rohner111OrientedSingle2019} and (110)~\cite{welterScanningNitrogenvacancyCenter2022a} diamond, providing NV centers oriented either along the normal of the sample surface or in-plane respectively.

\subsubsection{\label{sec:nv_calib} Calibration of the distance between the NV center and the surface}

Finally, as we will show in detail in section~\ref{sec:quantitative}, knowing the distance $d_\text{NV}$ between the NV sensor and the sample surface is crucial to quantitatively analyze experimental data. In addition, in a scanning NV microscopy experiment, $d_\text{NV}$ is the parameter which sets the spatial resolution. When studying antiferromagnetic textures, it is desirable to use diamond tips with NV-to-sample distances as small as possible, to be able to distinguish nanoscale features and to probe larger fields, as the stray field amplitude decreases fast with the distance to the sensor. Using commercial all-diamond probes, $d_\text{NV}$ values of \SI{50}{\nano\meter} can routinely be achieved.

The parameter $d_\text{NV}$ can be obtained by adding it as a fitting parameter during the analysis, but the most effective way to determine $d_\text{NV}$ is to calibrate it beforehand using a well-known reference sample. The most standard approach for this is presented in ref.~\onlinecite{hingantMeasuringMagneticMoment2015} and illustrated in Fig.~\ref{fig:tips}(c). Here, the calibration sample is a uniformly magnetized ferromagnetic CoFeB stripe of thickness $t=\SI{1}{\nano\meter}$ and width $l=\SI{2}{\micro\meter}$. The stray field produced at the edges of the stripe can be expressed analytically, and in the thin film limit $t \ll d_\text{NV}$, for an edge along the $y$ axis and placed at $x=0$, we can write the field $\vec{B}^\text{edge}$ across the stripe as:
\begin{equation}
  \label{eq:calib}
   \left \lbrace
   \begin{split}
     B_x^\text{edge}(x) & =  \frac{\mu_0 M_s  t}{2 \pi} \ \frac{d}{x^2+d^2}\\
     B_y^\text{edge}(x) & = 0 \\
     B_z^\text{edge}(x) & =  - \frac{\mu_0  M_s  t}{2 \pi} \ \frac{x}{x^2+d^2} \\
   \end{split}
   \right.
 \end{equation}
 where $M_s$ is the saturation magnetization of the ferromagnetic layer and $d$ the distance between the sensor and the sample surface. Using the knowledge of the angles $\theta$ and $\phi$ describing the orientation of the NV center, one can fit a measured line profile across the stripe to $B_\text{NV}^\text{edge}(x)-B_\text{NV}^\text{edge}(x+l)$, leaving only $d$ and $ M_s$ as free parameters. However, the topography of the sample has to be included in the calculation, as the NV center might not be located in the center of the diamond pillar and would therefore probe differently both edges. This would result in an asymmetric shape of the profile. In order to take this into account, $d$ has to be replaced by $d_\text{NV} + \text{topo}(x)$ during the fit, where $\text{topo}(x)$ is the measured topography with an offset such that $\text{topo}(x) = 0$ when the tip is located above the magnetic stripe~\cite{hingantMeasuringMagneticMoment2015}. Such a topography measurement is shown as the blue curve in the plot of Fig.~\ref{fig:tips}(c). 
 
This calibration method usually yields a typical uncertainty of about 10\% on the parameter $d_\text{NV}$. This uncertainty on $d_\text{NV}$ is the major source of error during the quantitative analysis of images, like those described in the next section. 


%% file: quantitative.tex
\subsection{Imaging the stray field from uncompensated antiferromagnetic textures}

Scanning NV microscopy offers the possibility to measure quantitatively magnetic fields with a nanoscale spatial resolution and thus to obtain maps of the stray field produced by various types of magnetic textures. Although a uniform and perfectly compensated antiferromagnet -- similarly to a completely uniform ferromagnet -- does not produce magnetic stray field, the presence of non-collinear textures generates a weak stray field.
In addition, it is not rare that antiferromagnetic samples exhibit an uncompensated magnetic moment, also leading to a stray field at edges or above non-collinear magnetic objects. Such small fields can be probed using standard NV magnetometry procedures as presented in section~\ref{sec:meas_principle}. The uncompensated moment can result either from a A-type antiferromagnetic state, like for example in \CO~\cite{heRobustIsothermalElectric2010a}, or from a canting of the magnetic moments induced by the presence of a Dzyaloshinskii-Moriya interaction (DMI) like in \hem~\cite{dzyaloshinskiiThermodynamicTheoryWeak1958}, \BFO~\cite{ramazanogluLocalWeakFerromagnetism2011} or Mn\textsubscript{3}Sn~\cite{tomiyoshiMagneticStructureWeak1982}.

\subsubsection{Antiferromagnets with uncompensated surface magnetic moments}

In A-type antiferromagnets, when the magnetic moments are arranged ferromagnetically in each plane parallel to the surface, the last layer of magnetic moments can be uncompensated and therefore result in the presence of a detectable stray field a few tens of nanometers away from the surface. With NV center magnetometry, the most studied material of this type so far is \CO, but similar effects are found in other compounds, and in particular bidimensional van der Waals magnets like CrI\textsubscript{3}~\cite{thielProbingMagnetism2D2019}. 

\CO is a magnetoelectric antiferromagnet, therefore attracting attention for potential applications like the design of electrically controlled memory elements~\cite{kosubPurelyAntiferromagneticMagnetoelectric2017}. In (0001) oriented \CO, below the ordering temperature $T_\text{Néel} = \SI{308}{\K}$, the symmetry breaking at the surface leads to a surface magnetic moment with a direction linked to the Néel vector below, enabling the NV imaging of the antiferromagnetic domains through the resulting stray field~\cite{appelNanomagnetismMagnetoelectricGranular2019, hedrichNanoscaleMechanicsAntiferromagnetic2021, wornleCoexistenceBlochEel2021, ericksonNanoscaleImagingAntiferromagnetic2022, makushkoFlexomagnetismVerticallyGraded2022}. This uncompensated surface magnetic moment is normal to the surface plane, similarly to the Néel vector.

The morphology of the antiferromagnetic domains in this material depends strongly on the nature of the sample: granular thin films exhibit irregular domains with a size of a few hundreds of nanometers whereas single crystals host large micron size domains. The first magnetic images of \CO were obtained on the latter in 1995 by second harmonic generation (SHG)~\cite{fiebigDomainTopographyAntiferromagnetic1995}. Besides scanning NV magnetometry, scanning probe techniques like Magnetic Force Microscopy (MFM)~\cite{wuImagingControlSurface2011} and Magnetoelectric Force Microscopy (MeFM)~\cite{schoenherrMagnetoelectricForceMicroscopy2017} have been used to image \CO thin films and single crystals respectively, as well as SHG~\cite{fiebigDomainTopographyAntiferromagnetic1995}, other optical techniques~\cite{hayashidaObservationAntiferromagneticDomains2022} and PhotoElectron Emission Microsocopy combined with X-ray Magnetic Circular Dichroism (XMCD-PEEM)~\cite{wuImagingControlSurface2011}. However, none of these techniques can provide both the nanoscale spatial resolution and high sensitivity offered by scanning NV center magnetometry, together with the ability to perform quantitative measurements.

Many quantitative information can be extracted from scanning NV center microscopy experiments on \CO. In thin films, the stray field maps allow the determination of the size of the antiferromagnetic domains, demonstrating the presence of intergranular coupling as the domains always comprise several structural grains~\cite{appelNanomagnetismMagnetoelectricGranular2019}. In addition, looking at strained films revealed an unconventional flexomagnetic effect, as the strain gradient induces a modification of the Néel temperature~\cite{makushkoFlexomagnetismVerticallyGraded2022}. This has been measured through the evolution of the stray field above the films at various temperatures across the transition. The surface magnetic moment density pattern can also be retrieved from stray field maps using Fourier reverse propagation methods~\cite{broadwayImprovedCurrentDensity2020a}. This analysis technique is explained in section~\ref{sssec:backprop}.

\begin{figure*}
  \centering
  \includegraphics[scale=1.7]{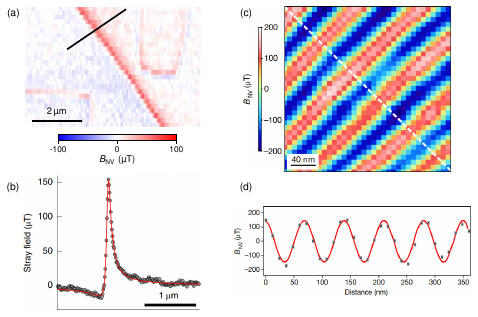}
  \caption{(a) Quantitative stray field map of a domain wall in a \CO single crystal. (b) Fit of the line profile extracted from (a) across the domain wall to the expected stray field profile of a Bloch wall. Adapted with permission from ref.~\onlinecite{hedrichNanoscaleMechanicsAntiferromagnetic2021}. (c) Quantitative stray field map from the antiferromagnetic cycloid in a \BFO thin film, showing the \SI{64}{\nano\meter} period of the cycloidal modulation. (d) Fit of the line profile marked with the white dashed line in (c) to an analytic model of the stray field produced by the spin density wave attached to the cycloid. Adapted from ref.~\onlinecite{grossRealspaceImagingNoncollinear2017}. }
   \label{fig:quanti}
 \end{figure*}

In the case of single crystals, the images are very different as only isolated domain walls separating micron size domains are present (see Fig~\ref{fig:quanti}(a)). This situation offers the opportunity to study the domain walls themselves, as it can be done in ferromagnets~\cite{tetienneNanoscaleImagingControl2014, tetienneNatureDomainWalls2015}, in particular the path they follow~\cite{hedrichNanoscaleMechanicsAntiferromagnetic2021} and it gives hints about their morphology~\cite{hedrichNanoscaleMechanicsAntiferromagnetic2021, wornleCoexistenceBlochEel2021}. Establishing if the domain walls are of Bloch or Néel type and determining their width is relevant for fundamental reasons, as it gives insight into the balance between the different magnetic energy contributions in the material, but also in the view of applications, since Bloch and Néel walls do not exhibit the same response to current-induced spin torques. This information could be extracted from stray field profiles measured across the domain walls, by fitting the experimental NV magnetometry data to analytical expressions of the stray field, as shown in Fig.~\ref{fig:quanti}~(b). In \CO, it requires however a very good accuracy of the measurement, as Bloch and Néel profiles are hardly distinguishable~\cite{hedrichNanoscaleMechanicsAntiferromagnetic2021, wornleCoexistenceBlochEel2021}. A detailed discussion about the quantitative analysis of stray field maps from analytical models can be found below, in section~\ref{sssec:model}.

 Synthetic antiferromagnets (SAF)~\cite{duineSyntheticAntiferromagneticSpintronics2018}, although they possess a fully compensated magnetic moment, can also be probed similarly, as some magnetic stray field is present above the sample surface. This happens because the distance between each ferromagnetic layer and the NV sensor is different. As a result, even if the total magnetic moment in the sample cancels, the stray fields from the different layers at the sensor position do not, enabling scanning NV microscopy imaging~\cite{fincoImagingNoncollinearAntiferromagnetic2021} while MFM experiments are challenging for a two-layers SAF~\cite{legrandRoomtemperatureStabilizationAntiferromagnetic2020}.

\subsubsection{Canted antiferromagnets}

Another mechanism leading to the presence of uncompensated magnetic moments in antiferromagnets is a DMI-induced canting, which is the origin of the \emph{weak ferromagnetism} observed in many antiferromagnetic oxides like \hem~\cite{dzyaloshinskiiThermodynamicTheoryWeak1958} or \BFO~\cite{ramazanogluLocalWeakFerromagnetism2011}. We will focus here on \BFO, which is the first canted antiferromagnet that has been studied with scanning NV center magnetometry~\cite{grossRealspaceImagingNoncollinear2017}, but other materials can be imaged using NV centers (\hem~\cite{guoCurrentinducedSwitchingThin2023, tanRevealingEmergentMagnetic2023, welterFastScanningNitrogenvacancy2022}, Mn\textsubscript{3}Sn~\cite{yanQuantumSensingImaging2022, liNanoscaleMagneticDomains2023}, etc.). In particular, very recent works demonstrated direct observation of electrically-driven domain switching in \hem~\cite{guoCurrentinducedSwitchingThin2023} and Mn\textsubscript{3}Sn~\cite{liNanoscaleMagneticDomains2023} using quantitative stray field maps measured with scanning NV magnetometry.

\BFO is multiferroic at room temperature and exhibits a G-type antiferromagnetic order with a cycloidal incommensurate modulation which is coupled to the ferroelectric polarization $\vec{P}$~\cite{catalanPhysicsApplicationsBismuth2009}. In bulk \BFO, the period of the cycloid is about \SI{64}{\nano\meter} and its propagation vector lies in the plane perpendicular to $\vec{P}$. However, this cycloidal modulation alone does not result in the creation of an uncompensated magnetic moment. The weak ferromagnetism in \BFO comes from an additional DMI-induced effect which tilts the magnetic moments out of the cycloid plane~\cite{ramazanogluLocalWeakFerromagnetism2011}. This canting can be described as a spin density wave which is locked onto the cycloid and therefore provides a way to image it owing to the stray field that it generates~\cite{fincoImagingTopologicalDefects2022}. Scanning NV center magnetometry is so far the only technique capable of imaging the cycloid in \BFO in real space, which is shown in Fig~\ref{fig:quanti}(c).

\BFO thin films are under intense study, as multiferroic materials at room temperature are especially rare. Imaging the cycloid directly with NV magnetometry lead to a clear demonstration that its propagation direction is coupled to $\vec{P}$ and can therefore be manipulated electrically~\cite{grossRealspaceImagingNoncollinear2017}. Further investigations about the effect of epitaxial strain on \BFO thin films helped to refine the phase diagram of \BFO under strain and confirmed that an exotic cycloid type is stabilized when the strain strengthens~\cite{haykalAntiferromagneticTexturesBiFeO2020, zhongQuantitativeImagingExotic2022}. Increasing the strain further destroys the cycloid and results in antiferromagnetic domains seen in SHG~\cite{chauleauMultistimuliManipulationAntiferromagnetic2017}, which also possess a small DMI-induced uncompensated moment and could thus be observed as well with scanning NV center magnetometry~\cite{haykalAntiferromagneticTexturesBiFeO2020}.

In addition, the ability to image directly the non-collinear magnetic texture of \BFO at the nanoscale allowed the discovery of exotic antiferromagnetic objects. At chiral ferroelectric domain walls between regions were the cycloid propagates in different directions, some magnetic hotspots appear in the stitching of the cycloidal pattern~\cite{chauleauElectricAntiferromagneticChiral2020}. These can be seen as the embryos of topological multi-$q$ states. In bulk \BFO single crystals, it is also possible to observe the antiferromagnetic cycloid and scanning NV center magnetometry revealed that, at the crystal surface, the direction of its wavevector rotates almost freely in the plane perpendicular to $\vec{P}$. As a result, some defects form in the cycloidal structure, which can be described as topological defects typically found in lamellar ordered media~\cite{fincoImagingTopologicalDefects2022}.

Finally, a quantitative analysis of the stray field maps can also be performed when studying the antiferromagnetic cycloid in \BFO, since the field originating from the cycloid and the spin density wave can be analytically computed. Owing to this type of analysis, one can in particular quantitatively extract the amplitude of the spin density wave. This topic is further discussed in the next section.

\subsection{Interpretation of the stray field maps}

Scanning NV center magnetometry provides a quantitative map of the magnetic field component along the NV axis (see Fig.~\ref{fig:NV}), at a fixed distance $d_\text{NV}$ from the sample surface. In order to extract relevant information like the magnetization intensity from these data, a careful calibration of the sensor is required. First, the orientation $(\theta, \phi)$ of the NV axis should be determined, ideally using a vector magnet to sweep the direction of an external magnetic field of fixed intensity while monitoring the ODMR spectrum. Then the distance between the NV center and the surface $d_\text{NV}$ has to be measured, usually by recording a stray field profile over a well-known sample like a ferromagnetic stripe~\cite{hingantMeasuringMagneticMoment2015}. However, such a calibration is not sufficient to derive unambiguously the magnetization distribution at the source of the measured stray field since there is no bijective correspondance between them. Interpreting the stray field map thus requires to make some assumptions about the magnetization configuration. Two approaches can be followed to analyze the data: comparing them to a model or reconstructing the magnetization pattern using reverse propagation methods~\cite{broadwayImprovedCurrentDensity2020a}.

\subsubsection{Comparison with a model}
\label{sssec:model}

When comparing stray field data with a model, there are two options. First, if a specific magnetic configuration is expected or if the magnetic system is known well enough to perform micromagnetic or atomistic simulations, a magnetic state can be predicted. From this prediction, it is then possible to compute the corresponding field map at a distance $d_\text{NV}$ from the surface and project it along the direction of the NV axis. The resulting image can be directly compared with the experimental data, as it was done to identify the chiral magnetic objects forming in the cycloid stitching at ferroelectric domain walls in \BFO~\cite{chauleauElectricAntiferromagneticChiral2020}. This is the most generic approach but the agreement between the model and the data will often be rather qualitative.

The other option is to use an analytical model of the stray field produced by the magnetic object of interest and fit it to the experiment in order to extract relevant parameters like the magnetization amplitude or the size of the object. This is the most accurate method to retrieve quantitative information from scanning NV center magnetometry measurements but it is limited to specific ``simple'' cases like step edges, domain walls between out-of-plane oriented domains or homogeneous spirals or spin density waves which are described with a sine function. We will detail two examples below: a domain wall in a \CO single crystal and the cycloid in a low strained \BFO thin film (see Fig.~\ref{fig:quanti}).

To model a Bloch domain wall in \CO, one can use the generic description of the evolution of the Néel vector $\vec{L}$ across such a wall parallel to the $y$ axis~\cite{belashchenkomagnetoelectric2016}:
\begin{equation}
  \label{eq:dw_cr2o3}
  L_x = 0, \ L_y = \sech\left(\frac{x}{w}\right), \ L_z = \tanh\left(\frac{x}{w}\right)
\end{equation}
where the wall width is $2w$. As discussed before, in \CO the uncompensated moment $\vec{m}$ follows $\vec{L}$. From this magnetization profile, the resulting stray field is obtained by going to Fourier space:
\begin{equation}
  \label{eq:B_cr2o3}
  \vec{B}(\vec{q}, d_\text{NV}) = D(\vec{q}, d_\text{NV}) \ \vec{m}(\vec{q})
\end{equation}
where $\vec{q}$ is a 2D wavevector and $D(\vec{q}, d)$ is the following propagator:
\begin{equation}
  \label{eq:propagator}
  \begin{split}
    D(\vec{q}, d) = & \ \frac{\mu_0}{2} \ (e^{-dq} - e^{-(d+t)q}) \\
    &  \begin{pmatrix}
    -\cos^2(\phi_q) & -\frac{1}{2}\sin(2\phi_q) & -i\cos(\phi_q)\\[1mm]
    -\frac{1}{2} \sin(2\phi_q) & -\sin^2(\phi_q) & -i\sin(\phi_q)\\[1mm]
    -i\cos(\phi_q) & -i\sin(\phi_q) & 1 \\
  \end{pmatrix}
  \end{split}
\end{equation}
with $t$ the film thickness and $\vec{q} = (q \cos\phi_q, \ q \sin\phi_q)$. An inverse Fourier transformation leads to the expected stray field profile of the wall~\cite{hedrichNanoscaleMechanicsAntiferromagnetic2021}:
\begin{equation}
  \label{eq:field_cr2o3}
 \left \lbrace
   \begin{split}
     B_x & = - \frac{\mu_0 m  t}{2 \pi^2 w} \ \mathrm{Re}\left[f(x, d_\text{NV}) \right]\\
     B_y & = 0 \\
     B_z & =  \frac{\mu_0  m  t}{2 \pi^2 w} \ \mathrm{Im}\left[f(x, d_\text{NV}) \right]\\
   \end{split}
   \right.
\end{equation}
where
\begin{equation}
  \label{eq:f_cr2o3}
  f(x, d) = g\left(\frac{2d + \pi w - 2ix}{2\pi w}\right) - g\left(\frac{2d + \pi w + 2ix}{2\pi w}\right)
\end{equation}
with $g$ the first derivative of the log gamma function. This analytical expression can then be fitted to the experimental data after projection along the NV center quantification axis, which provides the values of the wall width parameter $w$ and the magnetization $m$. If a preliminary calibration of $d_\text{NV}$ has not been performed, it is also possible to extract this value from the fitting procedure. However, this usually lowers the accuracy of the obtained values of $m$ and $w$. An example of the fit of a domain wall profile in \CO, from ref.~\onlinecite{hedrichNanoscaleMechanicsAntiferromagnetic2021} is shown in Fig.~\ref{fig:quanti}(b), and lead to a wall width of $w \lesssim \SI{32}{\nano\meter}$ and a surface magnetization $\sigma = m t = 2.1 \pm 0.3 \ \mu_B \ \si{\nano\meter^{-2}}$.

The same approach can be followed to analyze the data showing the cycloid in \BFO from Fig.~\ref{fig:quanti}(c). The observed stray field is produced by a spin density wave attached to the cycloid modulation, and which can be described as:
\begin{equation}
  \label{eq:sdw}
  \vec{m}(\vec{r}) = m_\text{DM} \cos\left(\vec{k} \cdot \vec{r}\right) \hat{v} 
\end{equation}
with $\vec{k} = \frac{k}{\sqrt{2}} (\hat{e_x} - \hat{e_y})$ the cycloid wavevector, $m_\text{DM}$ the amplitude of the spin density wave, and $\hat{v} = \frac{1}{\sqrt{6}} (\hat{e_x} + \hat{e_y} - 2\hat{e_z})$ the direction normal to the cycloid plane. Using again the Fourier propagator $D(\vec{q}, d_\text{NV})$, one gets the expression of the expected stray field after summing up over the whole thickness $t$ of the film:
\begin{equation}
  \label{eq:bfo_field}
  \left \lbrace
    \begin{aligned}
      B_{x} & = A \sin\left( \frac{k}{\sqrt{2}} (x-y)\right) \\
      B_{y} & = - A \sin\left( \frac{k}{\sqrt{2}} (x-y)\right) \\
       B_{z} & = \sqrt{2} A \cos\left( \frac{k}{\sqrt{2}} (x-y)\right) \\
    \end{aligned}
\right.
\end{equation}
where
\begin{equation}
  \label{eq:A}
 A = \frac{\mu_0 \ m_\text{DM}}{\sqrt{3} a^3}  e^{-k d_\text{NV}} \frac{1-e^{-kt}}{1-e^{-ka}} \sinh \left(\frac{ka}{2}\right)
\end{equation}
and $a$ is the cubic unit cell size of \BFO. The amplitude and the period of the measured sine profile from Fig.~\ref{fig:quanti}(d) can therefore be used to extract the cycloid period and the value of $m_\text{DM} = 0.22 \pm 0.09 \ \mu_B$ for this specific dataset~\cite{haykal2020}. Such a value is higher than what was measured with neutron scattering experiments on a single crystal~\cite{ramazanogluLocalWeakFerromagnetism2011}. As neutron scattering probes the bulk of the sample and scanning NV magnetometry only its surface, this discrepancy could indicate a surface effect modifying the tilt of the magnetic moments.

The error on the fitted parameters can be quite large. This is related to the imperfect knowledge of the distance $d_\text{NV}$ between the NV center and the surface.

\subsubsection{Reconstruction of the magnetization pattern}
\label{sssec:backprop}

\begin{figure*}
  \centering
  \includegraphics[scale=1]{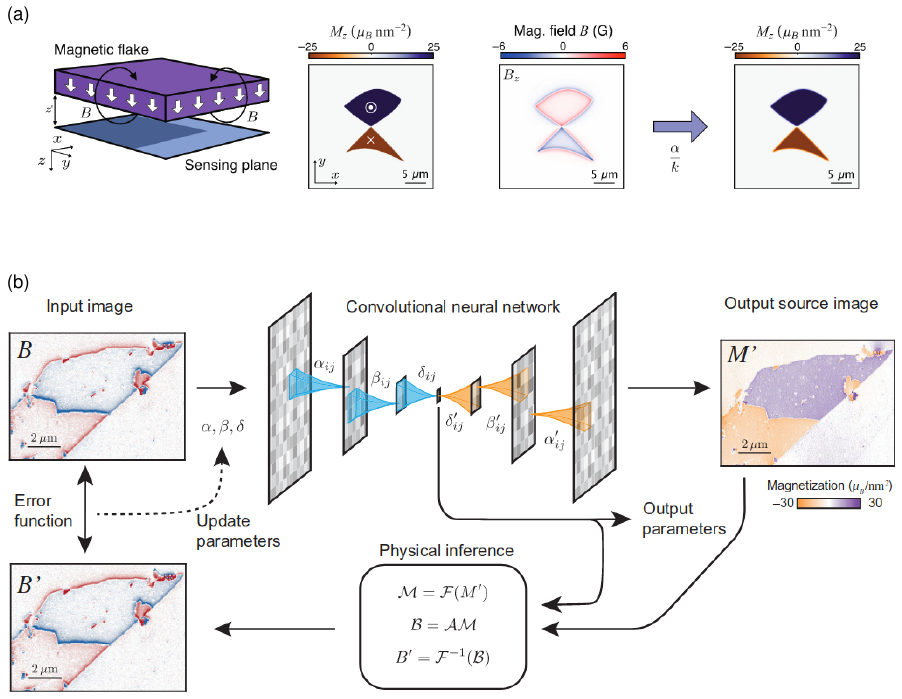}
  \caption{(a) Illustration of the Fourier method for reverse propagation, applied to a ferromagnetic out-of-plane magnetized flake. The sketch on the left describes the simulated sample, and the left $M_z$ map is used to compute the expected stray field map shown. Finally, the reverse propagation method is applied on this field map and yields the magnetization map shown on the left. Adapted with permission from ref.~\onlinecite{broadwayImprovedCurrentDensity2020a}. (b) Principle of the neural network method to achieve reverse propagation, on an in-plane magnetized flake. The input data is the measured stray field map. It is processed by the neural network to get a magnetization map as output. This magnetization configuration is then used to compute the corresponding stray field map, which is compared to the experimental data. The network is parametrized to minimize the error between this map and the experimental data. Adapted with permission from ref.~\onlinecite{duboisUntrainedPhysicallyInformed2022}.}
   \label{fig:backprop}
 \end{figure*}

Alternatively, based on assumptions on the magnetization distribution, it is possible to reverse propagate the measured magnetic field map to obtain a map of the magnetization density~\cite{broadwayImprovedCurrentDensity2020a}. This method is well-suited for ferromagnets and but can also be applied to antiferromagnets with uncompensated surface moments, like \CO~\cite{appelNanomagnetismMagnetoelectricGranular2019} or the van der Waals magnet CrI\textsubscript{3}~\cite{thielProbingMagnetism2D2019}. The principle of this method is to invert the procedure used in the previous section to compute the field from the magnetization distribution, also by going to Fourier space. However, this procedure cannot be properly inverted and therefore the reverse propagation approach requires some strong assumptions on the magnetic state of the sample: the magnetization $\vec{m}$ has to be confined to a plane, so it should depend only on $x$ and $y$ and its direction has to be assumed constant in the whole sample. Non-collinear magnetic textures cannot be retrieved from this method.

In Fourier space, the relationship between the magnetization $\vec{m}(q_x, q_y)$ and the magnetic field $\vec{B}(q_x, q_y, d_\text{NV})$ can be written as:
\begin{equation}
  \begin{pmatrix}
    B_x \\ B_y \\ B_z
  \end{pmatrix}
  = \frac{-\mu_0}{2 \ e^{qd_\text{NV}}}
  \begin{pmatrix}
    \frac{q_x^2}{q} & \frac{q_xq_y}{q} & iq_x \\[2mm]
    \frac{q_xq_y}{q} & \frac{q_y^2}{q} & iq_y \\[2mm]
    iq_x & iq_y & -q \\
  \end{pmatrix}
  \begin{pmatrix}
    m_x \\ m_y \\ m_z
  \end{pmatrix}
\end{equation}

The best performances are obtained for out-of-plane magnetized samples, like on the illustration in Fig.~\ref{fig:backprop}(a). With in-plane magnetized samples, specific attention has to be given to noise in the experimental data, as it is systematically amplified during the calculation\cite{broadwayImprovedCurrentDensity2020a}. The reverse propagation method was successfully applied to compute the surface magnetization of \CO, as well as to get the domain pattern in granular thin films~\cite{appelNanomagnetismMagnetoelectricGranular2019}. It has also been widely used in the study of van der Waals magnets~\cite{thielProbingMagnetism2D2019, sunMagneticDomainsDomain2021a, songDirectVisualizationMagnetic2021}.

To avoid numerical artifacts that arise from the direct inversion of the problem and get better results with in-plane magnetized samples, it has been proposed to use a neural network to compute the magnetization map~\cite{duboisUntrainedPhysicallyInformed2022}. A diagram representing the procedure is shown in Fig.~\ref{fig:backprop}(b). Here the network is not trained on a dataset, it rather minimizes an error function. This error function is obtained by computing the stray field expected from the output magnetization map and comparing it to the experimental data. This method has been successfully applied to in-plane magnetized samples, but is still limited to samples with a fixed magnetization direction so far.


%% file: relaxometry.tex
\subsection{\label{sec:saf_im} Imaging antiferromagnetic textures from spin wave noise}

For antiferromagnets without any small uncompensated moment, the quantitative measurement approach detailled above cannot be used, as the generated stray field is too small. A solution, which was proposed by Flebus \textit{et al.} in ref.~\onlinecite{flebusProposalDynamicImaging2018} is to rather probe the collective spin modes hosted by the antiferromagnetic textures like domain walls, using \emph{relaxometry}. In this measurement mode, fluctuating magnetic fields resonant with the spin transition of the NV center are detected because they modify its relaxation time $T_1$. The effect of a magnetic noise can therefore be described as~\cite{rolloQuantitativeStudyResponse2021a}:
\begin{equation}
  \label{eq:T1}
  \frac{1}{T_1} = \frac{1}{T_1^0} + 3 \gamma S_{B_\perp}(f_\pm) 
\end{equation}
where $\gamma = \SI{28}{\GHz\per\tesla}$ is the electron gyromagnetic ratio, $T_1^0$ is the intrinsic relaxation time of the NV center and $S_{B_\perp}(f_\pm)$ is the component of the field spectral density at the frequency $f_\pm$ perpendicular to the axis of the NV center. In the presence of magnetic noise at $f_\pm$, the photoluminescence emitted by the NV center decreases~\cite{rolloQuantitativeStudyResponse2021a}. Indeed, there is a competition between the optical pumping from the excitation laser, which polarizes the NV center in the bright state $m_s=0$ and the relaxation, which brings the system in thermal equilibrium, where all the states, dark and bright, are equally populated. The stronger the noise, the faster the relaxation, the NV center polarization decreases and the emitted photoluminescence drops. 

\begin{figure}[h]
  \centering
  \includegraphics[scale=0.95]{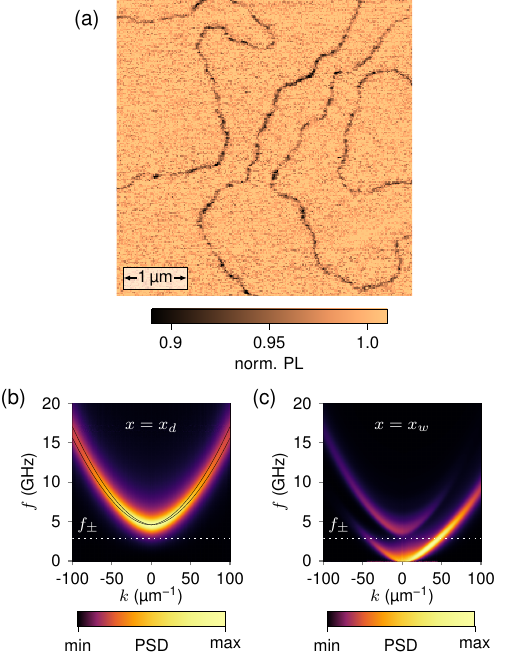}
  \caption{(a) NV photoluminescence map recorded above a synthetic antiferromagnet, showing domain walls as darker lines as a result from the presence of thermally activated spin waves with frequencies at $f_\pm$ inside the domain walls. (b) Spin wave dispersion computed in the bright domains of the map (a), showing a gap just above $f_\pm$: there are no thermally activated spin wave modes at $f_\pm$ in the domains. (c) Spin wave dispersion computed in the domain walls, showing that the gap closes and thus that spin wave modes at $f_\pm$ are present. Adapted from ref.~\onlinecite{fincoImagingNoncollinearAntiferromagnetic2021}.}
  \label{fig:saf}
\end{figure}

This property indicates that it is possible to localize sources of magnetic noise simply by monitoring the photoluminescence emitted by the NV center during a scan. An illustration of this is shown in Fig.~\ref{fig:saf}(a). It shows a photoluminescence map recorded above a synthetic antiferromagnet, a metallic multilayer stack including two ferromagnetic Co layers which are antiferromagnetically coupled through a Ru/Pt spacer~\cite{legrandRoomtemperatureStabilizationAntiferromagnetic2020}. Dark lines are visible on the image at the position of the domain walls in the magnetic layer, indicating an acceleration of the NV center relaxation. This is further confirmed by direct measurements of $T_1$ above the domains and above the domain walls, showing a significant drop, from about \SI{120}{\micro\second} to about \SI{20}{\micro\second} between both positions~\cite{fincoImagingNoncollinearAntiferromagnetic2021}. In addition, the observed photoluminescence contrast depends on the optical excitation power, which strengthens the NV center polarization in the bright state. The imaging contrast can thus be optimized by tuning the optical excitation power used. 

The origin of the measured magnetic noise is the presence of thermally activated spin waves in the synthetic antiferromagnet, with frequencies in the vicinity of $f_\pm$. Calculations of the dispersion of spin waves in this sample are shown in Fig.~\ref{fig:saf}(b) and (c). Panel (b) presents the spin wave dispersion in the domains, exhibiting a gap slightly larger than $f_\pm$. In contrast, the dispersion of spin waves channeled inside the domain walls is gapless, as displayed in panel (c). There are thus thermally activated modes at $f_\pm$ inside the domain walls which create a magnetic noise probed by the NV center. This noise accelerates the NV center relaxation, resulting in the observed drop of photoluminescence.

The closing of the gap in the spin wave dispersion at a domain wall is a general phenomenon resulting from the Goldstone theorem~\cite{goldstoneBrokenSymmetries1962}, which was predicted in ferromagnets~\cite{garcia-sanchezNarrowMagnonicWaveguides2015} and antiferromagnets~\cite{parkChannelingSpinWaves2021} and also measured using scanning transmission X-ray microscopy in synthetic antiferromagnets~\cite{slukaEmissionPropagation1D2019}. As a result, even though spin waves have usually much higher frequencies in antiferromagnets, rather in the \si{\THz} range which is not accessible directly with NV centers, noise from spin wave confined in domain walls should be detectable from any antiferromagnetic material.

\subsection{Measurement of spin transport properties}

\begin{figure}[h]
  \centering
  \includegraphics[scale=0.95]{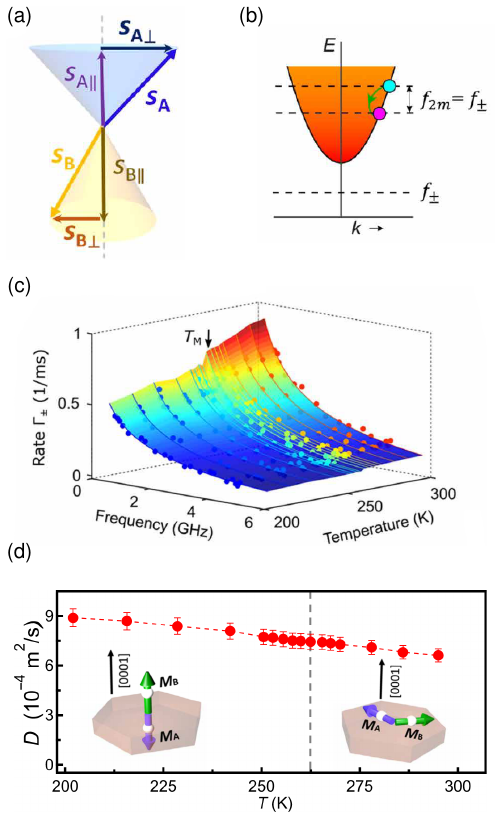}  
  \caption{(a) Sketch describing the transverse and longitudinal fluctuations of the Néel order in an antiferromagnet with two sublattices. (b) Sketch of the magnon dispersion in \hem, with 2 magnons processes indicated. (c) Measurement of the relaxation rate of the NV center as a function of the NV center transition frequency and the temperature, used to extract the spin diffusion constant in panel (d). (d) Temperature dependence of the spin diffusion constant in \hem measured across the Morin transition (dashed line). Adapted from ref.~\onlinecite{wangNoninvasiveMeasurementsSpin2022}.}
  \label{fig:diff}
\end{figure}

Besides magnetic imaging, NV-center-based relaxometry can also be used to probe other properties of antiferromagnets, and in particular spin transport, based on the fluctuation-dissipation theorem~\cite{flebusQuantumImpurityRelaxometryMagnetization2018}. NV centers can probe non-perturbatively the fluctuations of the spin density along the Néel order parameter and thus allow the extraction of the intrinsic spin diffusion constant. An experimental demonstration of this method has been reported in \hem~\cite{wangNoninvasiveMeasurementsSpin2022} (see Fig.~\ref{fig:diff}). 
While the imaging technique presented in section~\ref{sec:saf_im} relies on fluctuations of the transverse spin density $s_\perp$ (see Fig.~\ref{fig:diff}(a)), which are related to one-magnon scattering processes, here only the longitudinal fluctuations $s_\parallel$ are probed. Note that here we refer to the directions indicated in Fig.~\ref{fig:diff}(a), and the reference direction is the one of the Néel vector, not the NV center axis. The longitudinal fluctuations, and the resulting magnetic noise, correspond to two-magnon scattering processes, where the frequency difference $f_\text{2m}$ between the two magnons matches one of the magnetic resonance frequencies $f_\pm$ of the NV center (see Fig.~\ref{fig:diff}(b)). As a result, the frequency of the longitudinal spin noise can be arbitrarily small and does not depend on the magnon bandgap, and can thus be probed in the low \si{GHz} range using NV centers. Here, single NV centers inside a diamond nanobeam placed on the surface of the \hem sample are used, the sensor is thus stationary and probes a specific location on the sample.

In the presence of a magnetic field $H$, these longitudinal fluctuations can be related to the spin diffusion equation~\cite{flebusQuantumImpurityRelaxometryMagnetization2018, wangNoninvasiveMeasurementsSpin2022}:
\begin{equation}
   \left( \partial_t + \frac{1}{\tau_s} \right)  \vec{\mu} - D \ \nabla^2 \vec{\mu} = -  \partial_t \vec{H}
  \label{eq:spin_diff}
\end{equation}
with $\vec{\mu} = \nicefrac{\vec{s_\parallel}}{\chi_0} - \vec{H}$ and where $\chi_0$ is the static uniform longitudinal spin susceptibility, $D$ is the spin diffusion constant and $\tau_s$ is the coarse-grained spin relaxation time.

 By measuring the evolution of the relaxation rate $\Gamma$ of the NV center with the ESR frequency, i.e. under application of an external magnetic field, it is therefore possible to extract the value of $D$ (see Fig.~\ref{fig:diff}(c)-(d)), $\SI{6.6(0.4)e6}{\meter^2\per\second}$ at \SI{300}{\K}. This analysis was repeated at various temperatures, crossing at $T_\text{M}$ the Morin transition between on out-of-plane antiferromagnetic order below $T_\text{M}$ and an in-plane canted antiferromagnetic order above. The magnon gap is well above the NV detection frequencies on both sides of the transition, meaning that the analysis remains valid when changing the temperature. The data, presented in Fig.~\ref{fig:diff}(d), shows a smooth decrease of $D$ with temperature, without any significant feature at $T_\text{M}$. This decrease of $D$ is related to the increase of the inelastic scattering rate with temperature. The spin transport in equilibrium is mainly driven by thermal magnons governed by the exchange interaction. The changes in anisotropy and DMI occuring at the Morin transition correspond to small corrections, and are therefore not significantly affecting $D$ across the transition~\cite{wangNoninvasiveMeasurementsSpin2022}.

This type of experiments is especially interesting because they do not involve any currents or thermal gradients which can perturb the sample and be modified by interface effects. Probing spin fluctuations with NV centers gives thus access to fully intrinsic properties that can hardly be reached otherwise. Similar relaxometry experiments can also be achieved in a scanning probe microscope, offering the additionnal possibility to tune the distance $d_\text{NV}$ between the NV center and the surface of the magnetic sample. Measuring the NV center relaxation rate as a function of $d_\text{NV}$ should also provide insight into the spin diffusion length, without the need to apply a varying magnetic field during the experiment~\cite{flebusQuantumImpurityRelaxometryMagnetization2018}.


%% file: gradiometry.tex
In order to increase the sensitivity of the quantitative measurement presented in section~\ref{sec:quantitative}, one can use sequences of laser and microwave pulses to avoid power broadening of the ODMR spectra~\cite{dreauAvoidingPowerBroadening2011}. This measurement protocol called \emph{pulsed ODMR} provides an improvement up to one order of magnitude in magnetic field sensitivity compared to continuous ODMR and can easily be implemented in a scanning NV microscope~\cite{rondinMagnetometryNitrogenvacancyDefects2014}. It is also very helpful when working at cryogenic temperatures since it reduces the heating induced by the laser and microwave powers. In this context, it has recently successfully be applied to the imaging of ferromagnetic domains in the van der Waals magnet CrBr\textsubscript{3}~\cite{sunMagneticDomainsDomain2021a}.

\begin{figure*}
  \centering
  \includegraphics{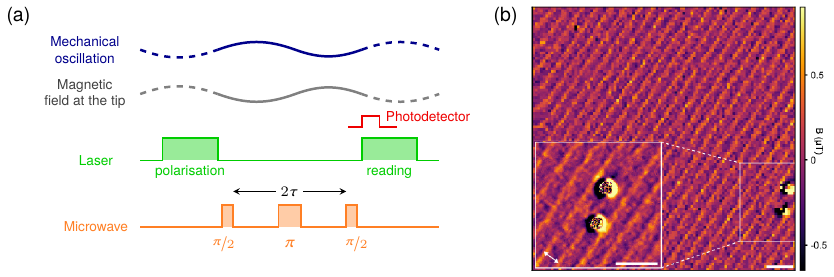}
  \caption{(a) Protocol for the detection of magnetic field gradients with spin echo. The mechanical oscillation of the tip results in a AC magnetic field at the NV center position in the presence of the spatial magnetic field gradient generated by the sample. This oscillation is synchronized with a spin echo sequence which measures the phase acquired by the NV spin during a period of oscillation of the magnetic field. (b) Image of the magnetic field gradient produced by atomic steps on the surface of \CO (0001), with a zoom-in around magnetic defects. The plotted quantitity is $x_\text{osc} \pdv{B}{x}$ and the scale bars correspond to \SI{1}{\micro\meter}. Adapted from ref.~\onlinecite{huxterScanningGradiometrySingle2022}.}
  \label{fig:gradio}
\end{figure*}

Another more efficient solution to increase the sensitivity of the measurement with the goal of imaging fully compensated antiferromagnets is to use AC sensing~\cite{voolImagingPhononmediatedHydrodynamic2021, palmImagingSubmicroampereCurrents2022}. When probing oscillating magnetic fields, dynamical decoupling protocols offer the possibilility to go beyond the sensitivity offered by pulsed ODMR~\cite{barrySensitivityOptimizationNVdiamond2020}, down to $\SI{50}{\nano\tesla\per\sqrt\Hz}$. However, very few systems produce a field that can be periodically modulated. A method to circumvent this issue was proposed~\cite{hongCoherentMechanicalControl2012, grinoldsNanoscaleMagneticImaging2013} and recently demonstrated on antiferromagnets by Huxter \emph{et al.} in ref.~\onlinecite{huxterScanningGradiometrySingle2022}: convert spatial gradients of the magnetic field into an oscillating signal. This conversion is performed using the oscillating movement of the tuning fork which provides feedback for the AFM. The resonance frequency $f_\text{TF}$ of these tuning forks lies around \SI{32}{\kHz}. The NV center is moving in a plane parallel to the sample surface, and the movement amplitude $x_\text{osc}$ is fixed and can be controlled by adjusting the input excitation signal sent to the tuning fork, in a range between 10 and \SI{70}{\nano\meter}. In the presence of a spatial gradient of magnetic field, the NV center experiences a periodically oscillating magnetic signal, which amplitude is the magnetic field gradient itself. The scan height $d_\text{NV}$ is thus fixed, meaning that this oscillation does not affects the spatial resolution of the measurement.

This signal can then be detected using a spin echo sequence which is synchronized with the tip oscillation (see Fig.~\ref{fig:gradio}(a)). The first laser pulse prepares the NV center in the bright state $m_s =0$. It is followed by a microwave $\nicefrac{\pi}{2}$  pulse in order to bring the spin in a superposition state. The system then acquires a phase in the presence of the oscillating magnetic field during a time $\tau$, before the application of microwave $\pi$ pulse to flip the spin. In the second part of the sequence, an additional phase is accumulated, as both the sign of the magnetic field and the spin evolution have been inverted. On the contrary, the effects of parasitic signals, static or at other frequencies, cancel out. Finally, the phase is extracted using another microwave $\nicefrac{\pi}{2}$ pulse and read using a last laser pulse. The accumulated phase $\varphi$ during the whole sequence realized at a position $x_0$ is given by:
\begin{equation}
  \label{eq:phase_gradio}
  \varphi = - 2\pi \gamma \ x_\text{osc} \eval{\pdv{B}{x}}_{x=x_0} \frac{2 \sin^2(\pi \tau f_\text{TF})}{\pi f_\text{TF}}
\end{equation}
where $\gamma = \SI{28}{\GHz\per\tesla}$ is the electronic gyromagnetic ratio. Measuring $\varphi$ gives thus a direct access to the magnetic field gradient.

First proof-of-principle experiments have been performed on the (0001) surface of a \CO single crystal~\cite{huxterScanningGradiometrySingle2022}, and the obtained gradient image is shown in Fig.~\ref{fig:gradio}(b). This image reveals a regular pattern of stripes which correspond to the stray field gradient generated by atomic step edges. Correlations of gradient and topography measurements even allow the identification of single and double atomic steps at the surface of the crystal. Such a measurement is reminiscent of images of the A-type antiferromagnetic state of a Cr(0001) single crystal imaged using spin-polarized scanning tunneling microscopy~\cite{hankeAbsenceSpinflipTransition2005}. Scanning NV gradiometry can thus provide access to atomic scale antiferromagnetism, which was so far restricted to spin-polarized scanning tunneling microscopy~\cite{wiesendangerSpinMappingNanoscale2009} or magnetic exchange force microscopy~\cite{kaiserMagneticExchangeForce2007}, techniques requiring ultra-high vacuum and mainly cryogenic conditions.

Finally, the gradiometry technique is not limited to the measurement of magnetic fields. NV centers are also sensitive to electric field via the Stark effect~\cite{doldeElectricfieldSensingUsing2011}, but their electric field sensitivity is much smaller than magnetic sensitivity. The sensitivity increase provided by the gradiometry approach, as well as its insensitivity to static parasitic fields, is an efficient technical solution. First measurements on patterned micron-sized electrodes~\cite{qiuNanoscaleElectricField2022a} and more recently on ferroelectric domains~\cite{huxterImagingFerroelectricDomains2023a} have shown that scanning NV electrometry can be achieved by gradiometry and open the perspective of combined non-perturbative nanoscale imaging of the antiferromagnetic and the ferroelectric order in multiferroics.


%% file: main.bbl
\begin{thebibliography}{93}%
\makeatletter
\providecommand \@ifxundefined [1]{%
 \@ifx{#1\undefined}
}%
\providecommand \@ifnum [1]{%
 \ifnum #1\expandafter \@firstoftwo
 \else \expandafter \@secondoftwo
 \fi
}%
\providecommand \@ifx [1]{%
 \ifx #1\expandafter \@firstoftwo
 \else \expandafter \@secondoftwo
 \fi
}%
\providecommand \natexlab [1]{#1}%
\providecommand \enquote  [1]{``#1''}%
\providecommand \bibnamefont  [1]{#1}%
\providecommand \bibfnamefont [1]{#1}%
\providecommand \citenamefont [1]{#1}%
\providecommand \href@noop [0]{\@secondoftwo}%
\providecommand \href [0]{\begingroup \@sanitize@url \@href}%
\providecommand \@href[1]{\@@startlink{#1}\@@href}%
\providecommand \@@href[1]{\endgroup#1\@@endlink}%
\providecommand \@sanitize@url [0]{\catcode `\\12\catcode `\$12\catcode
  `\&12\catcode `\#12\catcode `\^12\catcode `\_12\catcode `\%12\relax}%
\providecommand \@@startlink[1]{}%
\providecommand \@@endlink[0]{}%
\providecommand \url  [0]{\begingroup\@sanitize@url \@url }%
\providecommand \@url [1]{\endgroup\@href {#1}{\urlprefix }}%
\providecommand \urlprefix  [0]{URL }%
\providecommand \Eprint [0]{\href }%
\providecommand \doibase [0]{https://doi.org/}%
\providecommand \selectlanguage [0]{\@gobble}%
\providecommand \bibinfo  [0]{\@secondoftwo}%
\providecommand \bibfield  [0]{\@secondoftwo}%
\providecommand \translation [1]{[#1]}%
\providecommand \BibitemOpen [0]{}%
\providecommand \bibitemStop [0]{}%
\providecommand \bibitemNoStop [0]{.\EOS\space}%
\providecommand \EOS [0]{\spacefactor3000\relax}%
\providecommand \BibitemShut  [1]{\csname bibitem#1\endcsname}%
\let\auto@bib@innerbib\@empty
\bibitem [{\citenamefont {Cheong}\ \emph {et~al.}(2020)\citenamefont {Cheong},
  \citenamefont {Fiebig}, \citenamefont {Wu}, \citenamefont {Chapon},\ and\
  \citenamefont {Kiryukhin}}]{cheongSeeingBelievingVisualization2020}%
  \BibitemOpen
  \bibfield  {author} {\bibinfo {author} {\bibfnamefont {S.-W.}\ \bibnamefont
  {Cheong}}, \bibinfo {author} {\bibfnamefont {M.}~\bibnamefont {Fiebig}},
  \bibinfo {author} {\bibfnamefont {W.}~\bibnamefont {Wu}}, \bibinfo {author}
  {\bibfnamefont {L.}~\bibnamefont {Chapon}},\ and\ \bibinfo {author}
  {\bibfnamefont {V.}~\bibnamefont {Kiryukhin}},\ }\bibfield  {title} {\enquote
  {\bibinfo {title} {Seeing is believing: Visualization of antiferromagnetic
  domains},}\ }\href {https://doi.org/10.1038/s41535-019-0204-x} {\bibfield
  {journal} {\bibinfo  {journal} {npj Quantum Materials}\ }\textbf {\bibinfo
  {volume} {5}},\ \bibinfo {pages} {1--10} (\bibinfo {year}
  {2020})}\BibitemShut {NoStop}%
\bibitem [{\citenamefont {Jungwirth}\ \emph {et~al.}(2016)\citenamefont
  {Jungwirth}, \citenamefont {Marti}, \citenamefont {Wadley},\ and\
  \citenamefont {Wunderlich}}]{jungwirthAntiferromagneticSpintronics2016}%
  \BibitemOpen
  \bibfield  {author} {\bibinfo {author} {\bibfnamefont {T.}~\bibnamefont
  {Jungwirth}}, \bibinfo {author} {\bibfnamefont {X.}~\bibnamefont {Marti}},
  \bibinfo {author} {\bibfnamefont {P.}~\bibnamefont {Wadley}},\ and\ \bibinfo
  {author} {\bibfnamefont {J.}~\bibnamefont {Wunderlich}},\ }\bibfield  {title}
  {\enquote {\bibinfo {title} {Antiferromagnetic spintronics},}\ }\href
  {https://doi.org/10.1038/nnano.2016.18} {\bibfield  {journal} {\bibinfo
  {journal} {Nature Nanotechnology}\ }\textbf {\bibinfo {volume} {11}},\
  \bibinfo {pages} {231--241} (\bibinfo {year} {2016})}\BibitemShut {NoStop}%
\bibitem [{\citenamefont {Baltz}\ \emph {et~al.}(2018)\citenamefont {Baltz},
  \citenamefont {Manchon}, \citenamefont {Tsoi}, \citenamefont {Moriyama},
  \citenamefont {Ono},\ and\ \citenamefont
  {Tserkovnyak}}]{baltzAntiferromagneticSpintronics2018}%
  \BibitemOpen
  \bibfield  {author} {\bibinfo {author} {\bibfnamefont {V.}~\bibnamefont
  {Baltz}}, \bibinfo {author} {\bibfnamefont {A.}~\bibnamefont {Manchon}},
  \bibinfo {author} {\bibfnamefont {M.}~\bibnamefont {Tsoi}}, \bibinfo {author}
  {\bibfnamefont {T.}~\bibnamefont {Moriyama}}, \bibinfo {author}
  {\bibfnamefont {T.}~\bibnamefont {Ono}},\ and\ \bibinfo {author}
  {\bibfnamefont {Y.}~\bibnamefont {Tserkovnyak}},\ }\bibfield  {title}
  {\enquote {\bibinfo {title} {Antiferromagnetic spintronics},}\ }\href
  {https://doi.org/10.1103/RevModPhys.90.015005} {\bibfield  {journal}
  {\bibinfo  {journal} {Reviews of Modern Physics}\ }\textbf {\bibinfo {volume}
  {90}} (\bibinfo {year} {2018}),\ 10.1103/RevModPhys.90.015005}\BibitemShut
  {NoStop}%
\bibitem [{\citenamefont {Roth}(1960)}]{roth1960neutron}%
  \BibitemOpen
  \bibfield  {author} {\bibinfo {author} {\bibfnamefont {W.}~\bibnamefont
  {Roth}},\ }\bibfield  {title} {\enquote {\bibinfo {title} {Neutron and
  optical studies of domains in nio},}\ }\href
  {https://doi.org/10.1063/1.1735486} {\bibfield  {journal} {\bibinfo
  {journal} {Journal of Applied Physics}\ }\textbf {\bibinfo {volume} {31}},\
  \bibinfo {pages} {2000--2011} (\bibinfo {year} {1960})}\BibitemShut {NoStop}%
\bibitem [{\citenamefont {Krichevtsov}\ \emph {et~al.}(1993)\citenamefont
  {Krichevtsov}, \citenamefont {Pavlov}, \citenamefont {Pisarev},\ and\
  \citenamefont {Gridnev}}]{krichevtsov1993spontaneous}%
  \BibitemOpen
  \bibfield  {author} {\bibinfo {author} {\bibfnamefont {B.}~\bibnamefont
  {Krichevtsov}}, \bibinfo {author} {\bibfnamefont {V.}~\bibnamefont {Pavlov}},
  \bibinfo {author} {\bibfnamefont {R.}~\bibnamefont {Pisarev}},\ and\ \bibinfo
  {author} {\bibfnamefont {V.}~\bibnamefont {Gridnev}},\ }\bibfield  {title}
  {\enquote {\bibinfo {title} {Spontaneous non-reciprocal reflection of light
  from antiferromagnetic {Cr\textsubscript{2}O\textsubscript{3}}},}\ }\href
  {https://doi.org/10.1088/0953-8984/5/44/014} {\bibfield  {journal} {\bibinfo
  {journal} {Journal of Physics: Condensed Matter}\ }\textbf {\bibinfo {volume}
  {5}},\ \bibinfo {pages} {8233} (\bibinfo {year} {1993})}\BibitemShut
  {NoStop}%
\bibitem [{\citenamefont {Fiebig}\ \emph {et~al.}(1995)\citenamefont {Fiebig},
  \citenamefont {Fr{\"o}hlich}, \citenamefont {{Sluyterman v. L.}},\ and\
  \citenamefont {Pisarev}}]{fiebigDomainTopographyAntiferromagnetic1995}%
  \BibitemOpen
  \bibfield  {author} {\bibinfo {author} {\bibfnamefont {M.}~\bibnamefont
  {Fiebig}}, \bibinfo {author} {\bibfnamefont {D.}~\bibnamefont
  {Fr{\"o}hlich}}, \bibinfo {author} {\bibfnamefont {G.}~\bibnamefont
  {{Sluyterman v. L.}}},\ and\ \bibinfo {author} {\bibfnamefont {R.~V.}\
  \bibnamefont {Pisarev}},\ }\bibfield  {title} {\enquote {\bibinfo {title}
  {Domain topography of antiferromagnetic
  {{Cr\textsubscript{2}O\textsubscript{3}}} by second-harmonic generation},}\
  }\href {https://doi.org/10.1063/1.113699} {\bibfield  {journal} {\bibinfo
  {journal} {Applied Physics Letters}\ }\textbf {\bibinfo {volume} {66}},\
  \bibinfo {pages} {2906--2908} (\bibinfo {year} {1995})}\BibitemShut {NoStop}%
\bibitem [{\citenamefont {Figotin}\ and\ \citenamefont
  {Vitebsky}(2001)}]{figotin2001nonreciprocal}%
  \BibitemOpen
  \bibfield  {author} {\bibinfo {author} {\bibfnamefont {A.}~\bibnamefont
  {Figotin}}\ and\ \bibinfo {author} {\bibfnamefont {I.}~\bibnamefont
  {Vitebsky}},\ }\bibfield  {title} {\enquote {\bibinfo {title} {Nonreciprocal
  magnetic photonic crystals},}\ }\href
  {https://doi.org/10.1103/PhysRevE.63.066609} {\bibfield  {journal} {\bibinfo
  {journal} {Physical Review E}\ }\textbf {\bibinfo {volume} {63}},\ \bibinfo
  {pages} {066609} (\bibinfo {year} {2001})}\BibitemShut {NoStop}%
\bibitem [{\citenamefont {Chmiel}\ \emph {et~al.}(2018)\citenamefont {Chmiel},
  \citenamefont {Waterfield~Price}, \citenamefont {Johnson}, \citenamefont
  {Lamirand}, \citenamefont {Schad}, \citenamefont {van~der Laan},
  \citenamefont {Harris}, \citenamefont {Irwin}, \citenamefont {Rzchowski},
  \citenamefont {Eom} \emph {et~al.}}]{chmiel2018observation}%
  \BibitemOpen
  \bibfield  {author} {\bibinfo {author} {\bibfnamefont {F.~P.}\ \bibnamefont
  {Chmiel}}, \bibinfo {author} {\bibfnamefont {N.}~\bibnamefont
  {Waterfield~Price}}, \bibinfo {author} {\bibfnamefont {R.~D.}\ \bibnamefont
  {Johnson}}, \bibinfo {author} {\bibfnamefont {A.~D.}\ \bibnamefont
  {Lamirand}}, \bibinfo {author} {\bibfnamefont {J.}~\bibnamefont {Schad}},
  \bibinfo {author} {\bibfnamefont {G.}~\bibnamefont {van~der Laan}}, \bibinfo
  {author} {\bibfnamefont {D.~T.}\ \bibnamefont {Harris}}, \bibinfo {author}
  {\bibfnamefont {J.}~\bibnamefont {Irwin}}, \bibinfo {author} {\bibfnamefont
  {M.~S.}\ \bibnamefont {Rzchowski}}, \bibinfo {author} {\bibfnamefont {C.-B.}\
  \bibnamefont {Eom}}, \emph {et~al.},\ }\bibfield  {title} {\enquote {\bibinfo
  {title} {Observation of magnetic vortex pairs at room temperature in a planar
  $\alpha$-{Fe\textsubscript{2}O\textsubscript{3}/Co} heterostructure},}\
  }\href {https://doi.org/10.1038/s41563-018-0101-x} {\bibfield  {journal}
  {\bibinfo  {journal} {Nature Materials}\ }\textbf {\bibinfo {volume} {17}},\
  \bibinfo {pages} {581--585} (\bibinfo {year} {2018})}\BibitemShut {NoStop}%
\bibitem [{\citenamefont {Bode}\ \emph {et~al.}(2006)\citenamefont {Bode},
  \citenamefont {Vedmedenko}, \citenamefont {{von Bergmann}}, \citenamefont
  {Kubetzka}, \citenamefont {Ferriani}, \citenamefont {Heinze},\ and\
  \citenamefont {Wiesendanger}}]{bodeAtomicSpinStructure2006}%
  \BibitemOpen
  \bibfield  {author} {\bibinfo {author} {\bibfnamefont {M.}~\bibnamefont
  {Bode}}, \bibinfo {author} {\bibfnamefont {E.~Y.}\ \bibnamefont
  {Vedmedenko}}, \bibinfo {author} {\bibfnamefont {K.}~\bibnamefont {{von
  Bergmann}}}, \bibinfo {author} {\bibfnamefont {A.}~\bibnamefont {Kubetzka}},
  \bibinfo {author} {\bibfnamefont {P.}~\bibnamefont {Ferriani}}, \bibinfo
  {author} {\bibfnamefont {S.}~\bibnamefont {Heinze}},\ and\ \bibinfo {author}
  {\bibfnamefont {R.}~\bibnamefont {Wiesendanger}},\ }\bibfield  {title}
  {\enquote {\bibinfo {title} {Atomic spin structure of antiferromagnetic
  domain walls},}\ }\href {https://doi.org/10.1038/nmat1646} {\bibfield
  {journal} {\bibinfo  {journal} {Nature Materials}\ }\textbf {\bibinfo
  {volume} {5}},\ \bibinfo {pages} {477--481} (\bibinfo {year}
  {2006})}\BibitemShut {NoStop}%
\bibitem [{\citenamefont {Kaiser}, \citenamefont {Schwarz},\ and\ \citenamefont
  {Wiesendanger}(2007)}]{kaiserMagneticExchangeForce2007}%
  \BibitemOpen
  \bibfield  {author} {\bibinfo {author} {\bibfnamefont {U.}~\bibnamefont
  {Kaiser}}, \bibinfo {author} {\bibfnamefont {A.}~\bibnamefont {Schwarz}},\
  and\ \bibinfo {author} {\bibfnamefont {R.}~\bibnamefont {Wiesendanger}},\
  }\bibfield  {title} {\enquote {\bibinfo {title} {Magnetic exchange force
  microscopy with atomic resolution},}\ }\href
  {https://doi.org/10.1038/nature05617} {\bibfield  {journal} {\bibinfo
  {journal} {Nature}\ }\textbf {\bibinfo {volume} {446}},\ \bibinfo {pages}
  {522--525} (\bibinfo {year} {2007})}\BibitemShut {NoStop}%
\bibitem [{\citenamefont {Rondin}\ \emph {et~al.}(2014)\citenamefont {Rondin},
  \citenamefont {Tetienne}, \citenamefont {Hingant}, \citenamefont {Roch},
  \citenamefont {Maletinsky},\ and\ \citenamefont
  {Jacques}}]{rondinMagnetometryNitrogenvacancyDefects2014}%
  \BibitemOpen
  \bibfield  {author} {\bibinfo {author} {\bibfnamefont {L.}~\bibnamefont
  {Rondin}}, \bibinfo {author} {\bibfnamefont {J.-P.}\ \bibnamefont
  {Tetienne}}, \bibinfo {author} {\bibfnamefont {T.}~\bibnamefont {Hingant}},
  \bibinfo {author} {\bibfnamefont {J.-F.}\ \bibnamefont {Roch}}, \bibinfo
  {author} {\bibfnamefont {P.}~\bibnamefont {Maletinsky}},\ and\ \bibinfo
  {author} {\bibfnamefont {V.}~\bibnamefont {Jacques}},\ }\bibfield  {title}
  {\enquote {\bibinfo {title} {Magnetometry with nitrogen-vacancy defects in
  diamond},}\ }\href {https://doi.org/10.1088/0034-4885/77/5/056503} {\bibfield
   {journal} {\bibinfo  {journal} {Reports on Progress in Physics}\ }\textbf
  {\bibinfo {volume} {77}},\ \bibinfo {pages} {056503} (\bibinfo {year}
  {2014})}\BibitemShut {NoStop}%
\bibitem [{\citenamefont {Casola}, \citenamefont {{van der Sar}},\ and\
  \citenamefont {Yacoby}(2018)}]{casolaProbingCondensedMatter2018}%
  \BibitemOpen
  \bibfield  {author} {\bibinfo {author} {\bibfnamefont {F.}~\bibnamefont
  {Casola}}, \bibinfo {author} {\bibfnamefont {T.}~\bibnamefont {{van der
  Sar}}},\ and\ \bibinfo {author} {\bibfnamefont {A.}~\bibnamefont {Yacoby}},\
  }\bibfield  {title} {\enquote {\bibinfo {title} {Probing condensed matter
  physics with magnetometry based on nitrogen-vacancy centres in diamond},}\
  }\href {https://doi.org/10.1038/natrevmats.2017.88} {\bibfield  {journal}
  {\bibinfo  {journal} {Nature Reviews Materials}\ }\textbf {\bibinfo {volume}
  {3}},\ \bibinfo {pages} {17088} (\bibinfo {year} {2018})}\BibitemShut
  {NoStop}%
\bibitem [{\citenamefont {Xu}, \citenamefont {Zhang},\ and\ \citenamefont
  {Tian}(2023)}]{xuRecentAdvancesApplications2023}%
  \BibitemOpen
  \bibfield  {author} {\bibinfo {author} {\bibfnamefont {Y.}~\bibnamefont
  {Xu}}, \bibinfo {author} {\bibfnamefont {W.}~\bibnamefont {Zhang}},\ and\
  \bibinfo {author} {\bibfnamefont {C.}~\bibnamefont {Tian}},\ }\bibfield
  {title} {\enquote {\bibinfo {title} {Recent advances on applications of
  {{NV}}{\textsuperscript{-}} magnetometry in condensed matter physics},}\
  }\href {https://doi.org/10.1364/PRJ.471266} {\bibfield  {journal} {\bibinfo
  {journal} {Photonics Research}\ }\textbf {\bibinfo {volume} {11}},\ \bibinfo
  {pages} {393--412} (\bibinfo {year} {2023})}\BibitemShut {NoStop}%
\bibitem [{\citenamefont {Rondin}\ \emph {et~al.}(2013)\citenamefont {Rondin},
  \citenamefont {Tetienne}, \citenamefont {Rohart}, \citenamefont {Thiaville},
  \citenamefont {Hingant}, \citenamefont {Spinicelli}, \citenamefont {Roch},\
  and\ \citenamefont {Jacques}}]{rondinStrayfieldImagingMagnetic2013}%
  \BibitemOpen
  \bibfield  {author} {\bibinfo {author} {\bibfnamefont {L.}~\bibnamefont
  {Rondin}}, \bibinfo {author} {\bibfnamefont {J.-P.}\ \bibnamefont
  {Tetienne}}, \bibinfo {author} {\bibfnamefont {S.}~\bibnamefont {Rohart}},
  \bibinfo {author} {\bibfnamefont {A.}~\bibnamefont {Thiaville}}, \bibinfo
  {author} {\bibfnamefont {T.}~\bibnamefont {Hingant}}, \bibinfo {author}
  {\bibfnamefont {P.}~\bibnamefont {Spinicelli}}, \bibinfo {author}
  {\bibfnamefont {J.-F.}\ \bibnamefont {Roch}},\ and\ \bibinfo {author}
  {\bibfnamefont {V.}~\bibnamefont {Jacques}},\ }\bibfield  {title} {\enquote
  {\bibinfo {title} {Stray-field imaging of magnetic vortices with a single
  diamond spin},}\ }\href {https://doi.org/10.1038/ncomms3279} {\bibfield
  {journal} {\bibinfo  {journal} {Nature Communications}\ }\textbf {\bibinfo
  {volume} {4}},\ \bibinfo {pages} {2279} (\bibinfo {year} {2013})}\BibitemShut
  {NoStop}%
\bibitem [{\citenamefont {Tetienne}\ \emph {et~al.}(2014)\citenamefont
  {Tetienne}, \citenamefont {Hingant}, \citenamefont {Kim}, \citenamefont
  {Diez}, \citenamefont {Adam}, \citenamefont {Garcia}, \citenamefont {Roch},
  \citenamefont {Rohart}, \citenamefont {Thiaville}, \citenamefont
  {Ravelosona},\ and\ \citenamefont
  {Jacques}}]{tetienneNanoscaleImagingControl2014}%
  \BibitemOpen
  \bibfield  {author} {\bibinfo {author} {\bibfnamefont {J.-P.}\ \bibnamefont
  {Tetienne}}, \bibinfo {author} {\bibfnamefont {T.}~\bibnamefont {Hingant}},
  \bibinfo {author} {\bibfnamefont {J.-V.}\ \bibnamefont {Kim}}, \bibinfo
  {author} {\bibfnamefont {L.~H.}\ \bibnamefont {Diez}}, \bibinfo {author}
  {\bibfnamefont {J.-P.}\ \bibnamefont {Adam}}, \bibinfo {author}
  {\bibfnamefont {K.}~\bibnamefont {Garcia}}, \bibinfo {author} {\bibfnamefont
  {J.-F.}\ \bibnamefont {Roch}}, \bibinfo {author} {\bibfnamefont
  {S.}~\bibnamefont {Rohart}}, \bibinfo {author} {\bibfnamefont
  {A.}~\bibnamefont {Thiaville}}, \bibinfo {author} {\bibfnamefont
  {D.}~\bibnamefont {Ravelosona}},\ and\ \bibinfo {author} {\bibfnamefont
  {V.}~\bibnamefont {Jacques}},\ }\bibfield  {title} {\enquote {\bibinfo
  {title} {Nanoscale imaging and control of domain-wall hopping with a
  nitrogen-vacancy center microscope},}\ }\href
  {https://doi.org/10.1126/science.1250113} {\bibfield  {journal} {\bibinfo
  {journal} {Science}\ }\textbf {\bibinfo {volume} {344}},\ \bibinfo {pages}
  {1366--1369} (\bibinfo {year} {2014})}\BibitemShut {NoStop}%
\bibitem [{\citenamefont {Tetienne}\ \emph {et~al.}(2015)\citenamefont
  {Tetienne}, \citenamefont {Hingant}, \citenamefont {Mart{\'i}nez},
  \citenamefont {Rohart}, \citenamefont {Thiaville}, \citenamefont {Diez},
  \citenamefont {Garcia}, \citenamefont {Adam}, \citenamefont {Kim},
  \citenamefont {Roch}, \citenamefont {Miron}, \citenamefont {Gaudin},
  \citenamefont {Vila}, \citenamefont {Ocker}, \citenamefont {Ravelosona},\
  and\ \citenamefont {Jacques}}]{tetienneNatureDomainWalls2015}%
  \BibitemOpen
  \bibfield  {author} {\bibinfo {author} {\bibfnamefont {J.-P.}\ \bibnamefont
  {Tetienne}}, \bibinfo {author} {\bibfnamefont {T.}~\bibnamefont {Hingant}},
  \bibinfo {author} {\bibfnamefont {L.~J.}\ \bibnamefont {Mart{\'i}nez}},
  \bibinfo {author} {\bibfnamefont {S.}~\bibnamefont {Rohart}}, \bibinfo
  {author} {\bibfnamefont {A.}~\bibnamefont {Thiaville}}, \bibinfo {author}
  {\bibfnamefont {L.~H.}\ \bibnamefont {Diez}}, \bibinfo {author}
  {\bibfnamefont {K.}~\bibnamefont {Garcia}}, \bibinfo {author} {\bibfnamefont
  {J.-P.}\ \bibnamefont {Adam}}, \bibinfo {author} {\bibfnamefont {J.-V.}\
  \bibnamefont {Kim}}, \bibinfo {author} {\bibfnamefont {J.-F.}\ \bibnamefont
  {Roch}}, \bibinfo {author} {\bibfnamefont {I.~M.}\ \bibnamefont {Miron}},
  \bibinfo {author} {\bibfnamefont {G.}~\bibnamefont {Gaudin}}, \bibinfo
  {author} {\bibfnamefont {L.}~\bibnamefont {Vila}}, \bibinfo {author}
  {\bibfnamefont {B.}~\bibnamefont {Ocker}}, \bibinfo {author} {\bibfnamefont
  {D.}~\bibnamefont {Ravelosona}},\ and\ \bibinfo {author} {\bibfnamefont
  {V.}~\bibnamefont {Jacques}},\ }\bibfield  {title} {\enquote {\bibinfo
  {title} {The nature of domain walls in ultrathin ferromagnets revealed by
  scanning nanomagnetometry},}\ }\href {https://doi.org/10.1038/ncomms7733}
  {\bibfield  {journal} {\bibinfo  {journal} {Nature Communications}\ }\textbf
  {\bibinfo {volume} {6}},\ \bibinfo {pages} {6733} (\bibinfo {year}
  {2015})}\BibitemShut {NoStop}%
\bibitem [{\citenamefont {Dovzhenko}\ \emph {et~al.}(2018)\citenamefont
  {Dovzhenko}, \citenamefont {Casola}, \citenamefont {Schlotter}, \citenamefont
  {Zhou}, \citenamefont {B{\"u}ttner}, \citenamefont {Walsworth}, \citenamefont
  {Beach},\ and\ \citenamefont
  {Yacoby}}]{dovzhenkoMagnetostaticTwistsRoomtemperature2018}%
  \BibitemOpen
  \bibfield  {author} {\bibinfo {author} {\bibfnamefont {Y.}~\bibnamefont
  {Dovzhenko}}, \bibinfo {author} {\bibfnamefont {F.}~\bibnamefont {Casola}},
  \bibinfo {author} {\bibfnamefont {S.}~\bibnamefont {Schlotter}}, \bibinfo
  {author} {\bibfnamefont {T.~X.}\ \bibnamefont {Zhou}}, \bibinfo {author}
  {\bibfnamefont {F.}~\bibnamefont {B{\"u}ttner}}, \bibinfo {author}
  {\bibfnamefont {R.~L.}\ \bibnamefont {Walsworth}}, \bibinfo {author}
  {\bibfnamefont {G.~S.~D.}\ \bibnamefont {Beach}},\ and\ \bibinfo {author}
  {\bibfnamefont {A.}~\bibnamefont {Yacoby}},\ }\bibfield  {title} {\enquote
  {\bibinfo {title} {Magnetostatic twists in room-temperature skyrmions
  explored by nitrogen-vacancy center spin texture reconstruction},}\ }\href
  {https://doi.org/10.1038/s41467-018-05158-9} {\bibfield  {journal} {\bibinfo
  {journal} {Nature Communications}\ }\textbf {\bibinfo {volume} {9}},\
  \bibinfo {pages} {2712} (\bibinfo {year} {2018})}\BibitemShut {NoStop}%
\bibitem [{\citenamefont {Gross}\ \emph {et~al.}(2018)\citenamefont {Gross},
  \citenamefont {Akhtar}, \citenamefont {Hrabec}, \citenamefont {Sampaio},
  \citenamefont {Mart{\'i}nez}, \citenamefont {Chouaieb}, \citenamefont
  {Shields}, \citenamefont {Maletinsky}, \citenamefont {Thiaville},
  \citenamefont {Rohart},\ and\ \citenamefont
  {Jacques}}]{grossSkyrmionMorphologyUltrathin2018}%
  \BibitemOpen
  \bibfield  {author} {\bibinfo {author} {\bibfnamefont {I.}~\bibnamefont
  {Gross}}, \bibinfo {author} {\bibfnamefont {W.}~\bibnamefont {Akhtar}},
  \bibinfo {author} {\bibfnamefont {A.}~\bibnamefont {Hrabec}}, \bibinfo
  {author} {\bibfnamefont {J.}~\bibnamefont {Sampaio}}, \bibinfo {author}
  {\bibfnamefont {L.~J.}\ \bibnamefont {Mart{\'i}nez}}, \bibinfo {author}
  {\bibfnamefont {S.}~\bibnamefont {Chouaieb}}, \bibinfo {author}
  {\bibfnamefont {B.~J.}\ \bibnamefont {Shields}}, \bibinfo {author}
  {\bibfnamefont {P.}~\bibnamefont {Maletinsky}}, \bibinfo {author}
  {\bibfnamefont {A.}~\bibnamefont {Thiaville}}, \bibinfo {author}
  {\bibfnamefont {S.}~\bibnamefont {Rohart}},\ and\ \bibinfo {author}
  {\bibfnamefont {V.}~\bibnamefont {Jacques}},\ }\bibfield  {title} {\enquote
  {\bibinfo {title} {Skyrmion morphology in ultrathin magnetic films},}\ }\href
  {https://doi.org/10.1103/PhysRevMaterials.2.024406} {\bibfield  {journal}
  {\bibinfo  {journal} {Physical Review Materials}\ }\textbf {\bibinfo {volume}
  {2}},\ \bibinfo {pages} {024406} (\bibinfo {year} {2018})}\BibitemShut
  {NoStop}%
\bibitem [{\citenamefont {Pelliccione}\ \emph {et~al.}(2016)\citenamefont
  {Pelliccione}, \citenamefont {Jenkins}, \citenamefont {Ovartchaiyapong},
  \citenamefont {Reetz}, \citenamefont {Emmanouilidou}, \citenamefont {Ni},\
  and\ \citenamefont {Bleszynski~Jayich}}]{pelliccioneScannedProbeImaging2016}%
  \BibitemOpen
  \bibfield  {author} {\bibinfo {author} {\bibfnamefont {M.}~\bibnamefont
  {Pelliccione}}, \bibinfo {author} {\bibfnamefont {A.}~\bibnamefont
  {Jenkins}}, \bibinfo {author} {\bibfnamefont {P.}~\bibnamefont
  {Ovartchaiyapong}}, \bibinfo {author} {\bibfnamefont {C.}~\bibnamefont
  {Reetz}}, \bibinfo {author} {\bibfnamefont {E.}~\bibnamefont
  {Emmanouilidou}}, \bibinfo {author} {\bibfnamefont {N.}~\bibnamefont {Ni}},\
  and\ \bibinfo {author} {\bibfnamefont {A.~C.}\ \bibnamefont
  {Bleszynski~Jayich}},\ }\bibfield  {title} {\enquote {\bibinfo {title}
  {Scanned probe imaging of nanoscale magnetism at cryogenic temperatures with
  a single-spin quantum sensor},}\ }\href
  {https://doi.org/10.1038/nnano.2016.68} {\bibfield  {journal} {\bibinfo
  {journal} {Nature Nanotechnology}\ }\textbf {\bibinfo {volume} {11}},\
  \bibinfo {pages} {700--705} (\bibinfo {year} {2016})}\BibitemShut {NoStop}%
\bibitem [{\citenamefont {Thiel}\ \emph {et~al.}(2016)\citenamefont {Thiel},
  \citenamefont {Rohner}, \citenamefont {Ganzhorn}, \citenamefont {Appel},
  \citenamefont {Neu}, \citenamefont {M{\"u}ller}, \citenamefont {Kleiner},
  \citenamefont {Koelle},\ and\ \citenamefont
  {Maletinsky}}]{thielQuantitativeNanoscaleVortex2016}%
  \BibitemOpen
  \bibfield  {author} {\bibinfo {author} {\bibfnamefont {L.}~\bibnamefont
  {Thiel}}, \bibinfo {author} {\bibfnamefont {D.}~\bibnamefont {Rohner}},
  \bibinfo {author} {\bibfnamefont {M.}~\bibnamefont {Ganzhorn}}, \bibinfo
  {author} {\bibfnamefont {P.}~\bibnamefont {Appel}}, \bibinfo {author}
  {\bibfnamefont {E.}~\bibnamefont {Neu}}, \bibinfo {author} {\bibfnamefont
  {B.}~\bibnamefont {M{\"u}ller}}, \bibinfo {author} {\bibfnamefont
  {R.}~\bibnamefont {Kleiner}}, \bibinfo {author} {\bibfnamefont
  {D.}~\bibnamefont {Koelle}},\ and\ \bibinfo {author} {\bibfnamefont
  {P.}~\bibnamefont {Maletinsky}},\ }\bibfield  {title} {\enquote {\bibinfo
  {title} {Quantitative nanoscale vortex imaging using a cryogenic quantum
  magnetometer},}\ }\href {https://doi.org/10.1038/nnano.2016.63} {\bibfield
  {journal} {\bibinfo  {journal} {Nature Nanotechnology}\ }\textbf {\bibinfo
  {volume} {11}},\ \bibinfo {pages} {677--681} (\bibinfo {year}
  {2016})}\BibitemShut {NoStop}%
\bibitem [{\citenamefont {Wolfe}\ \emph {et~al.}(2014)\citenamefont {Wolfe},
  \citenamefont {Bhallamudi}, \citenamefont {Wang}, \citenamefont {Du},
  \citenamefont {Manuilov}, \citenamefont {{Teeling-Smith}}, \citenamefont
  {Berger}, \citenamefont {Adur}, \citenamefont {Yang},\ and\ \citenamefont
  {Hammel}}]{wolfeOffresonantManipulationSpins2014}%
  \BibitemOpen
  \bibfield  {author} {\bibinfo {author} {\bibfnamefont {C.~S.}\ \bibnamefont
  {Wolfe}}, \bibinfo {author} {\bibfnamefont {V.~P.}\ \bibnamefont
  {Bhallamudi}}, \bibinfo {author} {\bibfnamefont {H.~L.}\ \bibnamefont
  {Wang}}, \bibinfo {author} {\bibfnamefont {C.~H.}\ \bibnamefont {Du}},
  \bibinfo {author} {\bibfnamefont {S.}~\bibnamefont {Manuilov}}, \bibinfo
  {author} {\bibfnamefont {R.~M.}\ \bibnamefont {{Teeling-Smith}}}, \bibinfo
  {author} {\bibfnamefont {A.~J.}\ \bibnamefont {Berger}}, \bibinfo {author}
  {\bibfnamefont {R.}~\bibnamefont {Adur}}, \bibinfo {author} {\bibfnamefont
  {F.~Y.}\ \bibnamefont {Yang}},\ and\ \bibinfo {author} {\bibfnamefont
  {P.~C.}\ \bibnamefont {Hammel}},\ }\bibfield  {title} {\enquote {\bibinfo
  {title} {Off-resonant manipulation of spins in diamond via precessing
  magnetization of a proximal ferromagnet},}\ }\href
  {https://doi.org/10.1103/PhysRevB.89.180406} {\bibfield  {journal} {\bibinfo
  {journal} {Physical Review B}\ }\textbf {\bibinfo {volume} {89}},\ \bibinfo
  {pages} {180406} (\bibinfo {year} {2014})}\BibitemShut {NoStop}%
\bibitem [{\citenamefont {Du}\ \emph {et~al.}(2017)\citenamefont {Du},
  \citenamefont {{van der Sar}}, \citenamefont {Zhou}, \citenamefont
  {Upadhyaya}, \citenamefont {Casola}, \citenamefont {Zhang}, \citenamefont
  {Onbasli}, \citenamefont {Ross}, \citenamefont {Walsworth}, \citenamefont
  {Tserkovnyak},\ and\ \citenamefont {Yacoby}}]{duControlLocalMeasurement2017}%
  \BibitemOpen
  \bibfield  {author} {\bibinfo {author} {\bibfnamefont {C.}~\bibnamefont
  {Du}}, \bibinfo {author} {\bibfnamefont {T.}~\bibnamefont {{van der Sar}}},
  \bibinfo {author} {\bibfnamefont {T.~X.}\ \bibnamefont {Zhou}}, \bibinfo
  {author} {\bibfnamefont {P.}~\bibnamefont {Upadhyaya}}, \bibinfo {author}
  {\bibfnamefont {F.}~\bibnamefont {Casola}}, \bibinfo {author} {\bibfnamefont
  {H.}~\bibnamefont {Zhang}}, \bibinfo {author} {\bibfnamefont {M.~C.}\
  \bibnamefont {Onbasli}}, \bibinfo {author} {\bibfnamefont {C.~A.}\
  \bibnamefont {Ross}}, \bibinfo {author} {\bibfnamefont {R.~L.}\ \bibnamefont
  {Walsworth}}, \bibinfo {author} {\bibfnamefont {Y.}~\bibnamefont
  {Tserkovnyak}},\ and\ \bibinfo {author} {\bibfnamefont {A.}~\bibnamefont
  {Yacoby}},\ }\bibfield  {title} {\enquote {\bibinfo {title} {Control and
  local measurement of the spin chemical potential in a magnetic insulator},}\
  }\href {https://doi.org/10.1126/science.aak9611} {\bibfield  {journal}
  {\bibinfo  {journal} {Science}\ }\textbf {\bibinfo {volume} {357}},\ \bibinfo
  {pages} {195--198} (\bibinfo {year} {2017})}\BibitemShut {NoStop}%
\bibitem [{\citenamefont {{van der Sar}}\ \emph {et~al.}(2015)\citenamefont
  {{van der Sar}}, \citenamefont {Casola}, \citenamefont {Walsworth},\ and\
  \citenamefont {Yacoby}}]{vandersarNanometrescaleProbingSpin2015}%
  \BibitemOpen
  \bibfield  {author} {\bibinfo {author} {\bibfnamefont {T.}~\bibnamefont {{van
  der Sar}}}, \bibinfo {author} {\bibfnamefont {F.}~\bibnamefont {Casola}},
  \bibinfo {author} {\bibfnamefont {R.}~\bibnamefont {Walsworth}},\ and\
  \bibinfo {author} {\bibfnamefont {A.}~\bibnamefont {Yacoby}},\ }\bibfield
  {title} {\enquote {\bibinfo {title} {Nanometre-scale probing of spin waves
  using single electron spins},}\ }\href {https://doi.org/10.1038/ncomms8886}
  {\bibfield  {journal} {\bibinfo  {journal} {Nature Communications}\ }\textbf
  {\bibinfo {volume} {6}},\ \bibinfo {pages} {7886} (\bibinfo {year}
  {2015})}\BibitemShut {NoStop}%
\bibitem [{\citenamefont {Gross}\ \emph {et~al.}(2017)\citenamefont {Gross},
  \citenamefont {Akhtar}, \citenamefont {Garcia}, \citenamefont {Mart{\'i}nez},
  \citenamefont {Chouaieb}, \citenamefont {Garcia}, \citenamefont
  {Carr{\'e}t{\'e}ro}, \citenamefont {Barth{\'e}l{\'e}my}, \citenamefont
  {Appel}, \citenamefont {Maletinsky}, \citenamefont {Kim}, \citenamefont
  {Chauleau}, \citenamefont {Jaouen}, \citenamefont {Viret}, \citenamefont
  {Bibes}, \citenamefont {Fusil},\ and\ \citenamefont
  {Jacques}}]{grossRealspaceImagingNoncollinear2017}%
  \BibitemOpen
  \bibfield  {author} {\bibinfo {author} {\bibfnamefont {I.}~\bibnamefont
  {Gross}}, \bibinfo {author} {\bibfnamefont {W.}~\bibnamefont {Akhtar}},
  \bibinfo {author} {\bibfnamefont {V.}~\bibnamefont {Garcia}}, \bibinfo
  {author} {\bibfnamefont {L.~J.}\ \bibnamefont {Mart{\'i}nez}}, \bibinfo
  {author} {\bibfnamefont {S.}~\bibnamefont {Chouaieb}}, \bibinfo {author}
  {\bibfnamefont {K.}~\bibnamefont {Garcia}}, \bibinfo {author} {\bibfnamefont
  {C.}~\bibnamefont {Carr{\'e}t{\'e}ro}}, \bibinfo {author} {\bibfnamefont
  {A.}~\bibnamefont {Barth{\'e}l{\'e}my}}, \bibinfo {author} {\bibfnamefont
  {P.}~\bibnamefont {Appel}}, \bibinfo {author} {\bibfnamefont
  {P.}~\bibnamefont {Maletinsky}}, \bibinfo {author} {\bibfnamefont {J.-V.}\
  \bibnamefont {Kim}}, \bibinfo {author} {\bibfnamefont {J.~Y.}\ \bibnamefont
  {Chauleau}}, \bibinfo {author} {\bibfnamefont {N.}~\bibnamefont {Jaouen}},
  \bibinfo {author} {\bibfnamefont {M.}~\bibnamefont {Viret}}, \bibinfo
  {author} {\bibfnamefont {M.}~\bibnamefont {Bibes}}, \bibinfo {author}
  {\bibfnamefont {S.}~\bibnamefont {Fusil}},\ and\ \bibinfo {author}
  {\bibfnamefont {V.}~\bibnamefont {Jacques}},\ }\bibfield  {title} {\enquote
  {\bibinfo {title} {Real-space imaging of non-collinear antiferromagnetic
  order with a single-spin magnetometer},}\ }\href
  {https://doi.org/10.1038/nature23656} {\bibfield  {journal} {\bibinfo
  {journal} {Nature}\ }\textbf {\bibinfo {volume} {549}},\ \bibinfo {pages}
  {252--256} (\bibinfo {year} {2017})}\BibitemShut {NoStop}%
\bibitem [{\citenamefont {Kosub}\ \emph {et~al.}(2017)\citenamefont {Kosub},
  \citenamefont {Kopte}, \citenamefont {H{\"u}hne}, \citenamefont {Appel},
  \citenamefont {Shields}, \citenamefont {Maletinsky}, \citenamefont
  {H{\"u}bner}, \citenamefont {Liedke}, \citenamefont {Fassbender},
  \citenamefont {Schmidt},\ and\ \citenamefont
  {Makarov}}]{kosubPurelyAntiferromagneticMagnetoelectric2017}%
  \BibitemOpen
  \bibfield  {author} {\bibinfo {author} {\bibfnamefont {T.}~\bibnamefont
  {Kosub}}, \bibinfo {author} {\bibfnamefont {M.}~\bibnamefont {Kopte}},
  \bibinfo {author} {\bibfnamefont {R.}~\bibnamefont {H{\"u}hne}}, \bibinfo
  {author} {\bibfnamefont {P.}~\bibnamefont {Appel}}, \bibinfo {author}
  {\bibfnamefont {B.}~\bibnamefont {Shields}}, \bibinfo {author} {\bibfnamefont
  {P.}~\bibnamefont {Maletinsky}}, \bibinfo {author} {\bibfnamefont
  {R.}~\bibnamefont {H{\"u}bner}}, \bibinfo {author} {\bibfnamefont {M.~O.}\
  \bibnamefont {Liedke}}, \bibinfo {author} {\bibfnamefont {J.}~\bibnamefont
  {Fassbender}}, \bibinfo {author} {\bibfnamefont {O.~G.}\ \bibnamefont
  {Schmidt}},\ and\ \bibinfo {author} {\bibfnamefont {D.}~\bibnamefont
  {Makarov}},\ }\bibfield  {title} {\enquote {\bibinfo {title} {Purely
  antiferromagnetic magnetoelectric random access memory},}\ }\href
  {https://doi.org/10.1038/ncomms13985} {\bibfield  {journal} {\bibinfo
  {journal} {Nature Communications}\ }\textbf {\bibinfo {volume} {8}},\
  \bibinfo {pages} {13985} (\bibinfo {year} {2017})}\BibitemShut {NoStop}%
\bibitem [{\citenamefont {Flebus}\ and\ \citenamefont
  {Tserkovnyak}(2018)}]{flebusQuantumImpurityRelaxometryMagnetization2018}%
  \BibitemOpen
  \bibfield  {author} {\bibinfo {author} {\bibfnamefont {B.}~\bibnamefont
  {Flebus}}\ and\ \bibinfo {author} {\bibfnamefont {Y.}~\bibnamefont
  {Tserkovnyak}},\ }\bibfield  {title} {\enquote {\bibinfo {title}
  {Quantum-{{Impurity Relaxometry}} of {{Magnetization Dynamics}}},}\ }\href
  {https://doi.org/10.1103/PhysRevLett.121.187204} {\bibfield  {journal}
  {\bibinfo  {journal} {Physical Review Letters}\ }\textbf {\bibinfo {volume}
  {121}},\ \bibinfo {pages} {187204} (\bibinfo {year} {2018})}\BibitemShut
  {NoStop}%
\bibitem [{\citenamefont {Wang}\ \emph {et~al.}(2022)\citenamefont {Wang},
  \citenamefont {Zhang}, \citenamefont {McLaughlin}, \citenamefont {Flebus},
  \citenamefont {Huang}, \citenamefont {Xiao}, \citenamefont {Liu},
  \citenamefont {Wu}, \citenamefont {Fullerton}, \citenamefont {Tserkovnyak},\
  and\ \citenamefont {Du}}]{wangNoninvasiveMeasurementsSpin2022}%
  \BibitemOpen
  \bibfield  {author} {\bibinfo {author} {\bibfnamefont {H.}~\bibnamefont
  {Wang}}, \bibinfo {author} {\bibfnamefont {S.}~\bibnamefont {Zhang}},
  \bibinfo {author} {\bibfnamefont {N.~J.}\ \bibnamefont {McLaughlin}},
  \bibinfo {author} {\bibfnamefont {B.}~\bibnamefont {Flebus}}, \bibinfo
  {author} {\bibfnamefont {M.}~\bibnamefont {Huang}}, \bibinfo {author}
  {\bibfnamefont {Y.}~\bibnamefont {Xiao}}, \bibinfo {author} {\bibfnamefont
  {C.}~\bibnamefont {Liu}}, \bibinfo {author} {\bibfnamefont {M.}~\bibnamefont
  {Wu}}, \bibinfo {author} {\bibfnamefont {E.~E.}\ \bibnamefont {Fullerton}},
  \bibinfo {author} {\bibfnamefont {Y.}~\bibnamefont {Tserkovnyak}},\ and\
  \bibinfo {author} {\bibfnamefont {C.~R.}\ \bibnamefont {Du}},\ }\bibfield
  {title} {\enquote {\bibinfo {title} {Noninvasive measurements of spin
  transport properties of an antiferromagnetic insulator},}\ }\href
  {https://doi.org/10.1126/sciadv.abg8562} {\bibfield  {journal} {\bibinfo
  {journal} {Science Advances}\ } (\bibinfo {year} {2022}),\
  10.1126/sciadv.abg8562}\BibitemShut {NoStop}%
\bibitem [{\citenamefont {Degen}, \citenamefont {Reinhard},\ and\ \citenamefont
  {Cappellaro}(2017)}]{degenQuantumSensing2017}%
  \BibitemOpen
  \bibfield  {author} {\bibinfo {author} {\bibfnamefont {C.~L.}\ \bibnamefont
  {Degen}}, \bibinfo {author} {\bibfnamefont {F.}~\bibnamefont {Reinhard}},\
  and\ \bibinfo {author} {\bibfnamefont {P.}~\bibnamefont {Cappellaro}},\
  }\bibfield  {title} {\enquote {\bibinfo {title} {Quantum sensing},}\ }\href
  {https://doi.org/10.1103/RevModPhys.89.035002} {\bibfield  {journal}
  {\bibinfo  {journal} {Reviews of Modern Physics}\ }\textbf {\bibinfo {volume}
  {89}},\ \bibinfo {pages} {035002} (\bibinfo {year} {2017})}\BibitemShut
  {NoStop}%
\bibitem [{\citenamefont {Acosta}\ \emph {et~al.}(2009)\citenamefont {Acosta},
  \citenamefont {Bauch}, \citenamefont {Ledbetter}, \citenamefont {Santori},
  \citenamefont {Fu}, \citenamefont {Barclay}, \citenamefont {Beausoleil},
  \citenamefont {Linget}, \citenamefont {Roch}, \citenamefont {Treussart},
  \citenamefont {Chemerisov}, \citenamefont {Gawlik},\ and\ \citenamefont
  {Budker}}]{acostaDiamondsHighDensity2009}%
  \BibitemOpen
  \bibfield  {author} {\bibinfo {author} {\bibfnamefont {V.~M.}\ \bibnamefont
  {Acosta}}, \bibinfo {author} {\bibfnamefont {E.}~\bibnamefont {Bauch}},
  \bibinfo {author} {\bibfnamefont {M.~P.}\ \bibnamefont {Ledbetter}}, \bibinfo
  {author} {\bibfnamefont {C.}~\bibnamefont {Santori}}, \bibinfo {author}
  {\bibfnamefont {K.-M.~C.}\ \bibnamefont {Fu}}, \bibinfo {author}
  {\bibfnamefont {P.~E.}\ \bibnamefont {Barclay}}, \bibinfo {author}
  {\bibfnamefont {R.~G.}\ \bibnamefont {Beausoleil}}, \bibinfo {author}
  {\bibfnamefont {H.}~\bibnamefont {Linget}}, \bibinfo {author} {\bibfnamefont
  {J.~F.}\ \bibnamefont {Roch}}, \bibinfo {author} {\bibfnamefont
  {F.}~\bibnamefont {Treussart}}, \bibinfo {author} {\bibfnamefont
  {S.}~\bibnamefont {Chemerisov}}, \bibinfo {author} {\bibfnamefont
  {W.}~\bibnamefont {Gawlik}},\ and\ \bibinfo {author} {\bibfnamefont
  {D.}~\bibnamefont {Budker}},\ }\bibfield  {title} {\enquote {\bibinfo {title}
  {Diamonds with a high density of nitrogen-vacancy centers for magnetometry
  applications},}\ }\href {https://doi.org/10.1103/PhysRevB.80.115202}
  {\bibfield  {journal} {\bibinfo  {journal} {Physical Review B}\ }\textbf
  {\bibinfo {volume} {80}},\ \bibinfo {pages} {115202} (\bibinfo {year}
  {2009})}\BibitemShut {NoStop}%
\bibitem [{\citenamefont {Gruber}\ \emph {et~al.}(1997)\citenamefont {Gruber},
  \citenamefont {Dr{\"a}benstedt}, \citenamefont {Tietz}, \citenamefont
  {Fleury}, \citenamefont {Wrachtrup},\ and\ \citenamefont {von
  Borczyskowski}}]{gruberScanningConfocalOptical1997}%
  \BibitemOpen
  \bibfield  {author} {\bibinfo {author} {\bibfnamefont {A.}~\bibnamefont
  {Gruber}}, \bibinfo {author} {\bibfnamefont {A.}~\bibnamefont
  {Dr{\"a}benstedt}}, \bibinfo {author} {\bibfnamefont {C.}~\bibnamefont
  {Tietz}}, \bibinfo {author} {\bibfnamefont {L.}~\bibnamefont {Fleury}},
  \bibinfo {author} {\bibfnamefont {J.}~\bibnamefont {Wrachtrup}},\ and\
  \bibinfo {author} {\bibfnamefont {C.}~\bibnamefont {von Borczyskowski}},\
  }\bibfield  {title} {\enquote {\bibinfo {title} {Scanning {{Confocal Optical
  Microscopy}} and {{Magnetic Resonance}} on {{Single Defect Centers}}},}\
  }\href {https://doi.org/10.1126/science.276.5321.2012} {\bibfield  {journal}
  {\bibinfo  {journal} {Science}\ }\textbf {\bibinfo {volume} {276}},\ \bibinfo
  {pages} {2012--2014} (\bibinfo {year} {1997})}\BibitemShut {NoStop}%
\bibitem [{\citenamefont {Mittiga}\ \emph {et~al.}(2018)\citenamefont
  {Mittiga}, \citenamefont {Hsieh}, \citenamefont {Zu}, \citenamefont {Kobrin},
  \citenamefont {Machado}, \citenamefont {Bhattacharyya}, \citenamefont {Rui},
  \citenamefont {Jarmola}, \citenamefont {Choi}, \citenamefont {Budker},\ and\
  \citenamefont {Yao}}]{mittigaImagingLocalCharge2018}%
  \BibitemOpen
  \bibfield  {author} {\bibinfo {author} {\bibfnamefont {T.}~\bibnamefont
  {Mittiga}}, \bibinfo {author} {\bibfnamefont {S.}~\bibnamefont {Hsieh}},
  \bibinfo {author} {\bibfnamefont {C.}~\bibnamefont {Zu}}, \bibinfo {author}
  {\bibfnamefont {B.}~\bibnamefont {Kobrin}}, \bibinfo {author} {\bibfnamefont
  {F.}~\bibnamefont {Machado}}, \bibinfo {author} {\bibfnamefont
  {P.}~\bibnamefont {Bhattacharyya}}, \bibinfo {author} {\bibfnamefont {N.~Z.}\
  \bibnamefont {Rui}}, \bibinfo {author} {\bibfnamefont {A.}~\bibnamefont
  {Jarmola}}, \bibinfo {author} {\bibfnamefont {S.}~\bibnamefont {Choi}},
  \bibinfo {author} {\bibfnamefont {D.}~\bibnamefont {Budker}},\ and\ \bibinfo
  {author} {\bibfnamefont {N.~Y.}\ \bibnamefont {Yao}},\ }\bibfield  {title}
  {\enquote {\bibinfo {title} {Imaging the {{Local Charge Environment}} of
  {{Nitrogen-Vacancy Centers}} in {{Diamond}}},}\ }\href
  {https://doi.org/10.1103/PhysRevLett.121.246402} {\bibfield  {journal}
  {\bibinfo  {journal} {Physical Review Letters}\ }\textbf {\bibinfo {volume}
  {121}},\ \bibinfo {pages} {246402} (\bibinfo {year} {2018})}\BibitemShut
  {NoStop}%
\bibitem [{\citenamefont {Barry}\ \emph {et~al.}(2020)\citenamefont {Barry},
  \citenamefont {Schloss}, \citenamefont {Bauch}, \citenamefont {Turner},
  \citenamefont {Hart}, \citenamefont {Pham},\ and\ \citenamefont
  {Walsworth}}]{barrySensitivityOptimizationNVdiamond2020}%
  \BibitemOpen
  \bibfield  {author} {\bibinfo {author} {\bibfnamefont {J.~F.}\ \bibnamefont
  {Barry}}, \bibinfo {author} {\bibfnamefont {J.~M.}\ \bibnamefont {Schloss}},
  \bibinfo {author} {\bibfnamefont {E.}~\bibnamefont {Bauch}}, \bibinfo
  {author} {\bibfnamefont {M.~J.}\ \bibnamefont {Turner}}, \bibinfo {author}
  {\bibfnamefont {C.~A.}\ \bibnamefont {Hart}}, \bibinfo {author}
  {\bibfnamefont {L.~M.}\ \bibnamefont {Pham}},\ and\ \bibinfo {author}
  {\bibfnamefont {R.~L.}\ \bibnamefont {Walsworth}},\ }\bibfield  {title}
  {\enquote {\bibinfo {title} {Sensitivity optimization for {{NV-diamond}}
  magnetometry},}\ }\href {https://doi.org/10.1103/RevModPhys.92.015004}
  {\bibfield  {journal} {\bibinfo  {journal} {Reviews of Modern Physics}\
  }\textbf {\bibinfo {volume} {92}},\ \bibinfo {pages} {015004} (\bibinfo
  {year} {2020})}\BibitemShut {NoStop}%
\bibitem [{\citenamefont {Chernobrod}\ and\ \citenamefont
  {Berman}(2004)}]{chernobrodSpinMicroscopeBased2004}%
  \BibitemOpen
  \bibfield  {author} {\bibinfo {author} {\bibfnamefont {B.~M.}\ \bibnamefont
  {Chernobrod}}\ and\ \bibinfo {author} {\bibfnamefont {G.~P.}\ \bibnamefont
  {Berman}},\ }\bibfield  {title} {\enquote {\bibinfo {title} {Spin microscope
  based on optically detected magnetic resonance},}\ }\href
  {https://doi.org/10.1063/1.1829373} {\bibfield  {journal} {\bibinfo
  {journal} {Journal of Applied Physics}\ }\textbf {\bibinfo {volume} {97}},\
  \bibinfo {pages} {014903} (\bibinfo {year} {2004})}\BibitemShut {NoStop}%
\bibitem [{\citenamefont {Balasubramanian}\ \emph {et~al.}(2008)\citenamefont
  {Balasubramanian}, \citenamefont {Chan}, \citenamefont {Kolesov},
  \citenamefont {{Al-Hmoud}}, \citenamefont {Tisler}, \citenamefont {Shin},
  \citenamefont {Kim}, \citenamefont {Wojcik}, \citenamefont {Hemmer},
  \citenamefont {Krueger}, \citenamefont {Hanke}, \citenamefont
  {Leitenstorfer}, \citenamefont {Bratschitsch}, \citenamefont {Jelezko},\ and\
  \citenamefont {Wrachtrup}}]{balasubramanianNanoscaleImagingMagnetometry2008}%
  \BibitemOpen
  \bibfield  {author} {\bibinfo {author} {\bibfnamefont {G.}~\bibnamefont
  {Balasubramanian}}, \bibinfo {author} {\bibfnamefont {I.~Y.}\ \bibnamefont
  {Chan}}, \bibinfo {author} {\bibfnamefont {R.}~\bibnamefont {Kolesov}},
  \bibinfo {author} {\bibfnamefont {M.}~\bibnamefont {{Al-Hmoud}}}, \bibinfo
  {author} {\bibfnamefont {J.}~\bibnamefont {Tisler}}, \bibinfo {author}
  {\bibfnamefont {C.}~\bibnamefont {Shin}}, \bibinfo {author} {\bibfnamefont
  {C.}~\bibnamefont {Kim}}, \bibinfo {author} {\bibfnamefont {A.}~\bibnamefont
  {Wojcik}}, \bibinfo {author} {\bibfnamefont {P.~R.}\ \bibnamefont {Hemmer}},
  \bibinfo {author} {\bibfnamefont {A.}~\bibnamefont {Krueger}}, \bibinfo
  {author} {\bibfnamefont {T.}~\bibnamefont {Hanke}}, \bibinfo {author}
  {\bibfnamefont {A.}~\bibnamefont {Leitenstorfer}}, \bibinfo {author}
  {\bibfnamefont {R.}~\bibnamefont {Bratschitsch}}, \bibinfo {author}
  {\bibfnamefont {F.}~\bibnamefont {Jelezko}},\ and\ \bibinfo {author}
  {\bibfnamefont {J.}~\bibnamefont {Wrachtrup}},\ }\bibfield  {title} {\enquote
  {\bibinfo {title} {Nanoscale imaging magnetometry with diamond spins under
  ambient conditions},}\ }\href {https://doi.org/10.1038/nature07278}
  {\bibfield  {journal} {\bibinfo  {journal} {Nature}\ }\textbf {\bibinfo
  {volume} {455}},\ \bibinfo {pages} {648--651} (\bibinfo {year}
  {2008})}\BibitemShut {NoStop}%
\bibitem [{\citenamefont {Rondin}\ \emph {et~al.}(2012)\citenamefont {Rondin},
  \citenamefont {Tetienne}, \citenamefont {Spinicelli}, \citenamefont
  {Dal~Savio}, \citenamefont {Karrai}, \citenamefont {Dantelle}, \citenamefont
  {Thiaville}, \citenamefont {Rohart}, \citenamefont {Roch},\ and\
  \citenamefont {Jacques}}]{rondinNanoscaleMagneticField2012}%
  \BibitemOpen
  \bibfield  {author} {\bibinfo {author} {\bibfnamefont {L.}~\bibnamefont
  {Rondin}}, \bibinfo {author} {\bibfnamefont {J.-P.}\ \bibnamefont
  {Tetienne}}, \bibinfo {author} {\bibfnamefont {P.}~\bibnamefont
  {Spinicelli}}, \bibinfo {author} {\bibfnamefont {C.}~\bibnamefont
  {Dal~Savio}}, \bibinfo {author} {\bibfnamefont {K.}~\bibnamefont {Karrai}},
  \bibinfo {author} {\bibfnamefont {G.}~\bibnamefont {Dantelle}}, \bibinfo
  {author} {\bibfnamefont {A.}~\bibnamefont {Thiaville}}, \bibinfo {author}
  {\bibfnamefont {S.}~\bibnamefont {Rohart}}, \bibinfo {author} {\bibfnamefont
  {J.-F.}\ \bibnamefont {Roch}},\ and\ \bibinfo {author} {\bibfnamefont
  {V.}~\bibnamefont {Jacques}},\ }\bibfield  {title} {\enquote {\bibinfo
  {title} {Nanoscale magnetic field mapping with a single spin scanning probe
  magnetometer},}\ }\href {https://doi.org/10.1063/1.3703128} {\bibfield
  {journal} {\bibinfo  {journal} {Applied Physics Letters}\ }\textbf {\bibinfo
  {volume} {100}},\ \bibinfo {pages} {153118} (\bibinfo {year}
  {2012})}\BibitemShut {NoStop}%
\bibitem [{\citenamefont {Maletinsky}\ \emph {et~al.}(2012)\citenamefont
  {Maletinsky}, \citenamefont {Hong}, \citenamefont {Grinolds}, \citenamefont
  {Hausmann}, \citenamefont {Lukin}, \citenamefont {Walsworth}, \citenamefont
  {Loncar},\ and\ \citenamefont
  {Yacoby}}]{maletinskyRobustScanningDiamond2012}%
  \BibitemOpen
  \bibfield  {author} {\bibinfo {author} {\bibfnamefont {P.}~\bibnamefont
  {Maletinsky}}, \bibinfo {author} {\bibfnamefont {S.}~\bibnamefont {Hong}},
  \bibinfo {author} {\bibfnamefont {M.~S.}\ \bibnamefont {Grinolds}}, \bibinfo
  {author} {\bibfnamefont {B.}~\bibnamefont {Hausmann}}, \bibinfo {author}
  {\bibfnamefont {M.~D.}\ \bibnamefont {Lukin}}, \bibinfo {author}
  {\bibfnamefont {R.~L.}\ \bibnamefont {Walsworth}}, \bibinfo {author}
  {\bibfnamefont {M.}~\bibnamefont {Loncar}},\ and\ \bibinfo {author}
  {\bibfnamefont {A.}~\bibnamefont {Yacoby}},\ }\bibfield  {title} {\enquote
  {\bibinfo {title} {A robust scanning diamond sensor for nanoscale imaging
  with single nitrogen-vacancy centres},}\ }\href
  {https://doi.org/10.1038/nnano.2012.50} {\bibfield  {journal} {\bibinfo
  {journal} {Nature Nanotechnology}\ }\textbf {\bibinfo {volume} {7}},\
  \bibinfo {pages} {320--324} (\bibinfo {year} {2012})}\BibitemShut {NoStop}%
\bibitem [{\citenamefont {Appel}\ \emph {et~al.}(2016)\citenamefont {Appel},
  \citenamefont {Neu}, \citenamefont {Ganzhorn}, \citenamefont {Barfuss},
  \citenamefont {Batzer}, \citenamefont {Gratz}, \citenamefont {Tsch{\"o}pe},\
  and\ \citenamefont {Maletinsky}}]{appelFabricationAllDiamond2016}%
  \BibitemOpen
  \bibfield  {author} {\bibinfo {author} {\bibfnamefont {P.}~\bibnamefont
  {Appel}}, \bibinfo {author} {\bibfnamefont {E.}~\bibnamefont {Neu}}, \bibinfo
  {author} {\bibfnamefont {M.}~\bibnamefont {Ganzhorn}}, \bibinfo {author}
  {\bibfnamefont {A.}~\bibnamefont {Barfuss}}, \bibinfo {author} {\bibfnamefont
  {M.}~\bibnamefont {Batzer}}, \bibinfo {author} {\bibfnamefont
  {M.}~\bibnamefont {Gratz}}, \bibinfo {author} {\bibfnamefont
  {A.}~\bibnamefont {Tsch{\"o}pe}},\ and\ \bibinfo {author} {\bibfnamefont
  {P.}~\bibnamefont {Maletinsky}},\ }\bibfield  {title} {\enquote {\bibinfo
  {title} {Fabrication of all diamond scanning probes for nanoscale
  magnetometry},}\ }\href {https://doi.org/10.1063/1.4952953} {\bibfield
  {journal} {\bibinfo  {journal} {Review of Scientific Instruments}\ }\textbf
  {\bibinfo {volume} {87}},\ \bibinfo {pages} {063703} (\bibinfo {year}
  {2016})}\BibitemShut {NoStop}%
\bibitem [{\citenamefont {Hedrich}\ \emph {et~al.}(2020)\citenamefont
  {Hedrich}, \citenamefont {Rohner}, \citenamefont {Batzer}, \citenamefont
  {Maletinsky},\ and\ \citenamefont
  {Shields}}]{hedrichParabolicDiamondScanning2020}%
  \BibitemOpen
  \bibfield  {author} {\bibinfo {author} {\bibfnamefont {N.}~\bibnamefont
  {Hedrich}}, \bibinfo {author} {\bibfnamefont {D.}~\bibnamefont {Rohner}},
  \bibinfo {author} {\bibfnamefont {M.}~\bibnamefont {Batzer}}, \bibinfo
  {author} {\bibfnamefont {P.}~\bibnamefont {Maletinsky}},\ and\ \bibinfo
  {author} {\bibfnamefont {B.~J.}\ \bibnamefont {Shields}},\ }\bibfield
  {title} {\enquote {\bibinfo {title} {Parabolic {{Diamond Scanning Probes}}
  for {{Single-Spin Magnetic Field Imaging}}},}\ }\href
  {https://doi.org/10.1103/PhysRevApplied.14.064007} {\bibfield  {journal}
  {\bibinfo  {journal} {Physical Review Applied}\ }\textbf {\bibinfo {volume}
  {14}},\ \bibinfo {pages} {064007} (\bibinfo {year} {2020})}\BibitemShut
  {NoStop}%
\bibitem [{\citenamefont {Tetienne}\ \emph {et~al.}(2012)\citenamefont
  {Tetienne}, \citenamefont {Rondin}, \citenamefont {Spinicelli}, \citenamefont
  {Chipaux}, \citenamefont {Debuisschert}, \citenamefont {Roch},\ and\
  \citenamefont
  {Jacques}}]{tetienneMagneticfielddependentPhotodynamicsSingle2012}%
  \BibitemOpen
  \bibfield  {author} {\bibinfo {author} {\bibfnamefont {J.-P.}\ \bibnamefont
  {Tetienne}}, \bibinfo {author} {\bibfnamefont {L.}~\bibnamefont {Rondin}},
  \bibinfo {author} {\bibfnamefont {P.}~\bibnamefont {Spinicelli}}, \bibinfo
  {author} {\bibfnamefont {M.}~\bibnamefont {Chipaux}}, \bibinfo {author}
  {\bibfnamefont {T.}~\bibnamefont {Debuisschert}}, \bibinfo {author}
  {\bibfnamefont {J.-F.}\ \bibnamefont {Roch}},\ and\ \bibinfo {author}
  {\bibfnamefont {V.}~\bibnamefont {Jacques}},\ }\bibfield  {title} {\enquote
  {\bibinfo {title} {Magnetic-field-dependent photodynamics of single {{NV}}
  defects in diamond: An application to qualitative all-optical magnetic
  imaging},}\ }\href {https://doi.org/10.1088/1367-2630/14/10/103033}
  {\bibfield  {journal} {\bibinfo  {journal} {New Journal of Physics}\ }\textbf
  {\bibinfo {volume} {14}},\ \bibinfo {pages} {103033} (\bibinfo {year}
  {2012})}\BibitemShut {NoStop}%
\bibitem [{\citenamefont {Akhtar}\ \emph {et~al.}(2019)\citenamefont {Akhtar},
  \citenamefont {Hrabec}, \citenamefont {Chouaieb}, \citenamefont {Haykal},
  \citenamefont {Gross}, \citenamefont {Belmeguenai}, \citenamefont {Gabor},
  \citenamefont {Shields}, \citenamefont {Maletinsky}, \citenamefont
  {Thiaville}, \citenamefont {Rohart},\ and\ \citenamefont
  {Jacques}}]{akhtarCurrentInducedNucleationDynamics2019}%
  \BibitemOpen
  \bibfield  {author} {\bibinfo {author} {\bibfnamefont {W.}~\bibnamefont
  {Akhtar}}, \bibinfo {author} {\bibfnamefont {A.}~\bibnamefont {Hrabec}},
  \bibinfo {author} {\bibfnamefont {S.}~\bibnamefont {Chouaieb}}, \bibinfo
  {author} {\bibfnamefont {A.}~\bibnamefont {Haykal}}, \bibinfo {author}
  {\bibfnamefont {I.}~\bibnamefont {Gross}}, \bibinfo {author} {\bibfnamefont
  {M.}~\bibnamefont {Belmeguenai}}, \bibinfo {author} {\bibfnamefont
  {M.}~\bibnamefont {Gabor}}, \bibinfo {author} {\bibfnamefont
  {B.}~\bibnamefont {Shields}}, \bibinfo {author} {\bibfnamefont
  {P.}~\bibnamefont {Maletinsky}}, \bibinfo {author} {\bibfnamefont
  {A.}~\bibnamefont {Thiaville}}, \bibinfo {author} {\bibfnamefont
  {S.}~\bibnamefont {Rohart}},\ and\ \bibinfo {author} {\bibfnamefont
  {V.}~\bibnamefont {Jacques}},\ }\bibfield  {title} {\enquote {\bibinfo
  {title} {Current-{{Induced Nucleation}} and {{Dynamics}} of {{Skyrmions}} in
  a {{Co-based Heusler Alloy}}},}\ }\href
  {https://doi.org/10.1103/PhysRevApplied.11.034066} {\bibfield  {journal}
  {\bibinfo  {journal} {Physical Review Applied}\ }\textbf {\bibinfo {volume}
  {11}},\ \bibinfo {pages} {034066} (\bibinfo {year} {2019})}\BibitemShut
  {NoStop}%
\bibitem [{\citenamefont {Rohner}\ \emph {et~al.}(2019)\citenamefont {Rohner},
  \citenamefont {Happacher}, \citenamefont {Reiser}, \citenamefont {Tschudin},
  \citenamefont {Tallaire}, \citenamefont {Achard}, \citenamefont {Shields},\
  and\ \citenamefont {Maletinsky}}]{rohner111OrientedSingle2019}%
  \BibitemOpen
  \bibfield  {author} {\bibinfo {author} {\bibfnamefont {D.}~\bibnamefont
  {Rohner}}, \bibinfo {author} {\bibfnamefont {J.}~\bibnamefont {Happacher}},
  \bibinfo {author} {\bibfnamefont {P.}~\bibnamefont {Reiser}}, \bibinfo
  {author} {\bibfnamefont {M.~A.}\ \bibnamefont {Tschudin}}, \bibinfo {author}
  {\bibfnamefont {A.}~\bibnamefont {Tallaire}}, \bibinfo {author}
  {\bibfnamefont {J.}~\bibnamefont {Achard}}, \bibinfo {author} {\bibfnamefont
  {B.~J.}\ \bibnamefont {Shields}},\ and\ \bibinfo {author} {\bibfnamefont
  {P.}~\bibnamefont {Maletinsky}},\ }\bibfield  {title} {\enquote {\bibinfo
  {title} {(111)-oriented, single crystal diamond tips for nanoscale scanning
  probe imaging of out-of-plane magnetic fields},}\ }\href
  {https://doi.org/10.1063/1.5127101} {\bibfield  {journal} {\bibinfo
  {journal} {Applied Physics Letters}\ }\textbf {\bibinfo {volume} {115}},\
  \bibinfo {pages} {192401} (\bibinfo {year} {2019})}\BibitemShut {NoStop}%
\bibitem [{\citenamefont {Welter}\ \emph {et~al.}(2022)\citenamefont {Welter},
  \citenamefont {Rhensius}, \citenamefont {Morales}, \citenamefont
  {W{\"o}rnle}, \citenamefont {Lambert}, \citenamefont {{Puebla-Hellmann}},
  \citenamefont {Gambardella},\ and\ \citenamefont
  {Degen}}]{welterScanningNitrogenvacancyCenter2022a}%
  \BibitemOpen
  \bibfield  {author} {\bibinfo {author} {\bibfnamefont {P.}~\bibnamefont
  {Welter}}, \bibinfo {author} {\bibfnamefont {J.}~\bibnamefont {Rhensius}},
  \bibinfo {author} {\bibfnamefont {A.}~\bibnamefont {Morales}}, \bibinfo
  {author} {\bibfnamefont {M.~S.}\ \bibnamefont {W{\"o}rnle}}, \bibinfo
  {author} {\bibfnamefont {C.-H.}\ \bibnamefont {Lambert}}, \bibinfo {author}
  {\bibfnamefont {G.}~\bibnamefont {{Puebla-Hellmann}}}, \bibinfo {author}
  {\bibfnamefont {P.}~\bibnamefont {Gambardella}},\ and\ \bibinfo {author}
  {\bibfnamefont {C.~L.}\ \bibnamefont {Degen}},\ }\bibfield  {title} {\enquote
  {\bibinfo {title} {Scanning nitrogen-vacancy center magnetometry in large
  in-plane magnetic fields},}\ }\href {https://doi.org/10.1063/5.0084910}
  {\bibfield  {journal} {\bibinfo  {journal} {Applied Physics Letters}\
  }\textbf {\bibinfo {volume} {120}},\ \bibinfo {pages} {074003} (\bibinfo
  {year} {2022})}\BibitemShut {NoStop}%
\bibitem [{\citenamefont {Hingant}\ \emph {et~al.}(2015)\citenamefont
  {Hingant}, \citenamefont {Tetienne}, \citenamefont {Mart{\'i}nez},
  \citenamefont {Garcia}, \citenamefont {Ravelosona}, \citenamefont {Roch},\
  and\ \citenamefont {Jacques}}]{hingantMeasuringMagneticMoment2015}%
  \BibitemOpen
  \bibfield  {author} {\bibinfo {author} {\bibfnamefont {T.}~\bibnamefont
  {Hingant}}, \bibinfo {author} {\bibfnamefont {J.-P.}\ \bibnamefont
  {Tetienne}}, \bibinfo {author} {\bibfnamefont {L.~J.}\ \bibnamefont
  {Mart{\'i}nez}}, \bibinfo {author} {\bibfnamefont {K.}~\bibnamefont
  {Garcia}}, \bibinfo {author} {\bibfnamefont {D.}~\bibnamefont {Ravelosona}},
  \bibinfo {author} {\bibfnamefont {J.-F.}\ \bibnamefont {Roch}},\ and\
  \bibinfo {author} {\bibfnamefont {V.}~\bibnamefont {Jacques}},\ }\bibfield
  {title} {\enquote {\bibinfo {title} {Measuring the {{Magnetic Moment
  Density}} in {{Patterned Ultrathin Ferromagnets}} with {{Submicrometer
  Resolution}}},}\ }\href {https://doi.org/10.1103/PhysRevApplied.4.014003}
  {\bibfield  {journal} {\bibinfo  {journal} {Physical Review Applied}\
  }\textbf {\bibinfo {volume} {4}},\ \bibinfo {pages} {014003} (\bibinfo {year}
  {2015})}\BibitemShut {NoStop}%
\bibitem [{\citenamefont {He}\ \emph {et~al.}(2010)\citenamefont {He},
  \citenamefont {Wang}, \citenamefont {Wu}, \citenamefont {Caruso},
  \citenamefont {Vescovo}, \citenamefont {Belashchenko}, \citenamefont
  {Dowben},\ and\ \citenamefont {Binek}}]{heRobustIsothermalElectric2010a}%
  \BibitemOpen
  \bibfield  {author} {\bibinfo {author} {\bibfnamefont {X.}~\bibnamefont
  {He}}, \bibinfo {author} {\bibfnamefont {Y.}~\bibnamefont {Wang}}, \bibinfo
  {author} {\bibfnamefont {N.}~\bibnamefont {Wu}}, \bibinfo {author}
  {\bibfnamefont {A.~N.}\ \bibnamefont {Caruso}}, \bibinfo {author}
  {\bibfnamefont {E.}~\bibnamefont {Vescovo}}, \bibinfo {author} {\bibfnamefont
  {K.~D.}\ \bibnamefont {Belashchenko}}, \bibinfo {author} {\bibfnamefont
  {P.~A.}\ \bibnamefont {Dowben}},\ and\ \bibinfo {author} {\bibfnamefont
  {C.}~\bibnamefont {Binek}},\ }\bibfield  {title} {\enquote {\bibinfo {title}
  {Robust isothermal electric control of exchange bias at room temperature},}\
  }\href {https://doi.org/10.1038/nmat2785} {\bibfield  {journal} {\bibinfo
  {journal} {Nature Materials}\ }\textbf {\bibinfo {volume} {9}},\ \bibinfo
  {pages} {579--585} (\bibinfo {year} {2010})}\BibitemShut {NoStop}%
\bibitem [{\citenamefont
  {Dzyaloshinskii}(1958)}]{dzyaloshinskiiThermodynamicTheoryWeak1958}%
  \BibitemOpen
  \bibfield  {author} {\bibinfo {author} {\bibfnamefont {I.}~\bibnamefont
  {Dzyaloshinskii}},\ }\bibfield  {title} {\enquote {\bibinfo {title} {A
  thermodynamic theory of ``weak'' ferromagnetism of antiferromagnetics},}\
  }\href@noop {} {\bibfield  {journal} {\bibinfo  {journal} {Journal of Physics
  and Chemistry of Solids}\ }\textbf {\bibinfo {volume} {4}},\ \bibinfo {pages}
  {241--255} (\bibinfo {year} {1958})}\BibitemShut {NoStop}%
\bibitem [{\citenamefont {Ramazanoglu}\ \emph {et~al.}(2011)\citenamefont
  {Ramazanoglu}, \citenamefont {Laver}, \citenamefont {Ratcliff}, \citenamefont
  {Watson}, \citenamefont {Chen}, \citenamefont {Jackson}, \citenamefont
  {Kothapalli}, \citenamefont {Lee}, \citenamefont {Cheong},\ and\
  \citenamefont {Kiryukhin}}]{ramazanogluLocalWeakFerromagnetism2011}%
  \BibitemOpen
  \bibfield  {author} {\bibinfo {author} {\bibfnamefont {M.}~\bibnamefont
  {Ramazanoglu}}, \bibinfo {author} {\bibfnamefont {M.}~\bibnamefont {Laver}},
  \bibinfo {author} {\bibfnamefont {W.}~\bibnamefont {Ratcliff}}, \bibinfo
  {author} {\bibfnamefont {S.~M.}\ \bibnamefont {Watson}}, \bibinfo {author}
  {\bibfnamefont {W.~C.}\ \bibnamefont {Chen}}, \bibinfo {author}
  {\bibfnamefont {A.}~\bibnamefont {Jackson}}, \bibinfo {author} {\bibfnamefont
  {K.}~\bibnamefont {Kothapalli}}, \bibinfo {author} {\bibfnamefont
  {S.}~\bibnamefont {Lee}}, \bibinfo {author} {\bibfnamefont {S.-W.}\
  \bibnamefont {Cheong}},\ and\ \bibinfo {author} {\bibfnamefont
  {V.}~\bibnamefont {Kiryukhin}},\ }\bibfield  {title} {\enquote {\bibinfo
  {title} {Local {Weak Ferromagnetism} in {Single-Crystalline Ferroelectric}
  {BiFeO}\textsubscript{3}},}\ }\href
  {https://doi.org/10.1103/PhysRevLett.107.207206} {\bibfield  {journal}
  {\bibinfo  {journal} {Physical Review Letters}\ }\textbf {\bibinfo {volume}
  {107}},\ \bibinfo {pages} {207206} (\bibinfo {year} {2011})}\BibitemShut
  {NoStop}%
\bibitem [{\citenamefont {Tomiyoshi}\ and\ \citenamefont
  {Yamaguchi}(1982)}]{tomiyoshiMagneticStructureWeak1982}%
  \BibitemOpen
  \bibfield  {author} {\bibinfo {author} {\bibfnamefont {S.}~\bibnamefont
  {Tomiyoshi}}\ and\ \bibinfo {author} {\bibfnamefont {Y.}~\bibnamefont
  {Yamaguchi}},\ }\bibfield  {title} {\enquote {\bibinfo {title} {Magnetic
  {{Structure}} and {{Weak Ferromagnetism}} of {{Mn\textsubscript{3}Sn
  Studied}} by {{Polarized Neutron Diffraction}}},}\ }\href
  {https://doi.org/10.1143/JPSJ.51.2478} {\bibfield  {journal} {\bibinfo
  {journal} {Journal of the Physical Society of Japan}\ }\textbf {\bibinfo
  {volume} {51}},\ \bibinfo {pages} {2478--2486} (\bibinfo {year}
  {1982})}\BibitemShut {NoStop}%
\bibitem [{\citenamefont {Thiel}\ \emph {et~al.}(2019)\citenamefont {Thiel},
  \citenamefont {Wang}, \citenamefont {Tschudin}, \citenamefont {Rohner},
  \citenamefont {{Guti{\'e}rrez-Lezama}}, \citenamefont {Ubrig}, \citenamefont
  {Gibertini}, \citenamefont {Giannini}, \citenamefont {Morpurgo},\ and\
  \citenamefont {Maletinsky}}]{thielProbingMagnetism2D2019}%
  \BibitemOpen
  \bibfield  {author} {\bibinfo {author} {\bibfnamefont {L.}~\bibnamefont
  {Thiel}}, \bibinfo {author} {\bibfnamefont {Z.}~\bibnamefont {Wang}},
  \bibinfo {author} {\bibfnamefont {M.~A.}\ \bibnamefont {Tschudin}}, \bibinfo
  {author} {\bibfnamefont {D.}~\bibnamefont {Rohner}}, \bibinfo {author}
  {\bibfnamefont {I.}~\bibnamefont {{Guti{\'e}rrez-Lezama}}}, \bibinfo {author}
  {\bibfnamefont {N.}~\bibnamefont {Ubrig}}, \bibinfo {author} {\bibfnamefont
  {M.}~\bibnamefont {Gibertini}}, \bibinfo {author} {\bibfnamefont
  {E.}~\bibnamefont {Giannini}}, \bibinfo {author} {\bibfnamefont {A.~F.}\
  \bibnamefont {Morpurgo}},\ and\ \bibinfo {author} {\bibfnamefont
  {P.}~\bibnamefont {Maletinsky}},\ }\bibfield  {title} {\enquote {\bibinfo
  {title} {Probing magnetism in {{2D}} materials at the nanoscale with
  single-spin microscopy},}\ }\href {https://doi.org/10.1126/science.aav6926}
  {\bibfield  {journal} {\bibinfo  {journal} {Science}\ }\textbf {\bibinfo
  {volume} {364}},\ \bibinfo {pages} {973--976} (\bibinfo {year}
  {2019})}\BibitemShut {NoStop}%
\bibitem [{\citenamefont {Appel}\ \emph {et~al.}(2019)\citenamefont {Appel},
  \citenamefont {Shields}, \citenamefont {Kosub}, \citenamefont {Hedrich},
  \citenamefont {H{\"u}bner}, \citenamefont {Fa{\ss}bender}, \citenamefont
  {Makarov},\ and\ \citenamefont
  {Maletinsky}}]{appelNanomagnetismMagnetoelectricGranular2019}%
  \BibitemOpen
  \bibfield  {author} {\bibinfo {author} {\bibfnamefont {P.}~\bibnamefont
  {Appel}}, \bibinfo {author} {\bibfnamefont {B.~J.}\ \bibnamefont {Shields}},
  \bibinfo {author} {\bibfnamefont {T.}~\bibnamefont {Kosub}}, \bibinfo
  {author} {\bibfnamefont {N.}~\bibnamefont {Hedrich}}, \bibinfo {author}
  {\bibfnamefont {R.}~\bibnamefont {H{\"u}bner}}, \bibinfo {author}
  {\bibfnamefont {J.}~\bibnamefont {Fa{\ss}bender}}, \bibinfo {author}
  {\bibfnamefont {D.}~\bibnamefont {Makarov}},\ and\ \bibinfo {author}
  {\bibfnamefont {P.}~\bibnamefont {Maletinsky}},\ }\bibfield  {title}
  {\enquote {\bibinfo {title} {Nanomagnetism of {{Magnetoelectric Granular
  Thin-Film Antiferromagnets}}},}\ }\href
  {https://doi.org/10.1021/acs.nanolett.8b04681} {\bibfield  {journal}
  {\bibinfo  {journal} {Nano Letters}\ }\textbf {\bibinfo {volume} {19}},\
  \bibinfo {pages} {1682--1687} (\bibinfo {year} {2019})}\BibitemShut {NoStop}%
\bibitem [{\citenamefont {Hedrich}\ \emph {et~al.}(2021)\citenamefont
  {Hedrich}, \citenamefont {Wagner}, \citenamefont {Pylypovskyi}, \citenamefont
  {Shields}, \citenamefont {Kosub}, \citenamefont {Sheka}, \citenamefont
  {Makarov},\ and\ \citenamefont
  {Maletinsky}}]{hedrichNanoscaleMechanicsAntiferromagnetic2021}%
  \BibitemOpen
  \bibfield  {author} {\bibinfo {author} {\bibfnamefont {N.}~\bibnamefont
  {Hedrich}}, \bibinfo {author} {\bibfnamefont {K.}~\bibnamefont {Wagner}},
  \bibinfo {author} {\bibfnamefont {O.~V.}\ \bibnamefont {Pylypovskyi}},
  \bibinfo {author} {\bibfnamefont {B.~J.}\ \bibnamefont {Shields}}, \bibinfo
  {author} {\bibfnamefont {T.}~\bibnamefont {Kosub}}, \bibinfo {author}
  {\bibfnamefont {D.~D.}\ \bibnamefont {Sheka}}, \bibinfo {author}
  {\bibfnamefont {D.}~\bibnamefont {Makarov}},\ and\ \bibinfo {author}
  {\bibfnamefont {P.}~\bibnamefont {Maletinsky}},\ }\bibfield  {title}
  {\enquote {\bibinfo {title} {Nanoscale mechanics of antiferromagnetic domain
  walls},}\ }\href {https://doi.org/10.1038/s41567-020-01157-0} {\bibfield
  {journal} {\bibinfo  {journal} {Nature Physics}\ }\textbf {\bibinfo {volume}
  {17}},\ \bibinfo {pages} {574--577} (\bibinfo {year} {2021})}\BibitemShut
  {NoStop}%
\bibitem [{\citenamefont {Wörnle}\ \emph {et~al.}(2021)\citenamefont
  {Wörnle}, \citenamefont {Welter}, \citenamefont {Giraldo}, \citenamefont
  {Lottermoser}, \citenamefont {Fiebig}, \citenamefont {Gambardella},\ and\
  \citenamefont {Degen}}]{wornleCoexistenceBlochEel2021}%
  \BibitemOpen
  \bibfield  {author} {\bibinfo {author} {\bibfnamefont {M.~S.}\ \bibnamefont
  {Wörnle}}, \bibinfo {author} {\bibfnamefont {P.}~\bibnamefont {Welter}},
  \bibinfo {author} {\bibfnamefont {M.}~\bibnamefont {Giraldo}}, \bibinfo
  {author} {\bibfnamefont {T.}~\bibnamefont {Lottermoser}}, \bibinfo {author}
  {\bibfnamefont {M.}~\bibnamefont {Fiebig}}, \bibinfo {author} {\bibfnamefont
  {P.}~\bibnamefont {Gambardella}},\ and\ \bibinfo {author} {\bibfnamefont
  {C.~L.}\ \bibnamefont {Degen}},\ }\bibfield  {title} {\enquote {\bibinfo
  {title} {Coexistence of {{Bloch}} and {{Néel}} walls in a collinear
  antiferromagnet},}\ }\href {https://doi.org/10.1103/PhysRevB.103.094426}
  {\bibfield  {journal} {\bibinfo  {journal} {Physical Review B}\ }\textbf
  {\bibinfo {volume} {103}},\ \bibinfo {pages} {094426} (\bibinfo {year}
  {2021})}\BibitemShut {NoStop}%
\bibitem [{\citenamefont {Erickson}\ \emph {et~al.}(2022)\citenamefont
  {Erickson}, \citenamefont {Shah}, \citenamefont {Mahmood}, \citenamefont
  {Fescenko}, \citenamefont {Timalsina}, \citenamefont {Binek},\ and\
  \citenamefont {Laraoui}}]{ericksonNanoscaleImagingAntiferromagnetic2022}%
  \BibitemOpen
  \bibfield  {author} {\bibinfo {author} {\bibfnamefont {A.}~\bibnamefont
  {Erickson}}, \bibinfo {author} {\bibfnamefont {S.~Q.~A.}\ \bibnamefont
  {Shah}}, \bibinfo {author} {\bibfnamefont {A.}~\bibnamefont {Mahmood}},
  \bibinfo {author} {\bibfnamefont {I.}~\bibnamefont {Fescenko}}, \bibinfo
  {author} {\bibfnamefont {R.}~\bibnamefont {Timalsina}}, \bibinfo {author}
  {\bibfnamefont {C.}~\bibnamefont {Binek}},\ and\ \bibinfo {author}
  {\bibfnamefont {A.}~\bibnamefont {Laraoui}},\ }\bibfield  {title} {\enquote
  {\bibinfo {title} {Nanoscale imaging of antiferromagnetic domains in
  epitaxial films of {{Cr\textsubscript{2}O\textsubscript{3}}} via scanning
  diamond magnetic probe microscopy},}\ }\href
  {https://doi.org/10.1039/D2RA06440E} {\bibfield  {journal} {\bibinfo
  {journal} {RSC Advances}\ }\textbf {\bibinfo {volume} {13}},\ \bibinfo
  {pages} {178--185} (\bibinfo {year} {2022})}\BibitemShut {NoStop}%
\bibitem [{\citenamefont {Makushko}\ \emph {et~al.}(2022)\citenamefont
  {Makushko}, \citenamefont {Kosub}, \citenamefont {Pylypovskyi}, \citenamefont
  {Hedrich}, \citenamefont {Li}, \citenamefont {Pashkin}, \citenamefont
  {Avdoshenko}, \citenamefont {H{\"u}bner}, \citenamefont {Ganss},
  \citenamefont {Wolf}, \citenamefont {Lubk}, \citenamefont {Liedke},
  \citenamefont {Butterling}, \citenamefont {Wagner}, \citenamefont {Wagner},
  \citenamefont {Shields}, \citenamefont {Lehmann}, \citenamefont {Veremchuk},
  \citenamefont {Fassbender}, \citenamefont {Maletinsky},\ and\ \citenamefont
  {Makarov}}]{makushkoFlexomagnetismVerticallyGraded2022}%
  \BibitemOpen
  \bibfield  {author} {\bibinfo {author} {\bibfnamefont {P.}~\bibnamefont
  {Makushko}}, \bibinfo {author} {\bibfnamefont {T.}~\bibnamefont {Kosub}},
  \bibinfo {author} {\bibfnamefont {O.~V.}\ \bibnamefont {Pylypovskyi}},
  \bibinfo {author} {\bibfnamefont {N.}~\bibnamefont {Hedrich}}, \bibinfo
  {author} {\bibfnamefont {J.}~\bibnamefont {Li}}, \bibinfo {author}
  {\bibfnamefont {A.}~\bibnamefont {Pashkin}}, \bibinfo {author} {\bibfnamefont
  {S.}~\bibnamefont {Avdoshenko}}, \bibinfo {author} {\bibfnamefont
  {R.}~\bibnamefont {H{\"u}bner}}, \bibinfo {author} {\bibfnamefont
  {F.}~\bibnamefont {Ganss}}, \bibinfo {author} {\bibfnamefont
  {D.}~\bibnamefont {Wolf}}, \bibinfo {author} {\bibfnamefont {A.}~\bibnamefont
  {Lubk}}, \bibinfo {author} {\bibfnamefont {M.~O.}\ \bibnamefont {Liedke}},
  \bibinfo {author} {\bibfnamefont {M.}~\bibnamefont {Butterling}}, \bibinfo
  {author} {\bibfnamefont {A.}~\bibnamefont {Wagner}}, \bibinfo {author}
  {\bibfnamefont {K.}~\bibnamefont {Wagner}}, \bibinfo {author} {\bibfnamefont
  {B.~J.}\ \bibnamefont {Shields}}, \bibinfo {author} {\bibfnamefont
  {P.}~\bibnamefont {Lehmann}}, \bibinfo {author} {\bibfnamefont
  {I.}~\bibnamefont {Veremchuk}}, \bibinfo {author} {\bibfnamefont
  {J.}~\bibnamefont {Fassbender}}, \bibinfo {author} {\bibfnamefont
  {P.}~\bibnamefont {Maletinsky}},\ and\ \bibinfo {author} {\bibfnamefont
  {D.}~\bibnamefont {Makarov}},\ }\bibfield  {title} {\enquote {\bibinfo
  {title} {Flexomagnetism and vertically graded {{N\'eel}} temperature of
  antiferromagnetic {{Cr\textsubscript{2}O\textsubscript{3}}} thin films},}\
  }\href {https://doi.org/10.1038/s41467-022-34233-5} {\bibfield  {journal}
  {\bibinfo  {journal} {Nature Communications}\ }\textbf {\bibinfo {volume}
  {13}},\ \bibinfo {pages} {6745} (\bibinfo {year} {2022})}\BibitemShut
  {NoStop}%
\bibitem [{\citenamefont {Wu}\ \emph {et~al.}(2011)\citenamefont {Wu},
  \citenamefont {He}, \citenamefont {Wysocki}, \citenamefont {Lanke},
  \citenamefont {Komesu}, \citenamefont {Belashchenko}, \citenamefont {Binek},\
  and\ \citenamefont {Dowben}}]{wuImagingControlSurface2011}%
  \BibitemOpen
  \bibfield  {author} {\bibinfo {author} {\bibfnamefont {N.}~\bibnamefont
  {Wu}}, \bibinfo {author} {\bibfnamefont {X.}~\bibnamefont {He}}, \bibinfo
  {author} {\bibfnamefont {A.~L.}\ \bibnamefont {Wysocki}}, \bibinfo {author}
  {\bibfnamefont {U.}~\bibnamefont {Lanke}}, \bibinfo {author} {\bibfnamefont
  {T.}~\bibnamefont {Komesu}}, \bibinfo {author} {\bibfnamefont {K.~D.}\
  \bibnamefont {Belashchenko}}, \bibinfo {author} {\bibfnamefont
  {C.}~\bibnamefont {Binek}},\ and\ \bibinfo {author} {\bibfnamefont {P.~A.}\
  \bibnamefont {Dowben}},\ }\bibfield  {title} {\enquote {\bibinfo {title}
  {Imaging and {{Control}} of {{Surface Magnetization Domains}} in a
  {{Magnetoelectric Antiferromagnet}}},}\ }\href
  {https://doi.org/10.1103/PhysRevLett.106.087202} {\bibfield  {journal}
  {\bibinfo  {journal} {Physical Review Letters}\ }\textbf {\bibinfo {volume}
  {106}},\ \bibinfo {pages} {087202} (\bibinfo {year} {2011})}\BibitemShut
  {NoStop}%
\bibitem [{\citenamefont {Schoenherr}\ \emph {et~al.}(2017)\citenamefont
  {Schoenherr}, \citenamefont {Giraldo}, \citenamefont {Lilienblum},
  \citenamefont {Trassin}, \citenamefont {Meier},\ and\ \citenamefont
  {Fiebig}}]{schoenherrMagnetoelectricForceMicroscopy2017}%
  \BibitemOpen
  \bibfield  {author} {\bibinfo {author} {\bibfnamefont {P.}~\bibnamefont
  {Schoenherr}}, \bibinfo {author} {\bibfnamefont {L.~M.}\ \bibnamefont
  {Giraldo}}, \bibinfo {author} {\bibfnamefont {M.}~\bibnamefont {Lilienblum}},
  \bibinfo {author} {\bibfnamefont {M.}~\bibnamefont {Trassin}}, \bibinfo
  {author} {\bibfnamefont {D.}~\bibnamefont {Meier}},\ and\ \bibinfo {author}
  {\bibfnamefont {M.}~\bibnamefont {Fiebig}},\ }\bibfield  {title} {\enquote
  {\bibinfo {title} {Magnetoelectric {{Force Microscopy}} on
  {{Antiferromagnetic}} 180{$\circ$} {{Domains}} in
  {{Cr\textsubscript{2}O\textsubscript{3}}}},}\ }\href
  {https://doi.org/10.3390/ma10091051} {\bibfield  {journal} {\bibinfo
  {journal} {Materials}\ }\textbf {\bibinfo {volume} {10}},\ \bibinfo {pages}
  {1051} (\bibinfo {year} {2017})}\BibitemShut {NoStop}%
\bibitem [{\citenamefont {Hayashida}\ \emph {et~al.}(2022)\citenamefont
  {Hayashida}, \citenamefont {Arakawa}, \citenamefont {Oshima}, \citenamefont
  {Kimura},\ and\ \citenamefont
  {Kimura}}]{hayashidaObservationAntiferromagneticDomains2022}%
  \BibitemOpen
  \bibfield  {author} {\bibinfo {author} {\bibfnamefont {T.}~\bibnamefont
  {Hayashida}}, \bibinfo {author} {\bibfnamefont {K.}~\bibnamefont {Arakawa}},
  \bibinfo {author} {\bibfnamefont {T.}~\bibnamefont {Oshima}}, \bibinfo
  {author} {\bibfnamefont {K.}~\bibnamefont {Kimura}},\ and\ \bibinfo {author}
  {\bibfnamefont {T.}~\bibnamefont {Kimura}},\ }\bibfield  {title} {\enquote
  {\bibinfo {title} {Observation of antiferromagnetic domains in
  {C}r\textsubscript{2}{O}\textsubscript{3} using nonreciprocal optical
  effects},}\ }\href {https://doi.org/10.1103/PhysRevResearch.4.043063}
  {\bibfield  {journal} {\bibinfo  {journal} {Physical Review Research}\
  }\textbf {\bibinfo {volume} {4}},\ \bibinfo {pages} {043063} (\bibinfo {year}
  {2022})}\BibitemShut {NoStop}%
\bibitem [{\citenamefont {Broadway}\ \emph {et~al.}(2020)\citenamefont
  {Broadway}, \citenamefont {Lillie}, \citenamefont {Scholten}, \citenamefont
  {Rohner}, \citenamefont {Dontschuk}, \citenamefont {Maletinsky},
  \citenamefont {Tetienne},\ and\ \citenamefont
  {Hollenberg}}]{broadwayImprovedCurrentDensity2020a}%
  \BibitemOpen
  \bibfield  {author} {\bibinfo {author} {\bibfnamefont {D.}~\bibnamefont
  {Broadway}}, \bibinfo {author} {\bibfnamefont {S.}~\bibnamefont {Lillie}},
  \bibinfo {author} {\bibfnamefont {S.}~\bibnamefont {Scholten}}, \bibinfo
  {author} {\bibfnamefont {D.}~\bibnamefont {Rohner}}, \bibinfo {author}
  {\bibfnamefont {N.}~\bibnamefont {Dontschuk}}, \bibinfo {author}
  {\bibfnamefont {P.}~\bibnamefont {Maletinsky}}, \bibinfo {author}
  {\bibfnamefont {J.-P.}\ \bibnamefont {Tetienne}},\ and\ \bibinfo {author}
  {\bibfnamefont {L.}~\bibnamefont {Hollenberg}},\ }\bibfield  {title}
  {\enquote {\bibinfo {title} {Improved {{Current Density}} and {{Magnetization
  Reconstruction Through Vector Magnetic Field Measurements}}},}\ }\href
  {https://doi.org/10.1103/PhysRevApplied.14.024076} {\bibfield  {journal}
  {\bibinfo  {journal} {Physical Review Applied}\ }\textbf {\bibinfo {volume}
  {14}},\ \bibinfo {pages} {024076} (\bibinfo {year} {2020})}\BibitemShut
  {NoStop}%
\bibitem [{\citenamefont {Duine}\ \emph {et~al.}(2018)\citenamefont {Duine},
  \citenamefont {Lee}, \citenamefont {Parkin},\ and\ \citenamefont
  {Stiles}}]{duineSyntheticAntiferromagneticSpintronics2018}%
  \BibitemOpen
  \bibfield  {author} {\bibinfo {author} {\bibfnamefont {R.~A.}\ \bibnamefont
  {Duine}}, \bibinfo {author} {\bibfnamefont {K.-J.}\ \bibnamefont {Lee}},
  \bibinfo {author} {\bibfnamefont {S.~S.~P.}\ \bibnamefont {Parkin}},\ and\
  \bibinfo {author} {\bibfnamefont {M.~D.}\ \bibnamefont {Stiles}},\ }\bibfield
   {title} {\enquote {\bibinfo {title} {Synthetic antiferromagnetic
  spintronics},}\ }\href {https://doi.org/10.1038/s41567-018-0050-y} {\bibfield
   {journal} {\bibinfo  {journal} {Nature Physics}\ }\textbf {\bibinfo {volume}
  {14}},\ \bibinfo {pages} {217--219} (\bibinfo {year} {2018})}\BibitemShut
  {NoStop}%
\bibitem [{\citenamefont {Finco}\ \emph {et~al.}(2021)\citenamefont {Finco},
  \citenamefont {Haykal}, \citenamefont {Tanos}, \citenamefont {Fabre},
  \citenamefont {Chouaieb}, \citenamefont {Akhtar}, \citenamefont
  {{Robert-Philip}}, \citenamefont {Legrand}, \citenamefont {Ajejas},
  \citenamefont {Bouzehouane}, \citenamefont {Reyren}, \citenamefont
  {Devolder}, \citenamefont {Adam}, \citenamefont {Kim}, \citenamefont {Cros},\
  and\ \citenamefont
  {Jacques}}]{fincoImagingNoncollinearAntiferromagnetic2021}%
  \BibitemOpen
  \bibfield  {author} {\bibinfo {author} {\bibfnamefont {A.}~\bibnamefont
  {Finco}}, \bibinfo {author} {\bibfnamefont {A.}~\bibnamefont {Haykal}},
  \bibinfo {author} {\bibfnamefont {R.}~\bibnamefont {Tanos}}, \bibinfo
  {author} {\bibfnamefont {F.}~\bibnamefont {Fabre}}, \bibinfo {author}
  {\bibfnamefont {S.}~\bibnamefont {Chouaieb}}, \bibinfo {author}
  {\bibfnamefont {W.}~\bibnamefont {Akhtar}}, \bibinfo {author} {\bibfnamefont
  {I.}~\bibnamefont {{Robert-Philip}}}, \bibinfo {author} {\bibfnamefont
  {W.}~\bibnamefont {Legrand}}, \bibinfo {author} {\bibfnamefont
  {F.}~\bibnamefont {Ajejas}}, \bibinfo {author} {\bibfnamefont
  {K.}~\bibnamefont {Bouzehouane}}, \bibinfo {author} {\bibfnamefont
  {N.}~\bibnamefont {Reyren}}, \bibinfo {author} {\bibfnamefont
  {T.}~\bibnamefont {Devolder}}, \bibinfo {author} {\bibfnamefont {J.-P.}\
  \bibnamefont {Adam}}, \bibinfo {author} {\bibfnamefont {J.-V.}\ \bibnamefont
  {Kim}}, \bibinfo {author} {\bibfnamefont {V.}~\bibnamefont {Cros}},\ and\
  \bibinfo {author} {\bibfnamefont {V.}~\bibnamefont {Jacques}},\ }\bibfield
  {title} {\enquote {\bibinfo {title} {Imaging non-collinear antiferromagnetic
  textures via single spin relaxometry},}\ }\href
  {https://doi.org/10.1038/s41467-021-20995-x} {\bibfield  {journal} {\bibinfo
  {journal} {Nature Communications}\ }\textbf {\bibinfo {volume} {12}},\
  \bibinfo {pages} {767} (\bibinfo {year} {2021})}\BibitemShut {NoStop}%
\bibitem [{\citenamefont {Legrand}\ \emph {et~al.}(2020)\citenamefont
  {Legrand}, \citenamefont {Maccariello}, \citenamefont {Ajejas}, \citenamefont
  {Collin}, \citenamefont {Vecchiola}, \citenamefont {Bouzehouane},
  \citenamefont {Reyren}, \citenamefont {Cros},\ and\ \citenamefont
  {Fert}}]{legrandRoomtemperatureStabilizationAntiferromagnetic2020}%
  \BibitemOpen
  \bibfield  {author} {\bibinfo {author} {\bibfnamefont {W.}~\bibnamefont
  {Legrand}}, \bibinfo {author} {\bibfnamefont {D.}~\bibnamefont
  {Maccariello}}, \bibinfo {author} {\bibfnamefont {F.}~\bibnamefont {Ajejas}},
  \bibinfo {author} {\bibfnamefont {S.}~\bibnamefont {Collin}}, \bibinfo
  {author} {\bibfnamefont {A.}~\bibnamefont {Vecchiola}}, \bibinfo {author}
  {\bibfnamefont {K.}~\bibnamefont {Bouzehouane}}, \bibinfo {author}
  {\bibfnamefont {N.}~\bibnamefont {Reyren}}, \bibinfo {author} {\bibfnamefont
  {V.}~\bibnamefont {Cros}},\ and\ \bibinfo {author} {\bibfnamefont
  {A.}~\bibnamefont {Fert}},\ }\bibfield  {title} {\enquote {\bibinfo {title}
  {Room-temperature stabilization of antiferromagnetic skyrmions in synthetic
  antiferromagnets},}\ }\href {https://doi.org/10.1038/s41563-019-0468-3}
  {\bibfield  {journal} {\bibinfo  {journal} {Nature Materials}\ }\textbf
  {\bibinfo {volume} {19}},\ \bibinfo {pages} {34--42} (\bibinfo {year}
  {2020})}\BibitemShut {NoStop}%
\bibitem [{\citenamefont {Guo}\ \emph {et~al.}(2023)\citenamefont {Guo},
  \citenamefont {D'Addario}, \citenamefont {Cheng}, \citenamefont {Kline},
  \citenamefont {Gray}, \citenamefont {Cheung}, \citenamefont {Yang},
  \citenamefont {Nowack},\ and\ \citenamefont
  {Fuchs}}]{guoCurrentinducedSwitchingThin2023}%
  \BibitemOpen
  \bibfield  {author} {\bibinfo {author} {\bibfnamefont {Q.}~\bibnamefont
  {Guo}}, \bibinfo {author} {\bibfnamefont {A.}~\bibnamefont {D'Addario}},
  \bibinfo {author} {\bibfnamefont {Y.}~\bibnamefont {Cheng}}, \bibinfo
  {author} {\bibfnamefont {J.}~\bibnamefont {Kline}}, \bibinfo {author}
  {\bibfnamefont {I.}~\bibnamefont {Gray}}, \bibinfo {author} {\bibfnamefont
  {H.~F.~H.}\ \bibnamefont {Cheung}}, \bibinfo {author} {\bibfnamefont
  {F.}~\bibnamefont {Yang}}, \bibinfo {author} {\bibfnamefont {K.~C.}\
  \bibnamefont {Nowack}},\ and\ \bibinfo {author} {\bibfnamefont {G.~D.}\
  \bibnamefont {Fuchs}},\ }\bibfield  {title} {\enquote {\bibinfo {title}
  {Current-induced switching of thin film
  $\alpha$-{Fe\textsubscript{2}O\textsubscript{3}} devices imaged using a
  scanning single-spin microscope},}\ }\href
  {https://doi.org/10.1103/PhysRevMaterials.7.064402} {\bibfield  {journal}
  {\bibinfo  {journal} {Physical Review Materials}\ }\textbf {\bibinfo {volume}
  {7}},\ \bibinfo {pages} {064402} (\bibinfo {year} {2023})}\BibitemShut
  {NoStop}%
\bibitem [{\citenamefont {Tan}\ \emph {et~al.}(2023)\citenamefont {Tan},
  \citenamefont {Jani}, \citenamefont {H{\"o}gen}, \citenamefont {Stefan},
  \citenamefont {Castelnovo}, \citenamefont {Braund}, \citenamefont {Geim},
  \citenamefont {Feuer}, \citenamefont {Knowles}, \citenamefont {Ariando},
  \citenamefont {Radaelli},\ and\ \citenamefont
  {Atat{\"u}re}}]{tanRevealingEmergentMagnetic2023}%
  \BibitemOpen
  \bibfield  {author} {\bibinfo {author} {\bibfnamefont {A.~K.~C.}\
  \bibnamefont {Tan}}, \bibinfo {author} {\bibfnamefont {H.}~\bibnamefont
  {Jani}}, \bibinfo {author} {\bibfnamefont {M.}~\bibnamefont {H{\"o}gen}},
  \bibinfo {author} {\bibfnamefont {L.}~\bibnamefont {Stefan}}, \bibinfo
  {author} {\bibfnamefont {C.}~\bibnamefont {Castelnovo}}, \bibinfo {author}
  {\bibfnamefont {D.}~\bibnamefont {Braund}}, \bibinfo {author} {\bibfnamefont
  {A.}~\bibnamefont {Geim}}, \bibinfo {author} {\bibfnamefont {M.~S.~G.}\
  \bibnamefont {Feuer}}, \bibinfo {author} {\bibfnamefont {H.~S.}\ \bibnamefont
  {Knowles}}, \bibinfo {author} {\bibfnamefont {A.}~\bibnamefont {Ariando}},
  \bibinfo {author} {\bibfnamefont {P.~G.}\ \bibnamefont {Radaelli}},\ and\
  \bibinfo {author} {\bibfnamefont {M.}~\bibnamefont {Atat{\"u}re}},\
  }\bibfield  {title} {\enquote {\bibinfo {title} {Revealing {{Emergent
  Magnetic Charge}} in an {{Antiferromagnet}} with {{Diamond Quantum
  Magnetometry}}},}\ }\href {https://doi.org/10.48550/arXiv.2303.12125}
  {\bibfield  {journal} {\bibinfo  {journal} {arXiv:2303.12125}\ } (\bibinfo
  {year} {2023}),\ 10.48550/arXiv.2303.12125}\BibitemShut {NoStop}%
\bibitem [{\citenamefont {Welter}\ \emph {et~al.}(2023)\citenamefont {Welter},
  \citenamefont {J{\'o}steinsson}, \citenamefont {Josephy}, \citenamefont
  {Wittmann}, \citenamefont {Morales}, \citenamefont {{Puebla-Hellmann}},\ and\
  \citenamefont {Degen}}]{welterFastScanningNitrogenvacancy2022}%
  \BibitemOpen
  \bibfield  {author} {\bibinfo {author} {\bibfnamefont {P.}~\bibnamefont
  {Welter}}, \bibinfo {author} {\bibfnamefont {B.}~\bibnamefont
  {J{\'o}steinsson}}, \bibinfo {author} {\bibfnamefont {S.}~\bibnamefont
  {Josephy}}, \bibinfo {author} {\bibfnamefont {A.}~\bibnamefont {Wittmann}},
  \bibinfo {author} {\bibfnamefont {A.}~\bibnamefont {Morales}}, \bibinfo
  {author} {\bibfnamefont {G.}~\bibnamefont {{Puebla-Hellmann}}},\ and\
  \bibinfo {author} {\bibfnamefont {C.}~\bibnamefont {Degen}},\ }\bibfield
  {title} {\enquote {\bibinfo {title} {Fast {{Scanning Nitrogen-Vacancy
  Magnetometry}} by {{Spectrum Demodulation}}},}\ }\href
  {https://doi.org/10.1103/PhysRevApplied.19.034003} {\bibfield  {journal}
  {\bibinfo  {journal} {Physical Review Applied}\ }\textbf {\bibinfo {volume}
  {19}},\ \bibinfo {pages} {034003} (\bibinfo {year} {2023})}\BibitemShut
  {NoStop}%
\bibitem [{\citenamefont {Yan}\ \emph {et~al.}(2022)\citenamefont {Yan},
  \citenamefont {Li}, \citenamefont {Lu}, \citenamefont {Huang}, \citenamefont
  {Xiao}, \citenamefont {Wernert}, \citenamefont {Brock}, \citenamefont
  {Fullerton}, \citenamefont {Chen}, \citenamefont {Wang},\ and\ \citenamefont
  {Du}}]{yanQuantumSensingImaging2022}%
  \BibitemOpen
  \bibfield  {author} {\bibinfo {author} {\bibfnamefont {G.~Q.}\ \bibnamefont
  {Yan}}, \bibinfo {author} {\bibfnamefont {S.}~\bibnamefont {Li}}, \bibinfo
  {author} {\bibfnamefont {H.}~\bibnamefont {Lu}}, \bibinfo {author}
  {\bibfnamefont {M.}~\bibnamefont {Huang}}, \bibinfo {author} {\bibfnamefont
  {Y.}~\bibnamefont {Xiao}}, \bibinfo {author} {\bibfnamefont {L.}~\bibnamefont
  {Wernert}}, \bibinfo {author} {\bibfnamefont {J.~A.}\ \bibnamefont {Brock}},
  \bibinfo {author} {\bibfnamefont {E.~E.}\ \bibnamefont {Fullerton}}, \bibinfo
  {author} {\bibfnamefont {H.}~\bibnamefont {Chen}}, \bibinfo {author}
  {\bibfnamefont {H.}~\bibnamefont {Wang}},\ and\ \bibinfo {author}
  {\bibfnamefont {C.~R.}\ \bibnamefont {Du}},\ }\bibfield  {title} {\enquote
  {\bibinfo {title} {Quantum {{Sensing}} and {{Imaging}} of
  {{Spin}}\textendash{{Orbit-Torque-Driven Spin Dynamics}} in the
  {{Non-Collinear Antiferromagnet Mn\textsubscript{3}Sn}}},}\ }\href
  {https://doi.org/10.1002/adma.202200327} {\bibfield  {journal} {\bibinfo
  {journal} {Advanced Materials}\ }\textbf {\bibinfo {volume} {34}},\ \bibinfo
  {pages} {2200327} (\bibinfo {year} {2022})}\BibitemShut {NoStop}%
\bibitem [{\citenamefont {Li}\ \emph {et~al.}(2023)\citenamefont {Li},
  \citenamefont {Huang}, \citenamefont {Lu}, \citenamefont {McLaughlin},
  \citenamefont {Xiao}, \citenamefont {Zhou}, \citenamefont {Fullerton},
  \citenamefont {Chen}, \citenamefont {Wang},\ and\ \citenamefont
  {Du}}]{liNanoscaleMagneticDomains2023}%
  \BibitemOpen
  \bibfield  {author} {\bibinfo {author} {\bibfnamefont {S.}~\bibnamefont
  {Li}}, \bibinfo {author} {\bibfnamefont {M.}~\bibnamefont {Huang}}, \bibinfo
  {author} {\bibfnamefont {H.}~\bibnamefont {Lu}}, \bibinfo {author}
  {\bibfnamefont {N.~J.}\ \bibnamefont {McLaughlin}}, \bibinfo {author}
  {\bibfnamefont {Y.}~\bibnamefont {Xiao}}, \bibinfo {author} {\bibfnamefont
  {J.}~\bibnamefont {Zhou}}, \bibinfo {author} {\bibfnamefont {E.~E.}\
  \bibnamefont {Fullerton}}, \bibinfo {author} {\bibfnamefont {H.}~\bibnamefont
  {Chen}}, \bibinfo {author} {\bibfnamefont {H.}~\bibnamefont {Wang}},\ and\
  \bibinfo {author} {\bibfnamefont {C.~R.}\ \bibnamefont {Du}},\ }\bibfield
  {title} {\enquote {\bibinfo {title} {Nanoscale {{Magnetic Domains}} in
  {{Polycrystalline Mn\textsubscript{3}Sn Films Imaged}} by a {{Scanning
  Single-Spin Magnetometer}}},}\ }\href
  {https://doi.org/10.1021/acs.nanolett.3c01523} {\bibfield  {journal}
  {\bibinfo  {journal} {Nano Letters}\ }\textbf {\bibinfo {volume} {23}},\
  \bibinfo {pages} {5326--5333} (\bibinfo {year} {2023})}\BibitemShut {NoStop}%
\bibitem [{\citenamefont {Catalan}\ and\ \citenamefont
  {Scott}(2009)}]{catalanPhysicsApplicationsBismuth2009}%
  \BibitemOpen
  \bibfield  {author} {\bibinfo {author} {\bibfnamefont {G.}~\bibnamefont
  {Catalan}}\ and\ \bibinfo {author} {\bibfnamefont {J.~F.}\ \bibnamefont
  {Scott}},\ }\bibfield  {title} {\enquote {\bibinfo {title} {Physics and
  {{Applications}} of {{Bismuth Ferrite}}},}\ }\href
  {https://doi.org/10.1002/adma.200802849} {\bibfield  {journal} {\bibinfo
  {journal} {Advanced Materials}\ }\textbf {\bibinfo {volume} {21}},\ \bibinfo
  {pages} {2463--2485} (\bibinfo {year} {2009})}\BibitemShut {NoStop}%
\bibitem [{\citenamefont {Finco}\ \emph {et~al.}(2022)\citenamefont {Finco},
  \citenamefont {Haykal}, \citenamefont {Fusil}, \citenamefont {Kumar},
  \citenamefont {Dufour}, \citenamefont {Forget}, \citenamefont {Colson},
  \citenamefont {Chauleau}, \citenamefont {Viret}, \citenamefont {Jaouen},
  \citenamefont {Garcia},\ and\ \citenamefont
  {Jacques}}]{fincoImagingTopologicalDefects2022}%
  \BibitemOpen
  \bibfield  {author} {\bibinfo {author} {\bibfnamefont {A.}~\bibnamefont
  {Finco}}, \bibinfo {author} {\bibfnamefont {A.}~\bibnamefont {Haykal}},
  \bibinfo {author} {\bibfnamefont {S.}~\bibnamefont {Fusil}}, \bibinfo
  {author} {\bibfnamefont {P.}~\bibnamefont {Kumar}}, \bibinfo {author}
  {\bibfnamefont {P.}~\bibnamefont {Dufour}}, \bibinfo {author} {\bibfnamefont
  {A.}~\bibnamefont {Forget}}, \bibinfo {author} {\bibfnamefont
  {D.}~\bibnamefont {Colson}}, \bibinfo {author} {\bibfnamefont {J.-Y.}\
  \bibnamefont {Chauleau}}, \bibinfo {author} {\bibfnamefont {M.}~\bibnamefont
  {Viret}}, \bibinfo {author} {\bibfnamefont {N.}~\bibnamefont {Jaouen}},
  \bibinfo {author} {\bibfnamefont {V.}~\bibnamefont {Garcia}},\ and\ \bibinfo
  {author} {\bibfnamefont {V.}~\bibnamefont {Jacques}},\ }\bibfield  {title}
  {\enquote {\bibinfo {title} {Imaging {{Topological Defects}} in a
  {{Noncollinear Antiferromagnet}}},}\ }\href
  {https://doi.org/10.1103/PhysRevLett.128.187201} {\bibfield  {journal}
  {\bibinfo  {journal} {Physical Review Letters}\ }\textbf {\bibinfo {volume}
  {128}},\ \bibinfo {pages} {187201} (\bibinfo {year} {2022})}\BibitemShut
  {NoStop}%
\bibitem [{\citenamefont {Haykal}\ \emph {et~al.}(2020)\citenamefont {Haykal},
  \citenamefont {Fischer}, \citenamefont {Akhtar}, \citenamefont {Chauleau},
  \citenamefont {Sando}, \citenamefont {Finco}, \citenamefont {Godel},
  \citenamefont {Birkh{\"o}lzer}, \citenamefont {Carr{\'e}t{\'e}ro},
  \citenamefont {Jaouen}, \citenamefont {Bibes}, \citenamefont {Viret},
  \citenamefont {Fusil}, \citenamefont {Jacques},\ and\ \citenamefont
  {Garcia}}]{haykalAntiferromagneticTexturesBiFeO2020}%
  \BibitemOpen
  \bibfield  {author} {\bibinfo {author} {\bibfnamefont {A.}~\bibnamefont
  {Haykal}}, \bibinfo {author} {\bibfnamefont {J.}~\bibnamefont {Fischer}},
  \bibinfo {author} {\bibfnamefont {W.}~\bibnamefont {Akhtar}}, \bibinfo
  {author} {\bibfnamefont {J.-Y.}\ \bibnamefont {Chauleau}}, \bibinfo {author}
  {\bibfnamefont {D.}~\bibnamefont {Sando}}, \bibinfo {author} {\bibfnamefont
  {A.}~\bibnamefont {Finco}}, \bibinfo {author} {\bibfnamefont
  {F.}~\bibnamefont {Godel}}, \bibinfo {author} {\bibfnamefont {Y.~A.}\
  \bibnamefont {Birkh{\"o}lzer}}, \bibinfo {author} {\bibfnamefont
  {C.}~\bibnamefont {Carr{\'e}t{\'e}ro}}, \bibinfo {author} {\bibfnamefont
  {N.}~\bibnamefont {Jaouen}}, \bibinfo {author} {\bibfnamefont
  {M.}~\bibnamefont {Bibes}}, \bibinfo {author} {\bibfnamefont
  {M.}~\bibnamefont {Viret}}, \bibinfo {author} {\bibfnamefont
  {S.}~\bibnamefont {Fusil}}, \bibinfo {author} {\bibfnamefont
  {V.}~\bibnamefont {Jacques}},\ and\ \bibinfo {author} {\bibfnamefont
  {V.}~\bibnamefont {Garcia}},\ }\bibfield  {title} {\enquote {\bibinfo {title}
  {Antiferromagnetic textures in {{BiFeO}}\textsubscript{3} controlled by
  strain and electric field},}\ }\href
  {https://doi.org/10.1038/s41467-020-15501-8} {\bibfield  {journal} {\bibinfo
  {journal} {Nature Communications}\ }\textbf {\bibinfo {volume} {11}},\
  \bibinfo {pages} {1704} (\bibinfo {year} {2020})}\BibitemShut {NoStop}%
\bibitem [{\citenamefont {Zhong}\ \emph {et~al.}(2022)\citenamefont {Zhong},
  \citenamefont {Finco}, \citenamefont {Fischer}, \citenamefont {Haykal},
  \citenamefont {Bouzehouane}, \citenamefont {Carr{\'e}t{\'e}ro}, \citenamefont
  {Godel}, \citenamefont {Maletinsky}, \citenamefont {Munsch}, \citenamefont
  {Fusil}, \citenamefont {Jacques},\ and\ \citenamefont
  {Garcia}}]{zhongQuantitativeImagingExotic2022}%
  \BibitemOpen
  \bibfield  {author} {\bibinfo {author} {\bibfnamefont {H.}~\bibnamefont
  {Zhong}}, \bibinfo {author} {\bibfnamefont {A.}~\bibnamefont {Finco}},
  \bibinfo {author} {\bibfnamefont {J.}~\bibnamefont {Fischer}}, \bibinfo
  {author} {\bibfnamefont {A.}~\bibnamefont {Haykal}}, \bibinfo {author}
  {\bibfnamefont {K.}~\bibnamefont {Bouzehouane}}, \bibinfo {author}
  {\bibfnamefont {C.}~\bibnamefont {Carr{\'e}t{\'e}ro}}, \bibinfo {author}
  {\bibfnamefont {F.}~\bibnamefont {Godel}}, \bibinfo {author} {\bibfnamefont
  {P.}~\bibnamefont {Maletinsky}}, \bibinfo {author} {\bibfnamefont
  {M.}~\bibnamefont {Munsch}}, \bibinfo {author} {\bibfnamefont
  {S.}~\bibnamefont {Fusil}}, \bibinfo {author} {\bibfnamefont
  {V.}~\bibnamefont {Jacques}},\ and\ \bibinfo {author} {\bibfnamefont
  {V.}~\bibnamefont {Garcia}},\ }\bibfield  {title} {\enquote {\bibinfo {title}
  {Quantitative {{Imaging}} of {{Exotic Antiferromagnetic Spin Cycloids}} in
  {{BiFeO}}\textsubscript{3} {{Thin Films}}},}\ }\href
  {https://doi.org/10.1103/PhysRevApplied.17.044051} {\bibfield  {journal}
  {\bibinfo  {journal} {Physical Review Applied}\ }\textbf {\bibinfo {volume}
  {17}},\ \bibinfo {pages} {044051} (\bibinfo {year} {2022})}\BibitemShut
  {NoStop}%
\bibitem [{\citenamefont {Chauleau}\ \emph {et~al.}(2017)\citenamefont
  {Chauleau}, \citenamefont {Haltz}, \citenamefont {Carr{\'e}t{\'e}ro},
  \citenamefont {Fusil},\ and\ \citenamefont
  {Viret}}]{chauleauMultistimuliManipulationAntiferromagnetic2017}%
  \BibitemOpen
  \bibfield  {author} {\bibinfo {author} {\bibfnamefont {J.-Y.}\ \bibnamefont
  {Chauleau}}, \bibinfo {author} {\bibfnamefont {E.}~\bibnamefont {Haltz}},
  \bibinfo {author} {\bibfnamefont {C.}~\bibnamefont {Carr{\'e}t{\'e}ro}},
  \bibinfo {author} {\bibfnamefont {S.}~\bibnamefont {Fusil}},\ and\ \bibinfo
  {author} {\bibfnamefont {M.}~\bibnamefont {Viret}},\ }\bibfield  {title}
  {\enquote {\bibinfo {title} {Multi-stimuli manipulation of antiferromagnetic
  domains assessed by second-harmonic imaging},}\ }\href
  {https://doi.org/10.1038/nmat4899} {\bibfield  {journal} {\bibinfo  {journal}
  {Nature Materials}\ }\textbf {\bibinfo {volume} {16}},\ \bibinfo {pages}
  {803--807} (\bibinfo {year} {2017})}\BibitemShut {NoStop}%
\bibitem [{\citenamefont {Chauleau}\ \emph {et~al.}(2020)\citenamefont
  {Chauleau}, \citenamefont {Chirac}, \citenamefont {Fusil}, \citenamefont
  {Garcia}, \citenamefont {Akhtar}, \citenamefont {Tranchida}, \citenamefont
  {Thibaudeau}, \citenamefont {Gross}, \citenamefont {Blouzon}, \citenamefont
  {Finco}, \citenamefont {Bibes}, \citenamefont {Dkhil}, \citenamefont
  {Khalyavin}, \citenamefont {Manuel}, \citenamefont {Jacques}, \citenamefont
  {Jaouen},\ and\ \citenamefont
  {Viret}}]{chauleauElectricAntiferromagneticChiral2020}%
  \BibitemOpen
  \bibfield  {author} {\bibinfo {author} {\bibfnamefont {J.-Y.}\ \bibnamefont
  {Chauleau}}, \bibinfo {author} {\bibfnamefont {T.}~\bibnamefont {Chirac}},
  \bibinfo {author} {\bibfnamefont {S.}~\bibnamefont {Fusil}}, \bibinfo
  {author} {\bibfnamefont {V.}~\bibnamefont {Garcia}}, \bibinfo {author}
  {\bibfnamefont {W.}~\bibnamefont {Akhtar}}, \bibinfo {author} {\bibfnamefont
  {J.}~\bibnamefont {Tranchida}}, \bibinfo {author} {\bibfnamefont
  {P.}~\bibnamefont {Thibaudeau}}, \bibinfo {author} {\bibfnamefont
  {I.}~\bibnamefont {Gross}}, \bibinfo {author} {\bibfnamefont
  {C.}~\bibnamefont {Blouzon}}, \bibinfo {author} {\bibfnamefont
  {A.}~\bibnamefont {Finco}}, \bibinfo {author} {\bibfnamefont
  {M.}~\bibnamefont {Bibes}}, \bibinfo {author} {\bibfnamefont
  {B.}~\bibnamefont {Dkhil}}, \bibinfo {author} {\bibfnamefont {D.~D.}\
  \bibnamefont {Khalyavin}}, \bibinfo {author} {\bibfnamefont {P.}~\bibnamefont
  {Manuel}}, \bibinfo {author} {\bibfnamefont {V.}~\bibnamefont {Jacques}},
  \bibinfo {author} {\bibfnamefont {N.}~\bibnamefont {Jaouen}},\ and\ \bibinfo
  {author} {\bibfnamefont {M.}~\bibnamefont {Viret}},\ }\bibfield  {title}
  {\enquote {\bibinfo {title} {Electric and antiferromagnetic chiral textures
  at multiferroic domain walls},}\ }\href
  {https://doi.org/10.1038/s41563-019-0516-z} {\bibfield  {journal} {\bibinfo
  {journal} {Nature Materials}\ }\textbf {\bibinfo {volume} {19}},\ \bibinfo
  {pages} {386--390} (\bibinfo {year} {2020})}\BibitemShut {NoStop}%
\bibitem [{\citenamefont {Belashchenko}\ \emph {et~al.}(2016)\citenamefont
  {Belashchenko}, \citenamefont {Tchernyshyov}, \citenamefont {Kovalev},\ and\
  \citenamefont {Tretiakov}}]{belashchenkomagnetoelectric2016}%
  \BibitemOpen
  \bibfield  {author} {\bibinfo {author} {\bibfnamefont {K.~D.}\ \bibnamefont
  {Belashchenko}}, \bibinfo {author} {\bibfnamefont {O.}~\bibnamefont
  {Tchernyshyov}}, \bibinfo {author} {\bibfnamefont {A.~A.}\ \bibnamefont
  {Kovalev}},\ and\ \bibinfo {author} {\bibfnamefont {O.~A.}\ \bibnamefont
  {Tretiakov}},\ }\bibfield  {title} {\enquote {\bibinfo {title}
  {{Magnetoelectric domain wall dynamics and its implications for
  magnetoelectric memory}},}\ }\href {https://doi.org/10.1063/1.4944996}
  {\bibfield  {journal} {\bibinfo  {journal} {Applied Physics Letters}\
  }\textbf {\bibinfo {volume} {108}} (\bibinfo {year} {2016}),\
  10.1063/1.4944996}\BibitemShut {NoStop}%
\bibitem [{\citenamefont {Haykal}(2020)}]{haykal2020}%
  \BibitemOpen
  \bibfield  {author} {\bibinfo {author} {\bibfnamefont {A.}~\bibnamefont
  {Haykal}},\ }\emph {\bibinfo {title} {Exploring the antiferromagnetic order
  with a single-spin magnetometer}},\ \href {http://www.theses.fr/2020MONTS057}
  {Ph.D. thesis},\ \bibinfo  {school} {Université de Montpellier} (\bibinfo
  {year} {2020})\BibitemShut {NoStop}%
\bibitem [{\citenamefont {Dubois}\ \emph {et~al.}(2022)\citenamefont {Dubois},
  \citenamefont {Broadway}, \citenamefont {Stark}, \citenamefont {Tschudin},
  \citenamefont {Healey}, \citenamefont {Huber}, \citenamefont {Tetienne},
  \citenamefont {Greplova},\ and\ \citenamefont
  {Maletinsky}}]{duboisUntrainedPhysicallyInformed2022}%
  \BibitemOpen
  \bibfield  {author} {\bibinfo {author} {\bibfnamefont {A.}~\bibnamefont
  {Dubois}}, \bibinfo {author} {\bibfnamefont {D.}~\bibnamefont {Broadway}},
  \bibinfo {author} {\bibfnamefont {A.}~\bibnamefont {Stark}}, \bibinfo
  {author} {\bibfnamefont {M.}~\bibnamefont {Tschudin}}, \bibinfo {author}
  {\bibfnamefont {A.}~\bibnamefont {Healey}}, \bibinfo {author} {\bibfnamefont
  {S.}~\bibnamefont {Huber}}, \bibinfo {author} {\bibfnamefont {J.-P.}\
  \bibnamefont {Tetienne}}, \bibinfo {author} {\bibfnamefont {E.}~\bibnamefont
  {Greplova}},\ and\ \bibinfo {author} {\bibfnamefont {P.}~\bibnamefont
  {Maletinsky}},\ }\bibfield  {title} {\enquote {\bibinfo {title} {Untrained
  {{Physically Informed Neural Network}} for {{Image Reconstruction}} of
  {{Magnetic Field Sources}}},}\ }\href
  {https://doi.org/10.1103/PhysRevApplied.18.064076} {\bibfield  {journal}
  {\bibinfo  {journal} {Physical Review Applied}\ }\textbf {\bibinfo {volume}
  {18}},\ \bibinfo {pages} {064076} (\bibinfo {year} {2022})}\BibitemShut
  {NoStop}%
\bibitem [{\citenamefont {Sun}\ \emph {et~al.}(2021)\citenamefont {Sun},
  \citenamefont {Song}, \citenamefont {Anderson}, \citenamefont {Brunner},
  \citenamefont {F{\"o}rster}, \citenamefont {Shalomayeva}, \citenamefont
  {Taniguchi}, \citenamefont {Watanabe}, \citenamefont {Gr{\"a}fe},
  \citenamefont {St{\"o}hr}, \citenamefont {Xu},\ and\ \citenamefont
  {Wrachtrup}}]{sunMagneticDomainsDomain2021a}%
  \BibitemOpen
  \bibfield  {author} {\bibinfo {author} {\bibfnamefont {Q.-C.}\ \bibnamefont
  {Sun}}, \bibinfo {author} {\bibfnamefont {T.}~\bibnamefont {Song}}, \bibinfo
  {author} {\bibfnamefont {E.}~\bibnamefont {Anderson}}, \bibinfo {author}
  {\bibfnamefont {A.}~\bibnamefont {Brunner}}, \bibinfo {author} {\bibfnamefont
  {J.}~\bibnamefont {F{\"o}rster}}, \bibinfo {author} {\bibfnamefont
  {T.}~\bibnamefont {Shalomayeva}}, \bibinfo {author} {\bibfnamefont
  {T.}~\bibnamefont {Taniguchi}}, \bibinfo {author} {\bibfnamefont
  {K.}~\bibnamefont {Watanabe}}, \bibinfo {author} {\bibfnamefont
  {J.}~\bibnamefont {Gr{\"a}fe}}, \bibinfo {author} {\bibfnamefont
  {R.}~\bibnamefont {St{\"o}hr}}, \bibinfo {author} {\bibfnamefont
  {X.}~\bibnamefont {Xu}},\ and\ \bibinfo {author} {\bibfnamefont
  {J.}~\bibnamefont {Wrachtrup}},\ }\bibfield  {title} {\enquote {\bibinfo
  {title} {Magnetic domains and domain wall pinning in atomically thin
  {{CrBr\textsubscript{3}}} revealed by nanoscale imaging},}\ }\href
  {https://doi.org/10.1038/s41467-021-22239-4} {\bibfield  {journal} {\bibinfo
  {journal} {Nature Communications}\ }\textbf {\bibinfo {volume} {12}},\
  \bibinfo {pages} {1989} (\bibinfo {year} {2021})}\BibitemShut {NoStop}%
\bibitem [{\citenamefont {Song}\ \emph {et~al.}(2021)\citenamefont {Song},
  \citenamefont {Sun}, \citenamefont {Anderson}, \citenamefont {Wang},
  \citenamefont {Qian}, \citenamefont {Taniguchi}, \citenamefont {Watanabe},
  \citenamefont {McGuire}, \citenamefont {St{\"o}hr}, \citenamefont {Xiao},
  \citenamefont {Cao}, \citenamefont {Wrachtrup},\ and\ \citenamefont
  {Xu}}]{songDirectVisualizationMagnetic2021}%
  \BibitemOpen
  \bibfield  {author} {\bibinfo {author} {\bibfnamefont {T.}~\bibnamefont
  {Song}}, \bibinfo {author} {\bibfnamefont {Q.-C.}\ \bibnamefont {Sun}},
  \bibinfo {author} {\bibfnamefont {E.}~\bibnamefont {Anderson}}, \bibinfo
  {author} {\bibfnamefont {C.}~\bibnamefont {Wang}}, \bibinfo {author}
  {\bibfnamefont {J.}~\bibnamefont {Qian}}, \bibinfo {author} {\bibfnamefont
  {T.}~\bibnamefont {Taniguchi}}, \bibinfo {author} {\bibfnamefont
  {K.}~\bibnamefont {Watanabe}}, \bibinfo {author} {\bibfnamefont {M.~A.}\
  \bibnamefont {McGuire}}, \bibinfo {author} {\bibfnamefont {R.}~\bibnamefont
  {St{\"o}hr}}, \bibinfo {author} {\bibfnamefont {D.}~\bibnamefont {Xiao}},
  \bibinfo {author} {\bibfnamefont {T.}~\bibnamefont {Cao}}, \bibinfo {author}
  {\bibfnamefont {J.}~\bibnamefont {Wrachtrup}},\ and\ \bibinfo {author}
  {\bibfnamefont {X.}~\bibnamefont {Xu}},\ }\bibfield  {title} {\enquote
  {\bibinfo {title} {Direct visualization of magnetic domains and moir\'e
  magnetism in twisted {{2D}} magnets},}\ }\href
  {https://doi.org/10.1126/science.abj7478} {\bibfield  {journal} {\bibinfo
  {journal} {Science}\ }\textbf {\bibinfo {volume} {374}},\ \bibinfo {pages}
  {1140--1144} (\bibinfo {year} {2021})}\BibitemShut {NoStop}%
\bibitem [{\citenamefont {Flebus}\ \emph {et~al.}(2018)\citenamefont {Flebus},
  \citenamefont {Ochoa}, \citenamefont {Upadhyaya},\ and\ \citenamefont
  {Tserkovnyak}}]{flebusProposalDynamicImaging2018}%
  \BibitemOpen
  \bibfield  {author} {\bibinfo {author} {\bibfnamefont {B.}~\bibnamefont
  {Flebus}}, \bibinfo {author} {\bibfnamefont {H.}~\bibnamefont {Ochoa}},
  \bibinfo {author} {\bibfnamefont {P.}~\bibnamefont {Upadhyaya}},\ and\
  \bibinfo {author} {\bibfnamefont {Y.}~\bibnamefont {Tserkovnyak}},\
  }\bibfield  {title} {\enquote {\bibinfo {title} {Proposal for dynamic imaging
  of antiferromagnetic domain wall via quantum-impurity relaxometry},}\ }\href
  {https://doi.org/10.1103/PhysRevB.98.180409} {\bibfield  {journal} {\bibinfo
  {journal} {Physical Review B}\ }\textbf {\bibinfo {volume} {98}},\ \bibinfo
  {pages} {180409} (\bibinfo {year} {2018})}\BibitemShut {NoStop}%
\bibitem [{\citenamefont {Rollo}\ \emph {et~al.}(2021)\citenamefont {Rollo},
  \citenamefont {Finco}, \citenamefont {Tanos}, \citenamefont {Fabre},
  \citenamefont {Devolder}, \citenamefont {{Robert-Philip}},\ and\
  \citenamefont {Jacques}}]{rolloQuantitativeStudyResponse2021a}%
  \BibitemOpen
  \bibfield  {author} {\bibinfo {author} {\bibfnamefont {M.}~\bibnamefont
  {Rollo}}, \bibinfo {author} {\bibfnamefont {A.}~\bibnamefont {Finco}},
  \bibinfo {author} {\bibfnamefont {R.}~\bibnamefont {Tanos}}, \bibinfo
  {author} {\bibfnamefont {F.}~\bibnamefont {Fabre}}, \bibinfo {author}
  {\bibfnamefont {T.}~\bibnamefont {Devolder}}, \bibinfo {author}
  {\bibfnamefont {I.}~\bibnamefont {{Robert-Philip}}},\ and\ \bibinfo {author}
  {\bibfnamefont {V.}~\bibnamefont {Jacques}},\ }\bibfield  {title} {\enquote
  {\bibinfo {title} {Quantitative study of the response of a single {{NV}}
  defect in diamond to magnetic noise},}\ }\href
  {https://doi.org/10.1103/PhysRevB.103.235418} {\bibfield  {journal} {\bibinfo
   {journal} {Physical Review B}\ }\textbf {\bibinfo {volume} {103}},\ \bibinfo
  {pages} {235418} (\bibinfo {year} {2021})}\BibitemShut {NoStop}%
\bibitem [{\citenamefont {Goldstone}, \citenamefont {Salam},\ and\
  \citenamefont {Weinberg}(1962)}]{goldstoneBrokenSymmetries1962}%
  \BibitemOpen
  \bibfield  {author} {\bibinfo {author} {\bibfnamefont {J.}~\bibnamefont
  {Goldstone}}, \bibinfo {author} {\bibfnamefont {A.}~\bibnamefont {Salam}},\
  and\ \bibinfo {author} {\bibfnamefont {S.}~\bibnamefont {Weinberg}},\
  }\bibfield  {title} {\enquote {\bibinfo {title} {Broken {{Symmetries}}},}\
  }\href {https://doi.org/10.1103/PhysRev.127.965} {\bibfield  {journal}
  {\bibinfo  {journal} {Physical Review}\ }\textbf {\bibinfo {volume} {127}},\
  \bibinfo {pages} {965--970} (\bibinfo {year} {1962})}\BibitemShut {NoStop}%
\bibitem [{\citenamefont {{Garcia-Sanchez}}\ \emph {et~al.}(2015)\citenamefont
  {{Garcia-Sanchez}}, \citenamefont {Borys}, \citenamefont {Soucaille},
  \citenamefont {Adam}, \citenamefont {Stamps},\ and\ \citenamefont
  {Kim}}]{garcia-sanchezNarrowMagnonicWaveguides2015}%
  \BibitemOpen
  \bibfield  {author} {\bibinfo {author} {\bibfnamefont {F.}~\bibnamefont
  {{Garcia-Sanchez}}}, \bibinfo {author} {\bibfnamefont {P.}~\bibnamefont
  {Borys}}, \bibinfo {author} {\bibfnamefont {R.}~\bibnamefont {Soucaille}},
  \bibinfo {author} {\bibfnamefont {J.-P.}\ \bibnamefont {Adam}}, \bibinfo
  {author} {\bibfnamefont {R.~L.}\ \bibnamefont {Stamps}},\ and\ \bibinfo
  {author} {\bibfnamefont {J.-V.}\ \bibnamefont {Kim}},\ }\bibfield  {title}
  {\enquote {\bibinfo {title} {Narrow {{Magnonic Waveguides Based}} on {{Domain
  Walls}}},}\ }\href {https://doi.org/10.1103/PhysRevLett.114.247206}
  {\bibfield  {journal} {\bibinfo  {journal} {Physical Review Letters}\
  }\textbf {\bibinfo {volume} {114}},\ \bibinfo {pages} {247206} (\bibinfo
  {year} {2015})}\BibitemShut {NoStop}%
\bibitem [{\citenamefont {Park}\ and\ \citenamefont
  {Kim}(2021)}]{parkChannelingSpinWaves2021}%
  \BibitemOpen
  \bibfield  {author} {\bibinfo {author} {\bibfnamefont {H.-K.}\ \bibnamefont
  {Park}}\ and\ \bibinfo {author} {\bibfnamefont {S.-K.}\ \bibnamefont {Kim}},\
  }\bibfield  {title} {\enquote {\bibinfo {title} {Channeling of spin waves in
  antiferromagnetic domain walls},}\ }\href
  {https://doi.org/10.1103/PhysRevB.103.214420} {\bibfield  {journal} {\bibinfo
   {journal} {Physical Review B}\ }\textbf {\bibinfo {volume} {103}},\ \bibinfo
  {pages} {214420} (\bibinfo {year} {2021})}\BibitemShut {NoStop}%
\bibitem [{\citenamefont {Sluka}\ \emph {et~al.}(2019)\citenamefont {Sluka},
  \citenamefont {Schneider}, \citenamefont {Gallardo}, \citenamefont
  {K{\'a}kay}, \citenamefont {Weigand}, \citenamefont {Warnatz}, \citenamefont
  {Mattheis}, \citenamefont {{Rold{\'a}n-Molina}}, \citenamefont {Landeros},
  \citenamefont {Tiberkevich}, \citenamefont {Slavin}, \citenamefont
  {Sch{\"u}tz}, \citenamefont {Erbe}, \citenamefont {Deac}, \citenamefont
  {Lindner}, \citenamefont {Raabe}, \citenamefont {Fassbender},\ and\
  \citenamefont {Wintz}}]{slukaEmissionPropagation1D2019}%
  \BibitemOpen
  \bibfield  {author} {\bibinfo {author} {\bibfnamefont {V.}~\bibnamefont
  {Sluka}}, \bibinfo {author} {\bibfnamefont {T.}~\bibnamefont {Schneider}},
  \bibinfo {author} {\bibfnamefont {R.~A.}\ \bibnamefont {Gallardo}}, \bibinfo
  {author} {\bibfnamefont {A.}~\bibnamefont {K{\'a}kay}}, \bibinfo {author}
  {\bibfnamefont {M.}~\bibnamefont {Weigand}}, \bibinfo {author} {\bibfnamefont
  {T.}~\bibnamefont {Warnatz}}, \bibinfo {author} {\bibfnamefont
  {R.}~\bibnamefont {Mattheis}}, \bibinfo {author} {\bibfnamefont
  {A.}~\bibnamefont {{Rold{\'a}n-Molina}}}, \bibinfo {author} {\bibfnamefont
  {P.}~\bibnamefont {Landeros}}, \bibinfo {author} {\bibfnamefont
  {V.}~\bibnamefont {Tiberkevich}}, \bibinfo {author} {\bibfnamefont
  {A.}~\bibnamefont {Slavin}}, \bibinfo {author} {\bibfnamefont
  {G.}~\bibnamefont {Sch{\"u}tz}}, \bibinfo {author} {\bibfnamefont
  {A.}~\bibnamefont {Erbe}}, \bibinfo {author} {\bibfnamefont {A.}~\bibnamefont
  {Deac}}, \bibinfo {author} {\bibfnamefont {J.}~\bibnamefont {Lindner}},
  \bibinfo {author} {\bibfnamefont {J.}~\bibnamefont {Raabe}}, \bibinfo
  {author} {\bibfnamefont {J.}~\bibnamefont {Fassbender}},\ and\ \bibinfo
  {author} {\bibfnamefont {S.}~\bibnamefont {Wintz}},\ }\bibfield  {title}
  {\enquote {\bibinfo {title} {Emission and propagation of {{1D}} and {{2D}}
  spin waves with nanoscale wavelengths in anisotropic spin textures},}\ }\href
  {https://doi.org/10.1038/s41565-019-0383-4} {\bibfield  {journal} {\bibinfo
  {journal} {Nature Nanotechnology}\ }\textbf {\bibinfo {volume} {14}},\
  \bibinfo {pages} {328--333} (\bibinfo {year} {2019})}\BibitemShut {NoStop}%
\bibitem [{\citenamefont {Dr{\'e}au}\ \emph {et~al.}(2011)\citenamefont
  {Dr{\'e}au}, \citenamefont {Lesik}, \citenamefont {Rondin}, \citenamefont
  {Spinicelli}, \citenamefont {Arcizet}, \citenamefont {Roch},\ and\
  \citenamefont {Jacques}}]{dreauAvoidingPowerBroadening2011}%
  \BibitemOpen
  \bibfield  {author} {\bibinfo {author} {\bibfnamefont {A.}~\bibnamefont
  {Dr{\'e}au}}, \bibinfo {author} {\bibfnamefont {M.}~\bibnamefont {Lesik}},
  \bibinfo {author} {\bibfnamefont {L.}~\bibnamefont {Rondin}}, \bibinfo
  {author} {\bibfnamefont {P.}~\bibnamefont {Spinicelli}}, \bibinfo {author}
  {\bibfnamefont {O.}~\bibnamefont {Arcizet}}, \bibinfo {author} {\bibfnamefont
  {J.-F.}\ \bibnamefont {Roch}},\ and\ \bibinfo {author} {\bibfnamefont
  {V.}~\bibnamefont {Jacques}},\ }\bibfield  {title} {\enquote {\bibinfo
  {title} {Avoiding power broadening in optically detected magnetic resonance
  of single {{NV}} defects for enhanced {{DC}} magnetic field sensitivity},}\
  }\href {https://doi.org/10.1103/PhysRevB.84.195204} {\bibfield  {journal}
  {\bibinfo  {journal} {Physical Review B}\ }\textbf {\bibinfo {volume} {84}},\
  \bibinfo {pages} {195204} (\bibinfo {year} {2011})}\BibitemShut {NoStop}%
\bibitem [{\citenamefont {Huxter}\ \emph {et~al.}(2022)\citenamefont {Huxter},
  \citenamefont {Palm}, \citenamefont {Davis}, \citenamefont {Welter},
  \citenamefont {Lambert}, \citenamefont {Trassin},\ and\ \citenamefont
  {Degen}}]{huxterScanningGradiometrySingle2022}%
  \BibitemOpen
  \bibfield  {author} {\bibinfo {author} {\bibfnamefont {W.~S.}\ \bibnamefont
  {Huxter}}, \bibinfo {author} {\bibfnamefont {M.~L.}\ \bibnamefont {Palm}},
  \bibinfo {author} {\bibfnamefont {M.~L.}\ \bibnamefont {Davis}}, \bibinfo
  {author} {\bibfnamefont {P.}~\bibnamefont {Welter}}, \bibinfo {author}
  {\bibfnamefont {C.-H.}\ \bibnamefont {Lambert}}, \bibinfo {author}
  {\bibfnamefont {M.}~\bibnamefont {Trassin}},\ and\ \bibinfo {author}
  {\bibfnamefont {C.~L.}\ \bibnamefont {Degen}},\ }\bibfield  {title} {\enquote
  {\bibinfo {title} {Scanning gradiometry with a single spin quantum
  magnetometer},}\ }\href {https://doi.org/10.1038/s41467-022-31454-6}
  {\bibfield  {journal} {\bibinfo  {journal} {Nature Communications}\ }\textbf
  {\bibinfo {volume} {13}},\ \bibinfo {pages} {3761} (\bibinfo {year}
  {2022})}\BibitemShut {NoStop}%
\bibitem [{\citenamefont {Vool}\ \emph {et~al.}(2021)\citenamefont {Vool},
  \citenamefont {Hamo}, \citenamefont {Varnavides}, \citenamefont {Wang},
  \citenamefont {Zhou}, \citenamefont {Kumar}, \citenamefont {Dovzhenko},
  \citenamefont {Qiu}, \citenamefont {Garcia}, \citenamefont {Pierce},
  \citenamefont {Gooth}, \citenamefont {Anikeeva}, \citenamefont {Felser},
  \citenamefont {Narang},\ and\ \citenamefont
  {Yacoby}}]{voolImagingPhononmediatedHydrodynamic2021}%
  \BibitemOpen
  \bibfield  {author} {\bibinfo {author} {\bibfnamefont {U.}~\bibnamefont
  {Vool}}, \bibinfo {author} {\bibfnamefont {A.}~\bibnamefont {Hamo}}, \bibinfo
  {author} {\bibfnamefont {G.}~\bibnamefont {Varnavides}}, \bibinfo {author}
  {\bibfnamefont {Y.}~\bibnamefont {Wang}}, \bibinfo {author} {\bibfnamefont
  {T.~X.}\ \bibnamefont {Zhou}}, \bibinfo {author} {\bibfnamefont
  {N.}~\bibnamefont {Kumar}}, \bibinfo {author} {\bibfnamefont
  {Y.}~\bibnamefont {Dovzhenko}}, \bibinfo {author} {\bibfnamefont
  {Z.}~\bibnamefont {Qiu}}, \bibinfo {author} {\bibfnamefont {C.~A.~C.}\
  \bibnamefont {Garcia}}, \bibinfo {author} {\bibfnamefont {A.~T.}\
  \bibnamefont {Pierce}}, \bibinfo {author} {\bibfnamefont {J.}~\bibnamefont
  {Gooth}}, \bibinfo {author} {\bibfnamefont {P.}~\bibnamefont {Anikeeva}},
  \bibinfo {author} {\bibfnamefont {C.}~\bibnamefont {Felser}}, \bibinfo
  {author} {\bibfnamefont {P.}~\bibnamefont {Narang}},\ and\ \bibinfo {author}
  {\bibfnamefont {A.}~\bibnamefont {Yacoby}},\ }\bibfield  {title} {\enquote
  {\bibinfo {title} {Imaging phonon-mediated hydrodynamic flow in
  {{WTe\textsubscript{2}}}},}\ }\href
  {https://doi.org/10.1038/s41567-021-01341-w} {\bibfield  {journal} {\bibinfo
  {journal} {Nature Physics}\ }\textbf {\bibinfo {volume} {17}},\ \bibinfo
  {pages} {1216--1220} (\bibinfo {year} {2021})}\BibitemShut {NoStop}%
\bibitem [{\citenamefont {Palm}\ \emph {et~al.}(2022)\citenamefont {Palm},
  \citenamefont {Huxter}, \citenamefont {Welter}, \citenamefont {Ernst},
  \citenamefont {Scheidegger}, \citenamefont {Diesch}, \citenamefont {Chang},
  \citenamefont {Rickhaus}, \citenamefont {Taniguchi}, \citenamefont
  {Watanabe}, \citenamefont {Ensslin},\ and\ \citenamefont
  {Degen}}]{palmImagingSubmicroampereCurrents2022}%
  \BibitemOpen
  \bibfield  {author} {\bibinfo {author} {\bibfnamefont {M.}~\bibnamefont
  {Palm}}, \bibinfo {author} {\bibfnamefont {W.}~\bibnamefont {Huxter}},
  \bibinfo {author} {\bibfnamefont {P.}~\bibnamefont {Welter}}, \bibinfo
  {author} {\bibfnamefont {S.}~\bibnamefont {Ernst}}, \bibinfo {author}
  {\bibfnamefont {P.}~\bibnamefont {Scheidegger}}, \bibinfo {author}
  {\bibfnamefont {S.}~\bibnamefont {Diesch}}, \bibinfo {author} {\bibfnamefont
  {K.}~\bibnamefont {Chang}}, \bibinfo {author} {\bibfnamefont
  {P.}~\bibnamefont {Rickhaus}}, \bibinfo {author} {\bibfnamefont
  {T.}~\bibnamefont {Taniguchi}}, \bibinfo {author} {\bibfnamefont
  {K.}~\bibnamefont {Watanabe}}, \bibinfo {author} {\bibfnamefont
  {K.}~\bibnamefont {Ensslin}},\ and\ \bibinfo {author} {\bibfnamefont
  {C.}~\bibnamefont {Degen}},\ }\bibfield  {title} {\enquote {\bibinfo {title}
  {Imaging of {{Submicroampere Currents}} in {{Bilayer Graphene Using}} a
  {{Scanning Diamond Magnetometer}}},}\ }\href
  {https://doi.org/10.1103/PhysRevApplied.17.054008} {\bibfield  {journal}
  {\bibinfo  {journal} {Physical Review Applied}\ }\textbf {\bibinfo {volume}
  {17}},\ \bibinfo {pages} {054008} (\bibinfo {year} {2022})}\BibitemShut
  {NoStop}%
\bibitem [{\citenamefont {Hong}\ \emph {et~al.}(2012)\citenamefont {Hong},
  \citenamefont {Grinolds}, \citenamefont {Maletinsky}, \citenamefont
  {Walsworth}, \citenamefont {Lukin},\ and\ \citenamefont
  {Yacoby}}]{hongCoherentMechanicalControl2012}%
  \BibitemOpen
  \bibfield  {author} {\bibinfo {author} {\bibfnamefont {S.}~\bibnamefont
  {Hong}}, \bibinfo {author} {\bibfnamefont {M.~S.}\ \bibnamefont {Grinolds}},
  \bibinfo {author} {\bibfnamefont {P.}~\bibnamefont {Maletinsky}}, \bibinfo
  {author} {\bibfnamefont {R.~L.}\ \bibnamefont {Walsworth}}, \bibinfo {author}
  {\bibfnamefont {M.~D.}\ \bibnamefont {Lukin}},\ and\ \bibinfo {author}
  {\bibfnamefont {A.}~\bibnamefont {Yacoby}},\ }\bibfield  {title} {\enquote
  {\bibinfo {title} {Coherent, {{Mechanical Control}} of a {{Single Electronic
  Spin}}},}\ }\href {https://doi.org/10.1021/nl300775c} {\bibfield  {journal}
  {\bibinfo  {journal} {Nano Letters}\ }\textbf {\bibinfo {volume} {12}},\
  \bibinfo {pages} {3920--3924} (\bibinfo {year} {2012})}\BibitemShut {NoStop}%
\bibitem [{\citenamefont {Grinolds}\ \emph {et~al.}(2013)\citenamefont
  {Grinolds}, \citenamefont {Hong}, \citenamefont {Maletinsky}, \citenamefont
  {Luan}, \citenamefont {Lukin}, \citenamefont {Walsworth},\ and\ \citenamefont
  {Yacoby}}]{grinoldsNanoscaleMagneticImaging2013}%
  \BibitemOpen
  \bibfield  {author} {\bibinfo {author} {\bibfnamefont {M.~S.}\ \bibnamefont
  {Grinolds}}, \bibinfo {author} {\bibfnamefont {S.}~\bibnamefont {Hong}},
  \bibinfo {author} {\bibfnamefont {P.}~\bibnamefont {Maletinsky}}, \bibinfo
  {author} {\bibfnamefont {L.}~\bibnamefont {Luan}}, \bibinfo {author}
  {\bibfnamefont {M.~D.}\ \bibnamefont {Lukin}}, \bibinfo {author}
  {\bibfnamefont {R.~L.}\ \bibnamefont {Walsworth}},\ and\ \bibinfo {author}
  {\bibfnamefont {A.}~\bibnamefont {Yacoby}},\ }\bibfield  {title} {\enquote
  {\bibinfo {title} {Nanoscale magnetic imaging of a single electron spin under
  ambient conditions},}\ }\href {https://doi.org/10.1038/nphys2543} {\bibfield
  {journal} {\bibinfo  {journal} {Nature Physics}\ }\textbf {\bibinfo {volume}
  {9}},\ \bibinfo {pages} {215--219} (\bibinfo {year} {2013})}\BibitemShut
  {NoStop}%
\bibitem [{\citenamefont {H{\"a}nke}\ \emph {et~al.}(2005)\citenamefont
  {H{\"a}nke}, \citenamefont {Krause}, \citenamefont {{Berbil-Bautista}},
  \citenamefont {Bode}, \citenamefont {Wiesendanger}, \citenamefont {Wagner},
  \citenamefont {Lott},\ and\ \citenamefont
  {Schreyer}}]{hankeAbsenceSpinflipTransition2005}%
  \BibitemOpen
  \bibfield  {author} {\bibinfo {author} {\bibfnamefont {T.}~\bibnamefont
  {H{\"a}nke}}, \bibinfo {author} {\bibfnamefont {S.}~\bibnamefont {Krause}},
  \bibinfo {author} {\bibfnamefont {L.}~\bibnamefont {{Berbil-Bautista}}},
  \bibinfo {author} {\bibfnamefont {M.}~\bibnamefont {Bode}}, \bibinfo {author}
  {\bibfnamefont {R.}~\bibnamefont {Wiesendanger}}, \bibinfo {author}
  {\bibfnamefont {V.}~\bibnamefont {Wagner}}, \bibinfo {author} {\bibfnamefont
  {D.}~\bibnamefont {Lott}},\ and\ \bibinfo {author} {\bibfnamefont
  {A.}~\bibnamefont {Schreyer}},\ }\bibfield  {title} {\enquote {\bibinfo
  {title} {Absence of spin-flip transition at the {{Cr}}(001) surface:{{A}}
  combined spin-polarized scanning tunneling microscopy and neutron scattering
  study},}\ }\href {https://doi.org/10.1103/PhysRevB.71.184407} {\bibfield
  {journal} {\bibinfo  {journal} {Physical Review B}\ }\textbf {\bibinfo
  {volume} {71}},\ \bibinfo {pages} {184407} (\bibinfo {year}
  {2005})}\BibitemShut {NoStop}%
\bibitem [{\citenamefont
  {Wiesendanger}(2009)}]{wiesendangerSpinMappingNanoscale2009}%
  \BibitemOpen
  \bibfield  {author} {\bibinfo {author} {\bibfnamefont {R.}~\bibnamefont
  {Wiesendanger}},\ }\bibfield  {title} {\enquote {\bibinfo {title} {Spin
  mapping at the nanoscale and atomic scale},}\ }\href
  {https://doi.org/10.1103/RevModPhys.81.1495} {\bibfield  {journal} {\bibinfo
  {journal} {Reviews of Modern Physics}\ }\textbf {\bibinfo {volume} {81}},\
  \bibinfo {pages} {1495--1550} (\bibinfo {year} {2009})}\BibitemShut {NoStop}%
\bibitem [{\citenamefont {Dolde}\ \emph {et~al.}(2011)\citenamefont {Dolde},
  \citenamefont {Fedder}, \citenamefont {Doherty}, \citenamefont {N{\"o}bauer},
  \citenamefont {Rempp}, \citenamefont {Balasubramanian}, \citenamefont {Wolf},
  \citenamefont {Reinhard}, \citenamefont {Hollenberg}, \citenamefont
  {Jelezko},\ and\ \citenamefont
  {Wrachtrup}}]{doldeElectricfieldSensingUsing2011}%
  \BibitemOpen
  \bibfield  {author} {\bibinfo {author} {\bibfnamefont {F.}~\bibnamefont
  {Dolde}}, \bibinfo {author} {\bibfnamefont {H.}~\bibnamefont {Fedder}},
  \bibinfo {author} {\bibfnamefont {M.~W.}\ \bibnamefont {Doherty}}, \bibinfo
  {author} {\bibfnamefont {T.}~\bibnamefont {N{\"o}bauer}}, \bibinfo {author}
  {\bibfnamefont {F.}~\bibnamefont {Rempp}}, \bibinfo {author} {\bibfnamefont
  {G.}~\bibnamefont {Balasubramanian}}, \bibinfo {author} {\bibfnamefont
  {T.}~\bibnamefont {Wolf}}, \bibinfo {author} {\bibfnamefont {F.}~\bibnamefont
  {Reinhard}}, \bibinfo {author} {\bibfnamefont {L.~C.~L.}\ \bibnamefont
  {Hollenberg}}, \bibinfo {author} {\bibfnamefont {F.}~\bibnamefont
  {Jelezko}},\ and\ \bibinfo {author} {\bibfnamefont {J.}~\bibnamefont
  {Wrachtrup}},\ }\bibfield  {title} {\enquote {\bibinfo {title}
  {Electric-field sensing using single diamond spins},}\ }\href
  {https://doi.org/10.1038/nphys1969} {\bibfield  {journal} {\bibinfo
  {journal} {Nature Physics}\ }\textbf {\bibinfo {volume} {7}},\ \bibinfo
  {pages} {459--463} (\bibinfo {year} {2011})}\BibitemShut {NoStop}%
\bibitem [{\citenamefont {Qiu}\ \emph {et~al.}(2022)\citenamefont {Qiu},
  \citenamefont {Hamo}, \citenamefont {Vool}, \citenamefont {Zhou},\ and\
  \citenamefont {Yacoby}}]{qiuNanoscaleElectricField2022a}%
  \BibitemOpen
  \bibfield  {author} {\bibinfo {author} {\bibfnamefont {Z.}~\bibnamefont
  {Qiu}}, \bibinfo {author} {\bibfnamefont {A.}~\bibnamefont {Hamo}}, \bibinfo
  {author} {\bibfnamefont {U.}~\bibnamefont {Vool}}, \bibinfo {author}
  {\bibfnamefont {T.~X.}\ \bibnamefont {Zhou}},\ and\ \bibinfo {author}
  {\bibfnamefont {A.}~\bibnamefont {Yacoby}},\ }\bibfield  {title} {\enquote
  {\bibinfo {title} {Nanoscale electric field imaging with an ambient scanning
  quantum sensor microscope},}\ }\href
  {https://doi.org/10.1038/s41534-022-00622-3} {\bibfield  {journal} {\bibinfo
  {journal} {npj Quantum Information}\ }\textbf {\bibinfo {volume} {8}},\
  \bibinfo {pages} {1--7} (\bibinfo {year} {2022})}\BibitemShut {NoStop}%
\bibitem [{\citenamefont {Huxter}\ \emph {et~al.}(2023)\citenamefont {Huxter},
  \citenamefont {Sarott}, \citenamefont {Trassin},\ and\ \citenamefont
  {Degen}}]{huxterImagingFerroelectricDomains2023a}%
  \BibitemOpen
  \bibfield  {author} {\bibinfo {author} {\bibfnamefont {W.~S.}\ \bibnamefont
  {Huxter}}, \bibinfo {author} {\bibfnamefont {M.~F.}\ \bibnamefont {Sarott}},
  \bibinfo {author} {\bibfnamefont {M.}~\bibnamefont {Trassin}},\ and\ \bibinfo
  {author} {\bibfnamefont {C.~L.}\ \bibnamefont {Degen}},\ }\bibfield  {title}
  {\enquote {\bibinfo {title} {Imaging ferroelectric domains with a single-spin
  scanning quantum sensor},}\ }\href
  {https://doi.org/10.1038/s41567-022-01921-4} {\bibfield  {journal} {\bibinfo
  {journal} {Nature Physics}\ }\textbf {\bibinfo {volume} {19}},\ \bibinfo
  {pages} {644--648} (\bibinfo {year} {2023})}\BibitemShut {NoStop}%
\end{thebibliography}
